\documentclass[a4paper,10pt]{article}

\usepackage[hyphens]{url}
\usepackage{mathtools,amssymb}
\usepackage[usenames,dvipsnames,table]{xcolor}
\usepackage[titletoc]{appendix}
\usepackage{mathtools,amssymb}
\usepackage[makeroom]{cancel}
\usepackage{graphicx}
\usepackage{epstopdf}
\usepackage{mathrsfs}
\usepackage{enumitem}
\usepackage{stackrel}
\usepackage{float}
\usepackage{booktabs}
\usepackage[normalem]{ulem}  %%% to use strikeout \sout{unwanted text}
\usepackage{cancel} %%% to use the user-defined cancel command in mathmode
\usepackage{hyperref}
\usepackage[final]{showkeys} %%% should be loaded after hyperref %%% add final to avoid printing the labels
\usepackage{longtable}
\usepackage{physics}
\usepackage{empheq}
\usepackage{cite} %to generate compact citations
\usepackage{comment}
\usepackage{tikz}
\usetikzlibrary{decorations.pathmorphing}
\usetikzlibrary{decorations.pathreplacing}
\usetikzlibrary{decorations.text}
\usepackage{braket}
\usepackage{fullpage}
\usepackage{subcaption}
\usepackage{pgfplots}
\DeclareMathAlphabet{\mathpzc}{OT1}{pzc}{m}{it}

\allowdisplaybreaks
\allowbreak

\definecolor{refkey}{gray}{0.75}
\definecolor{labelkey}{RGB}{155,48,48}
\renewcommand*\showkeyslabelformat[1]{%
	\fbox{\parbox[t]{0.8\marginparwidth}{\raggedright\normalfont\scriptsize\url{#1}}}}

\usepackage{etoolbox}
\makeatletter
\patchcmd{\hyper@makecurrent}{%
	\ifx\Hy@param\Hy@chapterstring
	\let\Hy@param\Hy@chapapp
	\fi
}{%
	\iftoggle{inappendix}{%true-branch
		\@checkappendixparam{chapter}%
		\@checkappendixparam{section}%
		\@checkappendixparam{subsection}%
		\@checkappendixparam{subsubsection}%
		\@checkappendixparam{paragraph}%
		\@checkappendixparam{subparagraph}%
	}{}%
}{}{ \errmessage{failed to patch}}

\newcommand*{\@checkappendixparam}[1]{%
	\def\@checkappendixparamtmp{#1}%
	\ifx\Hy@param\@checkappendixparamtmp
	\let\Hy@param\Hy@appendixstring
	\fi
}
\makeatletter

\newtoggle{inappendix}
\togglefalse{inappendix}

\apptocmd{\appendix}{\toggletrue{inappendix}}{}{\errmessage{failed to patch}}
\apptocmd{\subappendices}{\toggletrue{inappendix}}{}{\errmessage{failed to patch}}
\newcommand{\lsim}{\mathrel{\hbox{\rlap{\lower .55ex
				\hbox{$\sim$}} \kern-.3em \raise.4ex \hbox{$<$}}}}
\newcommand{\gsim}{\mathrel{\hbox{\rlap{\lower.55ex
				\hbox{$\sim$}} \kern-.3em \raise.4ex \hbox{$<$}}}}

\begin{document}
	
%%%%  Gautam's definitions    %%%%%%%
	
\newcommand{\partiald}[2]{\dfrac{\partial #1}{\partial #2}}
\newcommand{\be}{\begin{equation}}
\newcommand{\ee}{\end{equation}}
\newcommand{\f}{\frac}
\newcommand{\s}{\sqrt}
\newcommand{\lm}{\mathcal{L}}
\newcommand{\wm}{\mathcal{W}}
\newcommand{\om}{\omega}
\newcommand{\bea}{\begin{eqnarray}}
\newcommand{\eea}{\end{eqnarray}}
\newcommand{\ba}{\begin{align}}
\newcommand{\ea}{\end{align}}
\newcommand{\ep}{\epsilon}
\newcommand{\h}{\hat}
\def\ads{AdS$_{\text{2}}$~}
\def\gap#1{\vspace{#1 ex}}
\def\be{\begin{equation}}
\def\ee{\end{equation}}
\def\bal{\begin{array}{l}}
\def\ba#1{\begin{array}{#1}}  %% e.g. \ba{cc}
\def\ea{\end{array}}
\def\bea{\begin{eqnarray}}
\def\eea{\end{eqnarray}}
\def\beas{\begin{eqnarray*}}
\def\eeas{\end{eqnarray*}}
\def\del{\partial}
\def\eq#1{(\ref{#1})}
\def\fig#1{Fig \ref{#1}} 
\def\re#1{{\bf #1}}
\def\bull{$\bullet$}
\def\nn{\nonumber}
\def\ub{\underbar}
\def\nl{\hfill\break}
\def\ni{\noindent}
\def\bibi{\bibitem}
\def\vev#1{\langle #1 \rangle} 
\def\mattwo#1#2#3#4{\left(\begin{array}{cc}#1&#2\\#3&#4\end{array}\right)} 
\def\tgen#1{T^{#1}}
\def\half{\frac12}
\def\floor#1{{\lfloor #1 \rfloor}}
\def\ceil#1{{\lceil #1 \rceil}}
	
\def\Tr{{\rm Tr}}
		
\def\mysec#1{\gap1\ni{\bf #1}\gap1}
\def\mycap#1{\begin{quote}{\footnotesize #1}\end{quote}}
		
\def\Red#1{{\color{red}#1}}
\def\blue#1{{\color{blue}#1}}
\def\Om{\Omega}
\def\a{\alpha}
\def\b{\beta}
\def\l{\lambda}
\def\g{\gamma}
\def\ep{\epsilon}
\def\Si{\Sigma}
\def\p{\phi}
\def\z{\zeta}

\def\lan{\langle}
\def\ran{\rangle}

\def\bit{\begin{item}}
\def\eit{\end{item}}
\def\benu{\begin{enumerate}}
\def\eenu{\end{enumerate}}
\def\fr#1#2{{\frac{#1}{#2}}}
\def\gsq{{{\tilde g}^2}}
	
\def\tr{{\rm tr}}
\def\intk#1{{\int\kern-#1pt}}

%%%%%%%%%%%%%%%%%%%%%%%%%%%%%%%%%%%%%%%%%%%%%%%%%%%%%%%%%%%%%%%%%%%%%%%%%%%%%%%%%%%%%%%%%%%

\newcommand{\com}{\textcolor{red}}
\newcommand{\new}[1]{{\color[rgb]{1.0,0.,0}#1}}
\newcommand{\old}[1]{{\color[rgb]{0.7,0,0.7}\sout{#1}}}%%%End Comment commands
		
\renewcommand{\real}{\ensuremath{\mathbb{R}}}
%%%%%% COMMAND FOR A BIGGER DOT
		
\newcommand*{\Cdot}[1][1.25]{%
	\mathpalette{\CdotAux{#1}}\cdot%
		}
		\newdimen\CdotAxis
		\newcommand*{\CdotAux}[3]{%
			{%
				\settoheight\CdotAxis{$#2\vcenter{}$}%
				\sbox0{%
					\raisebox\CdotAxis{%
						\scalebox{#1}{%
							\raisebox{-\CdotAxis}{%
								$\mathsurround=0pt #2#3$%
							}%
						}%
					}%
				}%
				% Remove depth that arises from scaling.
				\dp0=0pt %
				% Decrease scaled height.
				\sbox2{$#2\bullet$}%
				\ifdim\ht2<\ht0 %
				\ht0=\ht2 %
				\fi
				% Use the same width as the original \cdot.
				\sbox2{$\mathsurround=0pt #2#3$}%
				\hbox to \wd2{\hss\usebox{0}\hss}%
			}%
		}
		
%%%%%% COMMAND TO CANCEL INSIDE THE MATHMODE (REQUIRES PACKAGE CANCEL)
\newcommand\hcancel[2][black]{\setbox0=\hbox{$#2$}%
\rlap{\raisebox{.45\ht0}{\textcolor{#1}{\rule{\wd0}{1pt}}}}#2} 
		
\renewcommand{\arraystretch}{2.5}%
\renewcommand{\floatpagefraction}{.8}%
	
%%%%%%%%%%  This paper %%%%%%%%%%%%%%%
\def\newthing{\marginpar{{\color{red}****}}}
\def\doubt{{\color{red}{***}}}
\reversemarginpar
		
%%%%%%%%%%%%%%  Adwait's defs   %%%%%%%%%%%%%%%%%%%
		
\def\tu{\tau}
\def\ze{z}
\def\d{\partial}
\def\L{\varphi}  % cf. gauge transformation
		
\DeclareRobustCommand{\rchi}{{\mathpalette\irchi\relax}}
\newcommand{\irchi}[2]{\raisebox{\depth}{$#1\chi$}}

%%%%%%%%%%%%%%%%%%%%%%%%%%%%%%%%%%%%%

\hypersetup{pageanchor=false}
\begin{titlepage}
%\thispagestyle{empty}
%\begin{flushright}	
%	{\color{red} TIFR/TH/22-43}
%\end{flushright}
			
\vspace{.4cm}
\begin{center}
\noindent{\Large \bf{A Microscopic Model of Black Hole Evaporation\\ in Two Dimensions}}\\
\vspace{1cm} 
Adwait Gaikwad$^{a,b}$\footnote{Part of the work was completed when the author was at {\it b}.}\footnote{adwaitgaikwad@gmail.com}, 
Anurag Kaushal$^{b,c}$\footnote{anuragkaushal314@gmail.com},
Gautam Mandal$^b$\footnote{mandal@theory.tifr.res.in}, and 
Spenta R. Wadia$^c$\footnote{spenta.wadia@icts.res.in}
\vspace{.3cm}
\begin{center}
{\it a. School of Physics and Astronomy,}\\
{\it Tel Aviv University, Ramat Aviv 69978, Israel}\\
\vspace{.3cm}
{\it b. Department of Theoretical Physics}\\
{\it Tata Institute of Fundamental Research, Mumbai 400005, India.}\\
\vspace{.3cm}
{\it c. International Centre for Theoretical Sciences}\\
{\it Tata Institute of Fundamental Research, Shivakote, Bengaluru 560089, India.}
\end{center}
				
\gap2
\end{center}
%%%%%%%%%%%%%%%%%%%%%%%%%%%%%%%
%%%%%%%%%%%%  ABSTRACT %%%%%%%%%%%%
%%%%%%%%%%%%%%%%%%%%%%%%%%%%%%%
\begin{abstract}

We present a microscopic model of black hole (BH) `evaporation' in asymptotically $AdS_2$ spacetimes dual to the low energy sector of the SYK model. To describe evaporation, the SYK model is coupled to a bath comprising of $N_f$ free scalar fields $\Phi_i$. We consider a linear combination of couplings of the form $O_{SYK}(t)\sum_i\Phi_i(0,t)$, where $O_{SYK}$ involves products of the Kourkoulou-Maldacena operator $i J/N\sum_{k=1}^{N/2}s'_k\psi_{2k-1}(t)\psi_{2k}(t)$ specified by a spin vector $s'$. We discuss the time evolution of a product of (i) a pure state of the SYK system, namely a BH microstate characterized by a spin vector $s$ and an effective BH temperature $T_{BH}$, and (ii) a Calabrese-Cardy state of the bath characterized by an effective temperature $T_{bath}$. We take $T_{bath}\ll T_{BH}$, and $T_{BH}$ much lower than the characteristic UV scale $J$ of the SYK model, allowing a description in terms of the time reparameterization mode. Tracing over the bath degrees of freedom leads to a Feynman-Vernon type effective action for the SYK model, which we study in the low energy limit. The leading large $N$ behaviour of the time reparameterization mode is found, as well as the $O(1/\sqrt N)$ fluctuations. The latter are characterized by a non-Markovian non-linear stochastic differential equation with non-local Gaussian noise. In a restricted range of couplings, we find two classes of solutions which asymptotically approach (a) a BH at a lower temperature, and (b) a horizonless geometry. We identify these with partial and complete BH evaporation, respectively. Importantly, the asymptotic solution in both cases involves the scalar product of the spin vectors $s.s'$, which carries some information about the initial state. By repeating the dynamical process $O(N^2)$ times with different choices of the spin vector $s'$, one can in principle reconstruct the initial BH microstate.

\end{abstract}
			
%\gap5

\end{titlepage}
		
\pagenumbering{roman}
\tableofcontents
		%\enlargethispage{1000pt}
		%\pagebreak
\pagenumbering{arabic}
\setcounter{page}{1}
		
%%%%%%%%%%%%%%%%%%%%%%%%%%%%%%%
%%%%%%%%% INTRODUCTION %%%%%%%%%%
%%%%%%%%%%%%%%%%%%%%%%%%%%%%%%%
\section{Introduction and Summary}\label{sec:intro}
		
In the past few years the Sachdev-Ye-Kitaev (SYK) model\cite{Sachdev:1992fk, Kitaev-talks:2015, Maldacena:2016hyu} has emerged as a soluble model at large N that is dual to two-dimensional gravity \cite{Sachdev:2015efa, Maldacena:2016upp} (see, e.g., \cite{Sarosi:2017ykf, Trunin:2020vwy} for reviews). At low energies compared to a characteristic scale $J$ that occurs in the SYK Hamiltonian, this model is near an infrared fixed point which is characterized by the group of time reparameterizations. The corresponding action functional is the Schwarzian of these one dimensional maps. The SYK model has important positivity properties; the 4-point function can be described in terms of  a discrete spectrum that is positive and bounded from below, together with a pseudo-continuum with level spacings of $O(e^{-N})$. In the large $N$ limit, this pseudo-continuum is described by the Schwarzian action and leads to a density of states $e^{Ns_0} \sinh(2\pi\sqrt{2NE/J})$ \cite{Stanford:2017thb, Iliesiu:2020qvm}. For many reasons the character of the low lying spectrum mentioned above is perhaps one of the most important lessons of the SYK model for black hole physics. 

Historically the work of Bekenstein and Hawking \cite{Bekenstein:1972tm, Bekenstein:1973ur, Hawking:1974rv, Hawking:1975vcx} described black holes as thermodynamic objects. The work of Strominger and Vafa \cite{Strominger:1996sh} that constructed and counted black hole micro-states from D-branes enabled a statistical treatment of black hole thermodynamics and Hawking radiation for a class of near extremal supersymmetric black holes \cite{David:2002wn}. Maldacena's AdS/CFT correspondence gave an in-principle framework for quantum gravity including black holes \cite{Maldacena:1997re, Gubser:1998bc, Witten:1998qj, Witten:1998zw}, but the characteristics of the black hole spectrum remained unanswered till one learned to solve the SYK model. Whether such characteristics persist for higher dimensional black holes is a fundamental question to answer in the future. The other related problem is the bulk description of a general non-extremal black hole state in terms of elementary degrees of freedom of string theory similar to the brane construction of Strominger and Vafa for supersymmetric extremal black holes which avoided dealing with the strong coupling problem.\\

Recently there has been a lot of progress in our understanding of the information loss problem, especially in the context of the semiclassical treatment of the generalized entanglement entropy of an evaporating black hole in contact with a large reservoir \cite{Penington:2019npb, Almheiri:2019psf}(see also \cite{Almheiri:2019hni, Almheiri:2019yqk, Rozali:2019day,  Penington:2019kki, Almheiri:2019qdq, Chen:2020jvn, Almheiri:2020cfm}). Some other relevant papers discussing this are \cite{Geng:2020qvw, Geng:2021hlu, Krishnan:2020fer, Ghosh:2021axl, Krishnan:2020oun, Krishnan:2021ffb, Raju:2021lwh, DeVuyst:2022bua, Bahiru:2022oas}. Most of these developments involve (proposed modifications of) semiclassical gravity calculations. The subtlety and novelty of these ideas underline the need of a dual QFT calculation. Some progress in such calculations has been made in the context of the SYK model coupled to a bath (see, e.g. \cite{Almheiri:2019jqq, Maldacena:2019ufo, Chen:2020wiq}). A common feature of these references is that the SYK model is treated in the bilocal $G$-$\Sigma$ formulation and is outside the scope of the Schwarzian description\footnote{As a reminder, the Schwarzian mode of the SYK model, with a $2p>2$-body interaction, is the pseudo Nambu-Goldstone mode which describes the spontaneously broken time-reparameterization symmetry $f(\tau): G(\tau_1, \tau_2) \to G^f(\tau_1, \tau_2) $= $G(f(\tau_1), f(\tau_2)) f'(\tau_1)^{1/2p}  f'(\tau_2)^{1/2p}$. The symmetry is also explicitly broken slightly at finite but large $J$, which defines the range of energies $\ll J$ at which the system is essentially described by the dynamics of $f(\tau)$. The time reparameterization symmetry can be directly identified in a dual bulk gravity model, namely JT gravity, for which $f(\tau)$ can be identified by a large diffeomorphism at the boundary or alternatively as the shape deformation of the UV boundary curve defined by a constant dilaton\label{ftnt:schwarzian}.} for part of the time development, so that a gravity description of the entire dynamics is not possible.

\vspace{1ex}

The motivation of our work is two-fold:

\begin{itemize}

\item{(i)} Can we have a model of black hole evaporation entirely within the Schwarzian description (which can, therefore, be equivalently interpreted in terms of JT gravity \cite{Teitelboim:1983ux, Jackiw:1984je})?

\item{(ii)} Can we retrieve, at the end of the evaporation process, some information about the initial pure state?

\end{itemize}

At the outset we must specify what we mean by `evaporation' in the boundary setup considered in this paper. The bath is coupled directly to the reparameterization mode whose dynamics determines the bulk geometry. An observer in the bath would see an influx of energy coming from its boundary at $x=0$ where the SYK model sits. This loss of energy of the SYK model is interpreted as evaporation of the black hole in the dual theory as the horizon vanishes.\footnote{In the bulk picture of \cite{Almheiri:2019psf}, demanding energy conservation between the $AdS_2$ and the bath leads to the equation: $\frac{d}{dt} E_{sch}(t) = f'(t)^2 (T_{\hat x^- \hat x^-} - T_{\hat x^+ \hat x^+}) = T^{bdy}_{tx}$. Here $\hat x^\pm= \hat t \pm \hat z$ are $AdS_2$ Poincar\'e coordinates \eqref{PoincareAdS}. A decrease of the Schwarzian can happen through a negative $T_{\hat x^- \hat x^-}$ \cite{Kourkoulou:2017zaj} or a positive $T_{\hat x^+ \hat x^+}$ \cite{Engelsoy:2016xyb}. In either case the horizon shrinks and we interpret this as black hole evaporation.} (See below and Section \ref{sec:bh-evap} for more details.)

%{\color{red} Note that what we are doing here is different from the physics of traversable wormholes in an essential way. There is a way to obtain the $|B_s(l)\rangle$ state by starting with the thermofield double state and measuring the complete set of spins on the left hand side, thereby knowing the initial state immediately \cite{Kourkoulou:2017zaj}. But this is not what we are doing here, as the spins $\{s_k\}$ of the initial pure state are unknown to the observer. It is only by repeating the experiment exponentially large number of times that the observer can infer the full set of spins (see section \ref{sec:recovery}).} \doubt Does this sound correct? \doubt

Our model is described in Section \ref{sec:model}. It consists of an SYK model (characterized by $N$, the number of fermions) and a  bath consisting of $N_F$ massless scalar fields $\Phi_i(x)$ on a half-line $x>0$. In the large $N$ limit, $N_F/N$ is held fixed. The bath is coupled to the SYK model at $x=0$, through a coupling of the form $\mathcal{O} \sum_i \Phi_i(0)$, where $\mathcal{O}$ is an operator of the SYK model \eqref{om}. The dynamics is computed starting from a product of an SYK pure state $\ket{B_s(l)}$ (see \cite{Kourkoulou:2017zaj}, also reviewed in Section \ref{sec:review}) and a pure state of the bath (of the Calabrese-Cardy type\cite{Calabrese:2007rg, Cardy:2014rqa}). An effective description of the SYK dynamics can be obtained by integrating out the bath. At low energies such dynamics can be entirely written in terms of the time reparameterization mode $f(t)$ \footnote{Sometimes this is also called the `Schwarzian mode'.} (which we will equivalently parameterize in terms of the Liouville mode $\phi(t)$ defined by $\dot f(t)= e^{\phi(t)}$). In the low-energy limit the coupling essentially becomes $\sum_i\Phi_i(0) \left[g e^{\phi(t)/2} - g' e^{\phi(t)}\right]$. We look at two models (which we call (a) and (b)), depending on the type of SYK operator used to couple to the bath. In Model (a), $g'=0$ and in Model (b), both $g$ and $g'$ are non-zero. We find that the $g-g'$ parameter space is restricted to a subregion, outside which the large $N$ equations of motion do not have well-behaved solutions (see section \ref{domain} for more details). \textit{Since the time reparameterization mode $f(t)$ equally well describes JT gravity, our low energy model is equivalently described in terms of JT gravity coupled to a bath.} We compute the classical solution $\phi(t)$ of the Liouville mode in the large $N$ limit as well as the mean-square fluctuation $\lan \delta \phi(t) \delta \phi(t') \ran$. The results of these computations are detailed in Sections \ref{sec:results} and \ref{sec:fluctuations-2pt}, while the JT interpretation in terms of black hole evaporation is described in Section \ref{sec:bh-evap}. 

\vspace{1ex}

Our results (see Sections \ref{sec:results}, \ref{sec:recovery} and \ref{sec:fluctuations-2pt}) are summarized below:

\begin{enumerate}

\item{\bf Long time dynamics}

As indicated above, in this paper we focus on two main models of coupling to the bath:\\   
(a) In the first model, which we call Model (a) (Section \ref{sec:relevant}), we couple the SYK system to the bath by an operator that is closely similar to the operator $g\, \mathcal{O}_{KM}$ introduced by Kourkoulou and Maldacena \cite{Kourkoulou:2017zaj}. In this case, the dynamics reaches a well-defined equilibrium when the coupling $g$ is below a certain critical value. The classical solution $\phi(t)$ asymptotically reaches a form which can be identified with a black hole solution with lower ADM energy $E_\infty < E_{\rm ini}$ where $E_{\rm ini}$ is original energy of the SYK before coupling to the bath ({\it i.e.} the initial ADM energy of the black hole microstate). The fluctuation $\sqrt{\lan \delta \phi(t)^2 \ran}$ in equilibrium is characterized by Brownian-type fluctuations at the bath temperature around the classical motion $\phi(t)$. \\
(b) In the second model which we call Model (b) (Section \ref{sec:relevant+marginal}), the bath is coupled to a linear combination of a relevant and a marginal SYK operator. After a characteristic time, the coupling is slowly switched off to prevent the low energy approximation from breaking down. Here, the final form of the solution $\phi(t)$ has a bounded oscillatory behaviour which can be identified in terms of a horizonless geometry. Such a solution represents a process of complete evaporation, in the sense we mentioned above. The fluctuation $\sqrt{\lan \delta \phi(t)^2 \ran}$ also exhibits oscillatory behavior characteristic of a Brownian particle in a bounded potential. \\ 
In both models, coupling to the bath, in effect, dynamically changes the KM deformation parameter (recall that this parameter was changed externally in \cite{Kourkoulou:2017zaj, Dhar:2018pii}). Other couplings to the bath are described in Appendix \ref{app:other-bath-couplings}.

\item{\bf Black hole evaporation}

In Section \ref{sec:bh-evap} we interpret some of the above results in terms of JT gravity. While Model (a) can be interpreted as incomplete black hole evaporation, Model (b), with a horizonless asymptotic geometry, can be interpreted as complete black hole evaporation. The horizon structure is described in detail in Section \ref{sec:horizon} and Appendix \ref{app:geom}. The final state in both models is described in Section \ref{sec:final-state}. 

\item {\bf Retention of memory of the initial state through the evaporation}

The initial pure state $\ket{B_s(l)}$ can be regarded as a black hole `microstate', where different microstates are distinguished from each other by the spin vector $s$. In \cite{Kourkoulou:2017zaj} the SYK model was deformed by an operator $\mathcal{O}_{KM}$ carrying a spin vector which was chosen to be identical to that of the pure state, similar to the idea of \cite{Papadodimas:2013jku, Papadodimas:2015jra}. In \cite{Dhar:2018pii} the computation was generalized to the case where the spin vector $s'$ of the operator $\mathcal{O}_{KM}$ and the spin vector $s$ of the pure state are different; the resulting dynamics carried information about the scalar product $s'\cdot s$. In the present work too, the SYK operators coupling to the bath carry a particular spin vector $s'$; the dynamics carries information about the scalar product  which, in fact, survives at asymptotic times as well, both in models (a) and (b). If we regard the coupling to the bath as a probe, and {\it if} we are able to repeat the experiment with various choices of  the spin vector $s'$, every choice of $s'$ gives us additional information about the spin $s$ which characterizes initial $\ket{B_s(l)}$ state. A step by step algorithm of how to do this is described in Section \ref{sec:recovery}, where we show that if we repeat these experiments $O(N^2)$ times, we can have an in principle reconstruction of the spin $s$ and consequently the initial state. Note that the $O(N^2)$ repetitions are consistent with the polynomial form of unrestricted complexity \cite{Brown:2019rox}, since our protocol involves the couplings which act both on the SYK and the bath\footnote{We thank Onkar Parrikar for a discussion on this point.}.

The discussion above is similar to the extraction of information about a black hole microstate from polarization tensors of the probe graviton in \cite{Mandal:1995qb} (see equations (2.5), (2.6); the scattering amplitude in equation (2.4) depends on the inner product of polarization tensor of the probe particle and that of the black hole microstate -- hence by repeating the scattering experiment one can extract more and more information about the polarization tensor of the microstate). Similar remarks apply to the analysis of \cite{Dhar:1996vu} where the decay amplitude of a near-BPS state depends on the specific microstate (see equation (8) of \cite{Dhar:1996vu}).\footnote{Note that while in these papers the properties of the probes can be tuned at low energies, in our current work, the specific coupling to the bath, and in particular, the information about the spin vector $s'$ is a microscopic (UV) property. This is elaborated further in Section \ref{sec:recovery}.}

We note that even the two-point correlators of the models discussed in this paper retain the information about the initial state. This is rather non-trivial since one normally associates the asymptotic  two-point fluctuation with equilibrium properties (e.g. for a normal Brownian particle) which does not have a memory of the initial state.

See Section \ref{sec:recovery} (also Section \ref{sec:G-Sigma}) for more details.

\end{enumerate}

%{\color{blue} Note that what we are doing here is different from physics of traversable wormholes in an essential way. There is a way to obtain the $|B_s(l)\rangle$ state by starting with the thermofield double state and measuring the complete set of spins on the left hand side, thereby knowing the initial state immediately. But this is not what we are doing as the spins $\{s_k\}$ of the initial pure state are unknown to the observer. It is only by repeating the experiment exponentially large number of times that the observer can infer the full set of spins.}
 
%%%%%%%%%%%%%%%%%%%%%%%%%%%%%%%%%%%%%%%%%%%%%%%%%%%%%%%%%%%%%%%%%%%%%%%%%%%
\section{Review : Thermal microstates  and off-diagonal operators} \label{sec:review}
%%%%%%%%%%
\subsubsection*{Thermal microstates}
In this section, we briefly review the relevant properties of the pure states of \cite{Kourkoulou:2017zaj}. We will start with the SYK Hamiltonian which is written in terms of $N$ Majorana fermions
\begin{equation}\label{eq:ham0}
	\text{H}_0 = \text{H}_{SYK} = - \sum_{1\le a<b<c<d \le N} j_{abcd}\ \psi_a\psi_b\psi_c\psi_d
\end{equation}
where the couplings $j_{abcd}$ are drawn randomly from a Gaussian distribution with zero mean and variance, $\braket{j^2_{abcd}} = 3! J^2/ N^3$. The equal time anticommutation relation of the Majorana fermions, $\{\psi_a, \psi_b\}= \delta_{ab}$, coincides with the $O(N)$ Clifford algebra. We will call the normalized states which provide a spinorial representation of the above algebra, $|B_s\ran$, where $s= (\pm, \pm,...)$ are $N/2$ dimensional `spin' vectors. The total number of such vectors is $2^{N/2}$. Ref. \cite{Kourkoulou:2017zaj} introduced a class of pure states (similar to the Calabrese-Cardy states \cite{Calabrese:2005in} that were introduced to model quantum quench) given by \footnote{Since the Hilbert states of the SYK model is a linear space, in principle one can consider low-energy properties of a linear combination of these states. In particular it will be interesting to know if all microstates are black holes. We leave this discussion for future work.}
\begin{equation}
	| B_s(l)\ran = \exp[-l H_0]|B_s\ran,
\end{equation}
which reproduce thermal properties for a large class of
observables, corresponding to a temperature $1/\b$ where $\beta=2l$. E.g.
\begin{equation}
	\overline{\lan B_s(l)| B_s(l) \ran} = \overline{\lan B_s|e^{-2l H_0}|B_s \ran} = 2^{-N/2}\ \overline{\Tr(e^{-\b H_0})}\equiv 2^{-N/2}\ \overline{Z(\beta)} ,
	\label{thermal-a}
\end{equation}
\begin{equation}
	\f{\overline{\lan B_s(l)|\psi_a(t) \psi_a(t')| B_s(l) \ran}}{\overline{\lan B_s(l)| B_s(l) \ran}}= \frac{\overline{\Tr(e^{-\b H_0}\psi_a(t) \psi_a(t'))}}{\overline{Z(\beta)}} = G_\b(t-t') .
	\label{thermal-b}
\end{equation}
%above equations \eq{thermal-a}, \eq{thermal-b} are exact at finite $N$. 
The `bar' on these expressions indicate a disorder average. We will be discussing the low-energy sector at large $N$ in this paper so henceforth we will drop the `bar' on the expressions. All the expectation values discussed will be disorder averaged expectation values.  The fermion bilinear can be replaced by any flip-symmetric\footnote{a discrete subgroup of O(N), for more details see \cite{Kourkoulou:2017zaj}.} operator. These equations show that the basic dynamical variable of the SYK model, the bilocal variable $G(t,t')= (1/N)\sum_a \psi_a(t) \psi_a(t')$ which describes the $O(N)$ invariant sector, does not distinguish between the pure states $|B_s(l)\ran$ and the thermal (mixed) state $\rho_\b = \f1{Z(\b)} \exp[-\b H_0]$, $\b=2l$.

Equalities like \eq{thermal-a} can be obtained from a path-integral \cite{Kourkoulou:2017zaj} (a detailed derivation is presented in the appendix of \cite{Dhar:2018pii}). For large enough $l$ ($lJ\gg1$), both sides of the equation \eq{thermal-a} can be expressed as follows in terms of a path-integral over the time-reparameterization or Schwarzian mode $f(\tau):= \f\pi{\b J^2} \tan(\pi \varphi(\tau) / \b)$ %{\color{blue} Is this still true at finite N?}
\begin{align}
	\int [Df] \exp[i S_0[f]], \quad S_0[f]= -\f{N \alpha_s}{J}
	\int_{-l}^{l} d\tau \, \{f,\tau \} = -\f{N \alpha_s}{J}
	\int_{-l}^{l} d\tau \,\left[ \{\varphi(\tau),\tau\} + \f{2\pi^2}{\beta^2}\, \dot{\varphi}(\tau)^2\right] .
	\label{path-integral-Z}
\end{align}
In the above we have used the notation for the Schwarzian of a function $f(\tau)$
\begin{equation}
	\{f,t\}\equiv \frac{\dddot f}{\dot f} - \frac{3}{2} \left(\frac{\ddot f}{\dot f} \right)^2 ,
\end{equation}
where dot denotes derivative with respect to $\tau$. The boundary condition for the path integral to describe the LHS of \eq{thermal-a} is appropriate for an interval (see \cite{Kourkoulou:2017zaj}):
\begin{align}
	&\varphi(-l)=-l,\quad \varphi(l)=l,\qquad  \dot{\varphi}(-l)= \dot{\varphi}(l)=1. \label{interval-bc}
\end{align}
The boundary condition for the circle is given by periodic
identification of $\tau=-l$ and $\tau=l$ and winding number 1. The saddle point solutions for the two boundary conditions are different; however, the classical action $S_0[f]$ evaluates to the same value in both cases (the two boundary conditions differ in the SL(2) zero modes, which are gauge modes and do not affect physical quantities). Both boundary conditions also lead to the same result for Green's functions e.g. \eq{thermal-b} or \eq{off-diagonal}, which explains why $|B_s(l)\ran$ states reproduce thermal properties despite different boundary conditions.\footnote{We acknowledge crucial discussions with Juan Maldacena regarding the issue of boundary conditions.}
\par The saddle point solution corresponding to the boundary condition \eq{interval-bc} is $\varphi(\tau) = \tau$. In terms of the $f$ variable the solution is 
\begin{equation}\label{saddle-interval}
	f(\tau) = \f{\pi}{\b J^2}\, \tan\left({\f{\pi}{\b} \tau}\right).
\end{equation}

%%%%%%%%%%
\subsubsection*{Liouville theory}

The Schwarzian action is a higher derivative action. To discuss Hamiltonian formulation of such an action we can introduce auxiliary variables in terms of which the action becomes 2nd order \cite{Engelsoy:2016xyb}. This can be done by introducing a field $\phi$ defined by $\dot{f} =: e^\phi$ through a Lagrange multiplier $\lambda$. Remember that the Schwarzian mode is a reparameterization of time and should be monotonic ($\dot{f}(t)>0$), this redefinition implicitly takes care of this. We will refer to $\phi(t)$ as a Liouville mode (the nomenclature will be clear from the form of the action). Introducing the Liouville mode, the Schwarzian action can be rewritten as\footnote{In arriving at this expression, one ignored a total derivative term of the form $\int dt \dot f$ \cite{Kourkoulou:2017zaj}, which contributes an unimportant constant to the path integral. Further $\lambda$, which is a constant by equation of motion, is set to $\lambda=-4J$ by a gauge choice which is also made in writing the Euclidean solution \eqref{saddle-interval}. \label{ftnt:lagrange}} (see, e.g. \cite{Kourkoulou:2017zaj}) %\footnote{Note that $f(t)$ is supposed to a monotonically increasing function of $t$, making $\phi$ well-defined.}
\begin{equation}
	S_0= \frac{N \alpha_s}{J} \int dt \left[\frac{\dot{\phi}^2}{2} -2 J^2 e^\phi \right], \quad \dot f(t) =: \exp[\phi] .
	\label{liouville}
\end{equation}
This is simply Liouville theory \cite{Bagrets:2016cdf, Engelsoy:2016xyb}, namely a particle moving in an exponential potential\footnote{It was shown in \cite{Bagrets:2016cdf} that the measure in terms of the Liouville variable is flat.}.

%%%%%%%%%%%%%%%
\subsubsection*{Off-diagonal operators}
Equation \eq{thermal-b} demonstrates diagonal observables which show thermal properties in the $|B_s(l)\ran$ states.  What about off-diagonal bilinears? An important relation (presented in \cite{Kourkoulou:2017zaj}) is %(here $Z=\langle B_s(l)| B_s(l) \rangle$)
\begin{align}
	\frac{1}{Z}\lan B_s(l)| s_k \psi_{2k-1}(t) \psi_{2k}(t') |B_s(l) \ran =2 i G_\b(t-il)
	G_\b(t'-il) + O(\frac{1}{N}), \qquad Z=\langle B_s(l)| B_s(l) \rangle ,
	\label{off-diagonal}
\end{align}
which shows that the off-diagonal bilinears $\psi_{2k-1}(t) \psi_{2k}(t')$ by themselves do {\it not} have a thermal form (and depend on the spin vector $s$); but when they are in the combination as shown above ($s_k \psi_{2k-1}(t) \psi_{2k}(t')$), then the RHS is a product of thermal Greens functions and the memory of the spin vector $s$ is erased on the RHS. This relation can be generalized, using similar methods as above, to the case where the spin vector in the operator and the state is not aligned, to
\begin{align}
	\frac{1}{Z} \lan B_{s}(l)| \sum_{k=1}^{N/2} s'_k \psi_{2k-1}(t) \psi_{2k}(t')| B_{s}(l) \ran =2i s\cdot s' G_\b(t-il) G_\b(t'-il) + O(\frac{1}{N}), \; 
	s\cdot s'= \sum_{k=1}^{N/2} s'_k s_k =: N/2\ \cos\theta .
	\label{cos-alpha}
\end{align}
This shows that unlike in \eq{off-diagonal}, if the spin vectors in the state and operator are not matched, some memory of both the spin vectors is retained.

%%%%%%%%%%
\subsubsection*{Modified SYK}

We would like to deform the SYK theory with similar off-diagonal operators. We can write down a list of operators similar to \eq{off-diagonal} and \eq{cos-alpha} which have a large $N$ expression in terms of fermion 2-point function $G_\beta$. Description of the UV operators in terms of bilocal variables $G_\beta$ is desirable to discuss the low-energy sector of SYK and also the SYK-JT duality. Before we present the explicit form of the operators in the UV, let us write down  the corresponding low energy path integral modified by such an operator in terms of the Schwarzian mode
\begin{equation}
	\frac{1}{Z} \langle B_s(l)| e^{-i\int dt\, \epsilon(t) O_{\Delta}(t)} |B_s(l)\rangle = \int [Df] \ e^{iS_0[f] - i \int dt\, \epsilon(t) \frac{1}{Z} \langle B_s(l)| O_{\Delta}(t) |B_s(l)\rangle^{(f)} } ,
\end{equation}
where the superscript $(f)$ denotes the reparameterization $t\rightarrow f(t)$. $\Delta$ here is the IR mass dimension of the operator. %{\color{red} Note that $O_{\Delta}$ is not a conformal primary, its expectation value in the ground state vanishes.} 
Above equation is clearly correct in the perturbative regime of $\ep(t)$ but this relation can be exactly derived by expanding the exponent and resuming the pieces which are leading order in $N$ (similar to the presentation in Appendix of \cite{Dhar:2018pii}).\\ 

Consider the following off-diagonal operators
\begin{equation}\label{om}
	\mathcal{O}^{\{s^{(1)}, s^{(2)}, ..., s^{(m)}\}}_m(t) = \mathcal{O}_{\Delta = 2m \Delta_f}(t) = (-1)^{m+1} J \prod_{j=1}^{m}\left(\f{i}{N} \sum_{k=1}^{N/2} s^{(j)}_k \psi_{2k-1}(t) \psi_{2k}(t)\right) .
\end{equation}
This operator is characterized by $m$ different spin-vectors $s^{(j)}$. %\footnote{More precisely we should write $O_m^{\{s^{(1)}, s^{(2)}, ..., s^{(m)}\}}$. \label{ftnt:spins}}
Here the factor of $J$ is multiplied to ensure that the coupling constant is dimensionless in the UV, and $\Delta_f$ is the conformal dimension of the SYK fermion near the IR fixed point which is equal to $\f14$.\footnote{We are considering a 4 body interaction SYK Hamiltonian, in the case of q-body SYK Hamiltonian, $\Delta_f = 1/q$.} This operator is composite of $2m$ fermions and hence has a mass dimension $\Delta = 2m \Delta_f$ in the IR. The $m=1$ operator is the one which was originally introduced in \cite{Kourkoulou:2017zaj}. The leading large $N$ expectation values of these operators in a thermal microstate are (see Appendix \ref{app:IRoperators} for the derivation)
\begin{equation}\label{14}
	\frac{1}{Z}\braket{B_s(l)|\mathcal{O}^{\{s^{(1)}, s^{(2)}, ..., s^{(m)}\}}_m(t)|B_s(l)} = - J \left(\prod_{j=1}^m \cos(\theta^{(j)})\right) G_{\beta}(t-il)^{2m}, \quad N\cos(\theta^{(j)})/2 = \sum_{k=1}^{N/2} s_k s^{(j)}_k .
\end{equation}
As indicated earlier, when the spin vectors in the state and operator are not matched, some information about the spin vector $s$ of the pure state can be recovered from choices of the spin vectors $s^{(j)}$ of the probe operators. Note that, near the IR fixed point, operators $\mathcal{O}_m$ with $m<1/2\Delta_f$ are relevant operators and with $ m = 1/2\Delta_f$ is a marginal operator. For $\Delta_f = 1/4$, we have a relevant ($m=1$) and a marginal operator ($m=2$) each. %{\color{red} Recall that the Schwarzian action is also a low energy projection of the operator $\psi_i\partial_t\psi_i$, its mass dimension is $3/2$ and it is an irrelevant operator at the IR fixed point. (CITE)}\\

The last step to get the interaction term of the action is to couple the Lorentzian Schwarzian mode $f(t):= \f\pi{\b J^2} \tanh(\pi \varphi(t)/\beta)$ to
\begin{equation}
	G_\beta(t-il) = \f{C_{\Delta_f}}{\left[\f{J\beta}{\pi}\cosh\left(\f{\pi t}{\beta}\right)\right]^{2\Delta_f}}, \qquad \text{with} \quad C_\Delta = \left[\left(\f{1}{2}-\Delta\right) \f{\tan\pi \Delta}{\pi}\right]^\Delta .
\end{equation}
In order to couple the Schwarzian mode $\varphi(t)$, we reparameterize above expression with it
\begin{equation}
	G^\varphi_{\beta}(t) = \dot{\varphi}(t)^{\Delta_f}\, G_\beta(\varphi(t)-\varphi(il))\, \dot{\varphi}(il)^{\Delta_f} = \f{C_{\Delta_f}\, \dot{\varphi} (t)^{\Delta_f}}{\left[\f{J\beta}{\pi} \cosh\left(\f{\pi \varphi(t)}{\beta}\right)\right]^{2\Delta_f}} = C_{\Delta_f}\, \dot{f}(t)^{\Delta_f} ,
\end{equation}
hence the interaction term in the action is
\begin{equation}
	\hat g_m \int dt\, \epsilon(t) \frac{1}{Z} \langle B_s(l)|\, \mathcal{O}_m^{\{s^{(1)}, s^{(2)},..., s^{(m)}\}} (t)\, |B_s(l)\rangle ^{(f)} = - g_m J \int dt\, \epsilon(t)\,   e^{m\phi(t)/2}, \qquad g_m = \hat g_m \prod_{i=1}^m \frac{\cos(\theta^{(j)})}{\sqrt{4\pi}} .
\end{equation} 
%{\color{blue} \doubt It is ok to introduce the most general operators but then we should stick to the operators we actually use for computation in eq. \eqref{relevant}, \eqref{mixed} namely with a single $\theta$. \doubt}
In the equation above we have used $\Delta_f = 1/4$ and use it in all the equations hereafter. For $m=1$, which corresponds to a relevant operator, the above equation looks like
\begin{align} 
	\hat g_1 \int dt\, \epsilon(t) \frac{1}{Z} \langle B_s(l)|\, \mathcal{O}_1^{s'}(t)\, |B_s(l)\rangle ^{(f)} = - g J \int dt\, \epsilon(t)\, e^{\phi(t)/2}, \qquad g = \hat g_1 \frac{\cos\theta}{\sqrt{4\pi}} , 
\label{m=1}
\end{align}
where we have written $g$ for $g_1$ and $\theta$ for $\theta^{(1)}$. For $m=2$, which corresponds to a marginal operator, we take the two spins to be the same
\begin{align}
	&\hat g_2 \int dt\, \epsilon(t) \frac{1}{Z} \langle B_s(l)|\, \mathcal{O}^{\{s',s' \}}_2(t)\, |B_s(l)\rangle ^{(f)} = g' J \int dt\, \epsilon(t)\, e^{\phi(t)}, \qquad g' = -\hat g_2 \left(\frac{\cos\theta}{\sqrt{4\pi}} \right)^2 , %\frac{\cos\tilde \theta_2}{\sqrt{4\pi}} 
	\label{m=2}
\end{align}
where we have written $g'$ for $-g_2$ and $\theta$ for $\theta^{(1)}$. %In the two equations above we have explicitly written the spins according to footnote \ref{ftnt:spins}.
Note that $\mathcal{O}_2^{\{s',s' \}}$ is the same as $\left(\mathcal{O}_1^{\{s' \}}\right)^2$.
 
We can also couple multiple operators in a similar way. 
%\begin{equation}\label{sint}
%	\int dt\, \epsilon(t)\, \sum_m \hat g_m \langle B_s(l)|\, \mathcal{O}{\{s^{(1)}, s^{(2)},..., s^{(m)}\}}_m(f(t))\, |B_s(l)\rangle = -g\int dt\,  \epsilon(t)\, F[\phi(t)],\qquad F[\phi(t)] = \f{J}g \sum_m g_m\ e^{m\phi(t)/2}
%\end{equation} 
When operators with only $m=1$ and $m=2$ are used
\begin{align}\label{sint-1-2}
	\int dt\, \epsilon(t) \frac{1}{Z} \langle B_s(l)|\, \hat g_1 \mathcal{O}_1^{\{s' \}}(t) + \hat g_2 \mathcal{O}_2^{\{s',s' \}}(t)\, |B_s(l)\rangle ^{(f)} &= -g\int dt\, \epsilon(t) F[\phi(t)] , \\
	g F[\phi(t)] &= J \left[g e^{\phi(t)/2} - g' e^{\phi(t)}\right] ,
\end{align} 
where $g,g'$ are given by \eq{m=1} and \eq{m=2}. Note that the information $s\cdot s'$ from the UV stays unaltered in the IR limit.

\section{The model}\label{sec:model}
		
\begin{figure}
	\centering
	\begin{tikzpicture}
	%\draw[help lines, very thin, gray] (-4,-3) grid (8,3);
	\draw[red, thick]  (0,2) -- (-2,0) -- (-0.1,-1.9)-- (-.1,0);
	\draw[gray,thick,dashed] (0,0) -- (0,2);
	\node[red] at (-.7,.3) {BH};
	\draw[blue, thick] (0,2) -- (2,0) -- (0.1,-1.9) -- (.1,0);
	\node[blue] at (.7,.3) {bath};
	\draw[very thick] (-2,-2) .. controls (-4,0) and (-4,0) .. (-2,2);
	\node[left] at (-3.5,0) {EWB};
	\draw[red, dashed] (-2,0) -- (-3,-1);
	\draw[red, dashed] (-2,0) -- (-3,1);
	\draw[green, thick] (-3.5,0) -- (2,0);
	\draw[green, thick] (4.8,0) -- (7,0);
	
	\draw[thin, gray] (3,0.1) -- (3.5,0.1);
	\draw[thin, gray] (3,-0.1) -- (3.5,-0.1);
	%\draw[thin, gray] (3,0) -- (3.5,0);
	%\draw[red, ultra thick] (4.9,-2) -- (4.9,0);
	%\draw[red, ultra thick] (4.9,0) -- (5,0);
	\draw[red, ultra thick] (4.8,-2) -- (4.8,2);
	\node[red, left] at (4.7,0) {SYK};
	\draw[blue, thick] (5,2) -- (7,0) -- (5,-2) -- (5,2);
	\node[blue, above] at (6,0) {bath};
	\draw[gray, thick] (4.8,.2) -- (5,.2);
	\draw[gray, thick] (4.8,.4) -- (5,.4);
	\draw[gray, thick] (4.8,.6) -- (5,.6);
	\draw[gray, thick] (4.8,.8) -- (5,.8);
	\draw[gray, thick] (4.8,1) -- (5,1);
	\draw[gray, thick] (4.8,1.2) -- (5,1.2);
	\draw[gray, thick] (4.8,1.4) -- (5,1.4);
	\draw[gray, thick] (4.8,1.6) -- (5,1.6);
	\draw[gray, thick] (4.8,1.8) -- (5,1.8);
	\draw[gray, thick] (4.8,2) -- (5,2);
	\end{tikzpicture}
	\caption{The field theory setup (right) is a 1-d holographic quantum mechanical system coupled at $t=0$ to a 1+1 dimensional CFT on a half Minkowski plane. The horizontal gray lines indicates interaction. The green line is the $t=0$ slice and where the theory is prepared in a pure state given in \eq{eq:t0state}. In the dual bulk theory (left), we simply replace the 1-d holographic system by its dual, as we have prepared it in a thermal microstate the dual geometry is a black hole containing an end of the world brane in the interior\cite{Kourkoulou:2017zaj}. The black hole is coupled to flat space at $t=0$ (green slice). }\label{fig:dotline}
\end{figure}
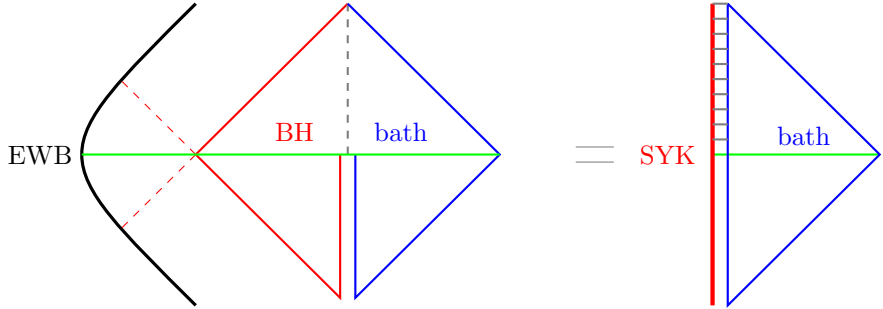

In this section, we will describe the coupled system consisting of $N_f$ free scalars on a half line ($x \in [0,\infty)$), modeling the bath, and the SYK theory at the boundary ($x=0$), modeling the quantum system dual to a black-hole (see Figure \ref{fig:dotline}). As mentioned in the introduction, we will focus on the IR regime of the SYK theory where it is described by a pseudo Nambu-Goldstone (NG) mode and the dynamics of this mode is governed by the Schwarzian action \eq{path-integral-Z},\eq{liouville}.

At $t=0$, the SYK and bath are prepared in a pure state of the form
\begin{equation}\label{eq:t0state}
	|B_s{(l)}\rangle \otimes |\Psi_0(L)\rangle .
\end{equation}
The geometry dual to the state $\ket{B_s(l)}$ is a black hole with inverse temperature $\beta =2l$, containing an end of the world brane (EWB) in the interior \cite{Kourkoulou:2017zaj} (see also \cite{Almheiri:2018ijj}). Further, the averaged expectation values of flip-symmetric operators in these pure states are the same as their averaged thermal expectation values with $\beta= 2l$. The bath state $|\Psi_0(L)\rangle$ is a Calabrese-Cardy (CC) state \cite{Calabrese:2005in}. 
\begin{equation}
	\ket{\Psi_0(L)} = e^{-L H_{bath}}\ket{Bd},\quad \text{with} \quad \Phi\ket{Bd}=0.
	\label{product-state}
\end{equation}
The CC states have a similar thermal character as $\ket{B_s(l)}$, to be precise, the expectation values of a string of local operators (confined to a subregion) are thermal at large times with an effective inverse temperature $\beta_b =4L$ \cite{Calabrese:2005in, Mandal:2015jla, Mandal:2015kxi, Banerjee:2019ilw}. Hence, the rational for preparing the coupled system in the \eq{eq:t0state} is to have an interpretation of coupling two systems with effective inverse temperature $2l$ (black-hole) and $4L$ (bath) respectively.     

In the bulk dual description, Hawking radiation from a black hole in an asymptotically AdS (AAdS) spacetime cannot escape and falls back in due to the effective AdS potential. As a result large black holes in AAdS spacetimes do not evaporate. One way to extract the radiation from the AdS region is to couple the AAdS geometry to an auxiliary non-gravitational system %\footnote{see \cite{Rocha:2008fe} for an early reference.}
with a large number of degrees of freedom \cite{Rocha:2008fe, Almheiri:2019psf, Penington:2019npb, Almheiri:2019hni, Almheiri:2019yqk}. In such a treatment of the evaporation process, the black hole is treated as an open system in both the bulk and boundary descriptions.%\\  

The action for the coupled system in the boundary description can be obtained as follows --  the interaction term $S_{int}$ is the same as \eqref{sint-1-2} with the external parameter $\ep(t)$ now replaced by the dynamical bath fields, i.e. $\ep(t) \rightarrow \sum_i \Phi_i(t,0)$
\begin{equation}
	S^{UV}_{int} = \int dt \sum_i \Phi_i(t,0) \left(\hat g_1 \mathcal{O}_1^{\{s' \}}(t) + \hat g_2 \mathcal{O}_2^{\{s',s' \}}(t) \right) \xrightarrow[]{ \;\text{IR}\; } S_{int} = -g\int dt \sum_i \Phi_i(t,0) F[\phi(t)] ,
	\label{IR-flow}
\end{equation}
with
\begin{equation}\label{F-final}
	F[\phi(t)] = J \left[e^{\phi(t)/2} - \frac{g'}{g} e^{\phi(t)}\right] .
\end{equation}
With this the full action reads
\begin{equation}\label{main-model}
	S= \underbrace{\frac{N \alpha_s}{J} \int dt \left(\frac{\dot{\phi}^2}{2} -2 J^2 e^\phi \right)}_{S_{0}} \quad \underbrace{-\ g\int dt\,  F\left[\phi(t)\right] \sum_{i=1}^{N_f}\Phi_i(t,0)}_{S_{int}} \quad +\  \underbrace{\frac{1}{2} \sum_{i=1}^{N_f}  \int dt \int_0^{\infty} dx\, (\partial \Phi_i)^2}_{S_{bath}} \,.
\end{equation}
We have already introduced $S_{0}$ \eq{liouville}. As mentioned earlier, our auxiliary bath system consist of $N_f$ free, massless scalars $\Phi_i$ on the half-line and its action is $S_{bath}$ in \eq{main-model}. The interaction between the bath and SYK is turned on at $t=0$ (see Figure \ref{fig:dotline}) and the interaction is localized at the boundary ($x=0$).

We will consider two sub-cases of $S_{int}$ in \eq{main-model} depending on whether $\hat g_2$ is zero or not: %\doubt {\color{blue} we will use only a single spin/$\theta$ everywhere as it is necessary for information recovery \doubt}
\begin{itemize}
	\item{(Model a)} Only a relevant coupling (see \eq{m=1}) 
\begin{equation}\label{relevant}
	F(\phi(t))/J = \exp[\phi(t)/2], \qquad  g= \frac{\hat g_1}{\sqrt{4\pi}} \cos\theta .
	%g F(\phi(t))/J &= g_2 \exp[\phi(t)] = \f{\hat g_2}{4\pi} \cos\tilde\theta_1 \cos\tilde\theta_2  \exp[\phi(t)] &\hbox{marginal} 
	%\label{marginal}\\
\end{equation}
\item{(Model b)} A mixture of relevant and marginal coupling (see \eq{sint-1-2})
\begin{equation}\label{mixed}
	F(\phi(t))/J = \exp[\phi(t)/2]- \f{g'}{g} \exp[\phi(t)], \qquad
	g= \f{\hat g_1}{\sqrt{4\pi}} \cos\theta, \qquad
	g'=-\f{\hat g_2}{4\pi} \cos^2\theta . %\cos\tilde\theta_2  
\end{equation}	
\end{itemize}
The black hole evaporation that we study in this paper occurs in a certain range of the parameters $g$ and $g'$ (see Figure \ref{g-gprime} and Section \ref{domain} for more details). One can make sure that $g$ and $g'$ are in this range by tuning $\hat g$ and $\hat g'$. The relative minus sign between the two couplings would be crucial to achieve complete black hole evaporation. In Model (b) we consider $g$ and $g'$ to be constant initially and to have a slow fall-off in time at late stages (see Section \ref{sec:relevant+marginal}). Note that $e^{\phi/2}$ is a relevant operator and $e^\phi$ is a marginal operator -- see comments below equation \eqref{14}.

\paragraph{Large $N$ limit}
The SYK action is order $N$, the interaction term and the bath action both are order $N_f$. We would take a large $N$ limit along with a large $N_f$ limit in such a way that  
\begin{align}
	n_f = N_f/N \label{nf-scaling}
\end{align}
is held fixed and small. This ensures that all the terms in equation \refeq{main-model} are comparable, of $O(N)$ and one can use the saddle point method. This scaling also ensures that the fraction of energy lost by the black hole due to evaporation is of the same order (in terms of $N$) as the initial energy, as we will see explicitly later in Section \ref{sec:results}. This feature is necessary to observe a time-dependent geometry due to backreaction from the radiation appropriate for an evaporating black hole.

We note here that the limit $n_f \to 0$ is interesting to study. One would expect to recover the usual quasi-static approximation of black hole radiation in this limit. It would be interesting to see if the energy flux in the bath discussed in Section \ref{sec:energy-gain} becomes a thermal flux in that limit. We wish to come back to this issue in the future.\footnote{We thank Shiraz Minwalla for making this suggestion.}

%%%%%%%%%%%%%%%%%%%%
\subsection{Schwinger-Keldysh Formalism and the Effective Action} \label{subsec:SKcontour}

Our primary aim is to obtain an evaporating black hole solution in terms of the Liouville mode $\phi(t)$. In the limits considered in \eq{nf-scaling}, the action is proportional to $N$ and the path integral can be computed by a saddle point approximation. To extract this solution which saturates the integral in real time, it is instructive to consider the expectation value of some operator $\mathcal{O}$ in the SYK theory at some arbitrary real time $t=T$ (see Figure \ref{fig:SK-contour}). These operators are order $1$ in $N$ scaling and won't change the solution of the Liouville mode which saturates the path integral. The expectation value is given by
\begin{equation}\label{ev}
	\langle \mathcal{O}(T)\rangle = \text{Tr}\left[\left( \hat{\rho}_{Sch} \otimes \hat{\rho}_{bath} \right) \, \mathcal{O}(T) \right]
	= \langle B_s(l) \otimes \Psi_0(L) |e^{i H T} \mathcal{O} e^{-i H T} |B_s(l) \otimes \Psi_0(L) \rangle ,
\end{equation}
here $H$ is the full Hamiltonian of SYK plus scalar field bath including the interaction. The initial state is taken to be a tensor product of \footnote{We can also consider the bath in a thermal state (see Appendix \ref{app:thermal}).}
\begin{equation}\label{dm}
	\hat\rho_{sch} = \ket{B_s(l)}\bra{B_s(l)}\quad \text{and} \quad \hat\rho_{bath} = \ket{\Psi_0(L)}\bra{\Psi_0(L)} .
\end{equation}
Note that the Euclidean evolution in the preparation of the state \eq{eq:t0state} (and \eq{dm}) is with the uncoupled Hamiltonian as opposed to the real time evolution which is with the full interacting Hamiltonian $H$. It is well-known (see \cite{kamenev2011field} etc.) that expectation values such as \eq{ev} involve contours which run back and forth in Lorentzian time, in addition to Euclidean parts for the specific pure states we mentioned above. Such contours are called Schwinger-Keldysh (SK) contours; they are depicted in Figure \ref{fig:SK-contour}.

%\vbox{
\begin{figure}[]
	\begin{minipage}{.5 \linewidth}
		\centering %change scale below to make figure bigger/smaller.
		\begin{tikzpicture}[scale=0.6, every node/.style={transform shape}]
			%\draw[help lines, very thin, gray] (-1,-6) grid (12,6);
			\draw[green, ultra thick] (0,2) -- (5,4);
			\draw[green, ultra thick] (0,-3) --(5,-.8);
			\draw[blue, very thick] (0,2) -- (0,0) -- (4,0);
			\draw[blue] (4,0) -- (9,2);
			\draw[blue, very thick] (0,-3) -- (0,-1) -- (4,-1);
			\draw[blue] (4,-1) -- (9,1);
			%\draw[blue, thick] (4,0) arc (90:-90:0.5);
			\draw[blue, very thick] (3.99,0) .. controls (4.5,-0.3) and (4.5,-0.7) .. (3.99,-1);
			\draw[blue] (9,2) .. controls (9.5,1.7) and (9.5,1.3) .. (9,1);
			
			\draw[blue] (0,0) -- (5,2) -- (5,4) -- (0,2);
			\draw[blue] (5,2) -- (9,2);
			\draw[blue] (0,-3) --(5,-.8) -- (5,-0.7);
			\draw[blue, dashed] (5,-0.7) -- (5,1) -- (9,1);
			\draw[blue, dashed] (0,-1) -- (5,1);
			
			\draw[red, very thick]  (-0.25,0.75) -- (-0.25,-0.25) -- (3.85,-0.25);
			\draw[red, very thick]  (-0.25,-1.75) -- (-0.25,-0.75) -- (3.85,-0.75);
			\draw[red, very thick] (3.84,-0.25).. controls (4.1,-0.4) and (4.1,-0.6) .. (3.84,-0.75);
			\draw[fill=purple] (4,-0.5) circle [radius=0.1];
			\draw[->,purple] (4.1,-0.6) .. controls (5,-0.9) and (4,-1.2) .. (5.5,-1.5); \node[right] at (5.5,-1.5) {$\mathcal{O}(T)$};
			\draw[fill=red] (-0.25,0.75) circle [radius=0.1];
			\draw[fill=red] (-0.25,-1.75) circle [radius=0.1];
			
			\draw[gray,->] (5,2) -- (7,2.8); \node[right] at (7,2.67) {$x$};
			\draw[->] (0.25,1.5) -- (0.25,0.5); \node[right] at (0.25,1) {$L$};
			\draw[->] (1.5,0.25) -- (3,0.25); \node[above] at (2.5,0.25) {$T$};
			\draw[gray,->] (-0.5,-0.5) -- (7,-0.5); \node[right] at (7,-0.5) {$t$};
			\draw[->] (-0.5,0.75) -- (-0.5,-0.25); \node[left] at (-0.5,0.5) {$l$};
			
			\draw[decoration={text along path, text={t=-iL}, text align={center}, raise=0.2cm},decorate] (0,2) -- (5,4);
			\draw[decoration={text along path, text={t=iL}, text align={center}, raise =.2cm},decorate] (0,-3) --(4,-1.4);
			\draw[decoration={text along path, text={t=0}, text align={center}, raise=0.2cm},decorate] (0,0) -- (5,2);
			\draw[decoration={text along path, text={t=T}, text align={center}, raise=0.2cm},decorate] (4,0) -- (9,2);
			
			\draw[gray] (-0.25,-0.25) -- (0,0);
			\draw[gray] (0.25,-0.25) -- (0.5,0);
			\draw[gray] (0.75,-0.25) -- (1,0);
			\draw[gray] (1.25,-0.25) -- (1.5,0);
			\draw[gray] (1.75,-0.25) -- (2,0);
			\draw[gray] (2.25,-0.25) -- (2.5,0);
			\draw[gray] (2.75,-0.25) -- (3,0);
			\draw[gray] (3.25,-0.25) -- (3.5,0);
			\draw[gray] (3.75,-0.25) -- (4,0);
			
			\draw[gray] (-0.25,-0.75) -- (0,-1);
			\draw[gray] (0.25,-0.75) -- (0.5,-1);
			\draw[gray] (0.75,-0.75) -- (1,-1);
			\draw[gray] (1.25,-0.75) -- (1.5,-1);
			\draw[gray] (1.75,-0.75) -- (2,-1);
			\draw[gray] (2.25,-0.75) -- (2.5,-1);
			\draw[gray] (2.75,-0.75) -- (3,-1);
			\draw[gray] (3.25,-0.75) -- (3.5,-1);
			\draw[gray] (3.75,-0.75) -- (4,-1);
			
			\draw[gray, thin] (4.25,-0.5) -- (4.5,-0.5);
		\end{tikzpicture}
		%	\caption{SK contour for the SYK and the bath in pure state. *** Show the SYK operator insertion at time $T$.}
		%	\label{fig:SK-contour-pure}
		%\end{figure}
	\end{minipage} \hfill
	\begin{minipage}{0.5 \linewidth}
		%\begin{figure}[H]
		\centering %change scale below to make figure bigger/smaller.
		\begin{tikzpicture}[scale=0.6, every node/.style={transform shape}]
			%\draw[help lines, very thin, gray] (-3,-6) grid (12,6);
			\draw[blue, very thick] (0,0) -- (4,0);
			\draw[blue] (4,0) -- (9,2);
			\draw[blue, very thick] (0,-1) -- (4,-1);
			\draw[blue] (4,-1) -- (9,1);
			\draw[blue, very thick] (0,0) arc(20:341:1.5);
			\draw[blue] (-2,0.88) -- (3,2.88);
			\draw[blue] (5,2) arc(20:113:1.5);
			\draw[blue, dashed] (3,2.88) arc(113:340:1.5);
			\draw[blue] (-1.0,-1.97) -- (1.4,-1);
			
			\draw[blue, very thick] (3.99,0) .. controls (4.5,-0.3) and (4.5,-0.7) .. (3.99,-1);
			\draw[blue] (9,2) .. controls (9.5,1.7) and (9.5,1.3) .. (9,1);
			
			\draw[blue] (0,0) -- (5,2);
			\draw[blue] (5,2) -- (9,2);
			\draw[blue, dashed] (5,1) -- (9,1);
			\draw[blue, dashed] (0,-1) -- (5,1);
			
			\draw[red, very thick]  (-0.25,0.75) -- (-0.25,-0.25) -- (3.85,-0.25);
			\draw[red, very thick]  (-0.25,-1.75) -- (-0.25,-0.75) -- (3.85,-0.75);
			\draw[red, very thick] (3.84,-0.25).. controls (4.1,-0.4) and (4.1,-0.6) .. (3.84,-0.75);
			\draw[fill=purple] (4,-0.5) circle [radius=0.1];
			\draw[->,purple] (4.1,-0.6) .. controls (5,-0.9) and (4,-1.2) .. (5.5,-1.5); \node[right] at (5.5,-1.5) {$\mathcal{O}(T)$};
			\draw[fill=red] (-0.25,0.75) circle [radius=0.1];
			\draw[fill=red] (-0.25,-1.75) circle [radius=0.1];
			
			\draw[gray,->] (5,2) -- (7,2.8); \node[right] at (7,2.67) {$x$};
			\draw[->] (1.5,0.25) -- (3,0.25); \node[above] at (2.5,0.25) {$T$};
			\draw[gray,->] (-0.5,-0.5) -- (7,-0.5); \node[right] at (7,-0.5) {$t$};
			\draw[->] (-0.5,0.75) -- (-0.5,-0.25); \node[left] at (-0.5,0.5) {$l$};
			\node[right] at (-2.9,-0.5) {$\beta_{b}$};
			
			%\draw[decoration={text along path, text={$\beta$}, text align={center}},decorate] (0,0) arc [start angle=10, end angle=350, x radius=1cm, y radius=3cm];
			\draw[decoration={text along path, text={t=0}, text align={left indent={0.6\dimexpr\pgfdecoratedpathlength\relax}}, raise=0.2cm},decorate] (0,0) -- (5,2);
			\draw[decoration={text along path, text={t=T}, text align={center}, raise=0.2cm},decorate] (4,0) -- (9,2);
			
			\draw[gray] (-0.25,-0.25) -- (0,0);
			\draw[gray] (0.25,-0.25) -- (0.5,0);
			\draw[gray] (0.75,-0.25) -- (1,0);
			\draw[gray] (1.25,-0.25) -- (1.5,0);
			\draw[gray] (1.75,-0.25) -- (2,0);
			\draw[gray] (2.25,-0.25) -- (2.5,0);
			\draw[gray] (2.75,-0.25) -- (3,0);
			\draw[gray] (3.25,-0.25) -- (3.5,0);
			\draw[gray] (3.75,-0.25) -- (4,0);
			
			\draw[gray] (-0.25,-0.75) -- (0,-1);
			\draw[gray] (0.25,-0.75) -- (0.5,-1);
			\draw[gray] (0.75,-0.75) -- (1,-1);
			\draw[gray] (1.25,-0.75) -- (1.5,-1);
			\draw[gray] (1.75,-0.75) -- (2,-1);
			\draw[gray] (2.25,-0.75) -- (2.5,-1);
			\draw[gray] (2.75,-0.75) -- (3,-1);
			\draw[gray] (3.25,-0.75) -- (3.5,-1);
			\draw[gray] (3.75,-0.75) -- (4,-1);
			
			\draw[gray, thin] (4.25,-0.5) -- (4.5,-0.5);
		\end{tikzpicture}
	\end{minipage}
	\caption{Schwinger-Keldysh contours: The horizontal direction is the real time and the transverse direction is the spatial extent of the bath theory. In the figure on the left, the vertical direction is the imaginary time direction. The bath (indicated by blue contours) is prepared in the Calabrese-Cardy state at $t=0$. It is prepared by a time evolution along the imaginary time, starting from $t=-iL$, indicated by the vertical blue lines. The green lines at $t = \pm i L$ are depicting the $\ket{Bd}$ ($\bra{Bd}$) state. The blue contours are joined up smoothly at the right end (at Lorentzian time $t=T$) as the operator which we have inserted there has support only on the SYK Hilbert space. The SYK contours (red) are similar, except for the insertion of an operator $\mathcal{O}$ at the right end ($t=T$), indicated by the purple dot. The figure on the right is similar to the left figure, except that the bath is in a thermal state at $t=0$ and we have depicted it by a closed circle of length $\beta_{b}$ in the imaginary time. In both the figures, gray lines between red and blue contours are indicating the interaction between the two systems and as depicted it is only present along the real time.}
	\label{fig:SK-contour}
\end{figure}
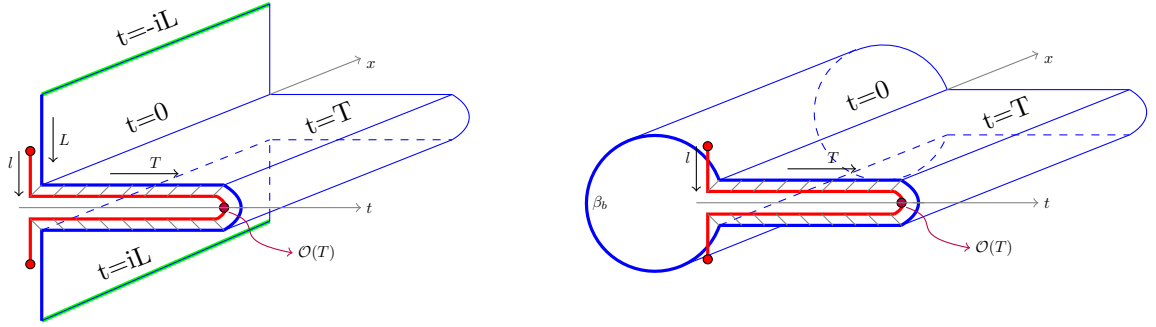
%}
\gap2
As we presently consider operators that have support only on the SYK Hilbert space, we might as well work with the reduced density matrix $RDM$ which can be obtained by taking a trace over the bath Hilbert space (see Figure \ref{SKc-main}):
\begin{equation}\label{RDM}
	\hat{\rho}(T) = \text{Tr}_{bath}\left[e^{-i H T}\, \hat{\rho}_{Sch} \otimes \hat{\rho}_{bath}\, e^{i H T}  \right] ,
\end{equation}
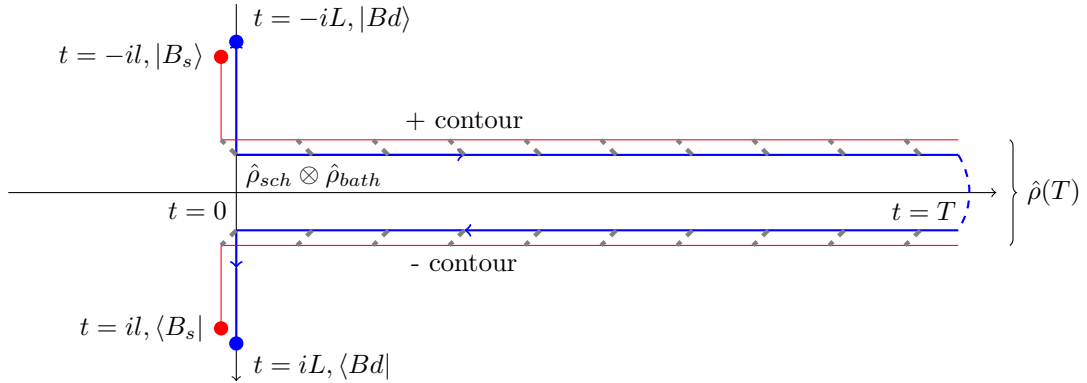
\begin{figure}[]
	\centering
	\begin{tikzpicture}
	\node[above right] at (3.1,2) {$t = -iL, \ket{Bd}$};
	\node[below right] at (3.1,-2) {$t = iL, \bra{Bd}$};
	\node[above left] at (2.7,1.5) {$t = -il, \ket{B_s}$};
	\node[below left] at (2.7,-1.5) {$t = il, \bra{B_s}$};
	
	\node[above right] at (3,-.05) {$\hat{\rho}_{sch}\otimes \hat{\rho}_{bath}$};
	%\node[below right] at (3,.05) {$\bra{B_s(l)}\otimes\bra{\Psi_0(L)}$};
	
	%\node[above right] at (12,1) {$e^{-iHT}\ket{B_s(l)}\otimes\ket{\Psi_0(L)}$};
	\node[above] at (6,0.7) {+ contour};
	\node[below] at (6,-0.7) {- contour};
	\draw[decoration={brace,raise=5pt},decorate] (13,.7) -- node[right=8pt] {$\hat{\rho}(T)$} (13,-.7);
	\draw [<-](3,-2.5)--(3,2.5);
	\draw [->,blue,thick] (3,2)--(3,2);
	%\draw [very thick,green] (2.8,2.9)--(3.2,2.9);
	\draw [blue,thick](3,2)--(3,0.5);
	\draw [->,blue,thick] (3,-.5)--(3,-1);
	\draw [blue,thick] (3,-1)--(3,-2);
	\draw [->](0,0)--(13,0);
	\draw [->,thick,blue](3,.5)--(6,.5);
	\draw [thick,blue](3,.5)--(12.5,.5);
	\draw [->,thick,blue](12.5,-.5)--(6,-.5);
	\draw [thick,blue](6,-.5)--(3,-.5);
	% \draw [dashed] (3,1)--(3,-1);
	%\draw [dashed] (5,1)--(5,-1);
	%\draw[very thick,green] (5,.3)--(5,.7);
	%\draw[very thick,green] (8,.3)--(8,.7);
	\draw[fill,blue,thick] (3,2) circle [radius=0.08];
	\draw[fill,blue,thick] (3,-2) circle [radius=0.08];
	%\draw[fill,green,thick] (3,2.9) circle [radius=0.08];
	%\draw[fill,green,thick] (5,.5) circle [radius=0.08];
	
	\draw [red](2.8,-1.8)--(2.8,-.7);
	\draw[red] (2.8,-.7)--(12.5,-.7);
	\draw [red](2.8,1.8)--(2.8,.7);
	\draw[red] (2.8,.7)--(12.5,.7);
	%\draw[dashed,red] (12.5,.7) to [out=315,in=45] (12.5,-.7);
	%\draw[green] (12.5,.7) -- (12.5,-.7);
	\draw[fill,red,thick] (2.8,-1.8) circle [radius=0.08];
	\draw[fill,red,thick] (2.8,1.8) circle [radius=0.08];
	\draw[gray,dashed,ultra thick] (3,.5)--(2.8,.7);
	\draw[gray,dashed,ultra thick] (4,.5)--(3.8,.7);
	\draw[gray,dashed,ultra thick] (5,.5)--(4.8,.7);
	\draw[gray,dashed,ultra thick] (6,.5)--(5.8,.7);
	\draw[gray,dashed,ultra thick] (7,.5)--(6.8,.7);
	\draw[gray,dashed,ultra thick] (8,.5)--(7.8,.7);
	\draw[gray,dashed,ultra thick] (9,.5)--(8.8,.7);
	\draw[gray,dashed,ultra thick] (10,.5)--(9.8,.7);
	\draw[gray,dashed,ultra thick] (11,.5)--(10.8,.7);
	\draw[gray,dashed,ultra thick] (12,.5)--(11.8,.7);
	%\draw[gray,dashed,ultra thick] (12.5,.5)--(12.3,.7);
	
	\draw[gray,dashed,ultra thick] (3,-.5)--(2.8,-.7);
	\draw[gray,dashed,ultra thick] (4,-.5)--(3.8,-.7);
	\draw[gray,dashed,ultra thick] (5,-.5)--(4.8,-.7);
	\draw[gray,dashed,ultra thick] (6,-.5)--(5.8,-.7);
	\draw[gray,dashed,ultra thick] (7,-.5)--(6.8,-.7);
	\draw[gray,dashed,ultra thick] (8,-.5)--(7.8,-.7);
	\draw[gray,dashed,ultra thick] (9,-.5)--(8.8,-.7);
	\draw[gray,dashed,ultra thick] (10,-.5)--(9.8,-.7);
	\draw[gray,dashed,ultra thick] (11,-.5)--(10.8,-.7);
	\draw[gray,dashed,ultra thick] (12,-.5)--(11.8,-.7);
	%\draw[gray,dashed,ultra thick] (12.5,-.5)--(12.3,-.7);
	\node [below left] at (3,0) {$ t =0$}; 
	%\draw [dashed] (8,1)--(8,-1);
	%\node [below right] at (8.1,0) {$t = t$};
	%\node [below right] at (5,0) {$t = t'$};
	\node [below right] at (11.5,0) {$t = T $};
	\draw[blue, thick, dashed] (12.5,.5) to [out=300,in=60] (12.5,-.5);
	%\draw[blue,very thick,dashed] (0,2)--(10,2);
	%\draw[blue,very thick,dashed] (0,-2)--(10,-2);
	%\draw[fill,green,thick] (8,.5) circle [radius=0.08];
	% \draw[decoration={brace,mirror,raise=5pt},decorate] (3,.6) -- node[right=8pt] {$l'$} (3,3);
	%\draw[decoration={brace,raise=5pt},decorate] (3,-.6) -- node[right=8pt] {$l$} (3,-2);
	\end{tikzpicture}
	\caption{ This is the same diagram as Figure \ref{fig:SK-contour} with the $x$ direction suppressed. The bath contour is depicted in blue and the SYK path is depicted by red. The blue and red dots on the top depict the kets $\ket{Bd}$ and $\ket{B_s}$ respectively and on the bottom depicts the bras $\bra{Bd}$ and $\bra{B_s}$ respectively. The state at $t=0$ is $\hat{\rho}_{sch}\otimes \hat{\rho}_{bath}$ and the state at $t=T$ is the reduced density matrix defined as $\hat{\rho}(T)=\text{Tr}_{bath}\left[e^{-iHT}\hat{\rho}_{sch}\otimes \hat{\rho}_{bath}e^{iHT}\right]$. The tracing out of the bath fields is depicted by the blue dashed line connected the upper and lower bath contour. The `+' (upper) contour and the `-' (lower) contour are at $\text{Im}(t)=0$, the gap in the above figure is only shown for better presentation, the value of real time along both contours is also the same. Hence, to distinguish fields on the upper contour from the fields on the lower contour we add a $\pm$ superscript.}
	\label{SKc-main}
\end{figure}
and the expectation value of the SYK operator can also computed from the $RDM$
\begin{equation}\label{rdmev}
	\langle \mathcal{O}(T)\rangle = \text{Tr}\left[ \hat{\rho}(T)\, \mathcal{O} \right] .
\end{equation}
As indicated earlier, our primary aim is to get the evaporating black hole solution in terms of the Liouville mode $\phi(t)$. We would like to write down a path integral expression for \eq{RDM} or \eq{rdmev}. The action for this path integral would be \eq{main-model} with the path integral contour described in Figure \ref{fig:SK-contour}. The SYK action in \eq{main-model} is already projected onto the low energy sector and the physics is described by the Liouville mode. We will also integrate out the bath fields from the path integral and obtain an effective action in which the Liouville mode will be the only dynamical variable. The details of the computation described above are presented in Appendix \ref{app:bath_details}. The final result is
\begin{align}
	\lan \mathcal{O} \ran &= \int d\phi^+_T d\phi^-_T\  \mathcal{O}(\phi_T^+,\phi^-_T)\  \int_{\phi_0}^{\phi^+_T} D\phi^+ \int_{\phi_0}^{\phi^-_T}\, D\phi^- \, e^{i S_{SYK}[\phi^+] - i S_{SYK}[{\phi^-}]} \,e^{N_f\,W[F(\phi^+),F({\phi^-})]} \nonumber \\
	& = \int d\phi^+_T d\phi^-_T\  \mathcal{O}(\phi_T^+,\phi^-_T)\  \int_{\phi_0}^{\phi^+_T} D\phi^+ \int_{\phi_0}^{\phi^-_T}\, D\phi^-\ e^{i S_{SK}[\phi^+,\phi^-]} 
	\label{ev-path-integral}
\end{align}
where $S_{SK}$ is given in \eq{S-sk} and the Feynman-Vernon influence functional $W[F(\phi^+),F(\phi^-)]$ appears as a result of integrating out the bath fields and is given by
\begin{align}\label{FVIF}
	W[F(\phi^+),F(\phi^-)] = g^2 & \Bigg[\int_0^T dt\, dt' F(\phi^+(t)) \kappa_{++}(t,t') F(\phi^+(t')) + \int_0^T dt\, dt' F(\phi^-(t)) \kappa_{--}(t,t') F(\phi^-(t'))\nonumber \\
	&+ 2\int_0^T dt\, dt' F(\phi^+(t)) \kappa_{+-}(t,t') F(\phi^-(t')) \Bigg].
\end{align}
The explicit form of the kernels is given in equation \eqref{kernels-pure} for the pure CC state and in equation \eqref{kernels-thermal} for the thermal state. The contour for the path integral is given in Figure \ref{SKc-main}. The field $\phi^+$($\phi^-$) lives on the `$+$' (`$-$') contour. 

The first thing to note is that the influence functional is non-local. The kernel $\kappa_{++}$ gives bilocal interactions on the `$+$' contour, $\kappa_{--}$ gives bilocal interactions on the `$-$' contour while $\kappa_{+-}$ leads to interactions with one leg each on the `$+$' and `$-$'  contours. This structure can be simply understood as arising after integrating a quadratic field leading to the standard functional form $e^{J.\Delta.J}$ for the `source' $J=F(\phi)$, which lives on both $+$ and $-$ contours. Details of this calculation are provided in Appendix \ref{app:effective_action}. In \eq{ev-path-integral}, $\mathcal{O} (\phi^+_T,\phi^-_T)$ = $\lan \phi^+_T | O_{\rm SYK} | \phi_T^-\ran$, representing the operator insertion at time $t=T$ (see Fig \ref{fig:SK-contour}). \\

Together with the Influence functional, the effective Schwarzian action on the Schwinger-Keldysh contour reads
\begin{align}\label{S-sk}
	S_{SK}[\phi^+,\phi^-]=& S_{Sch}[\phi^+] -S_{Sch}[\phi^-] -i N_f\, W[F(\phi^+),F(\phi^-)] \nonumber \\
	=& N \left\{\frac{\alpha_s}{J} \int_0^T dt \left[\frac{1}{2} \dot{\phi^+}^2 - V(\phi^+)\right] - \frac{\alpha_s}{J} \int_0^T dt \left[\frac{1}{2} \dot{\phi^-}^2 - V(\phi^-)\right] -i n_f W[F(\phi^+),F(\phi^-)] \right\} ,
\end{align}
with 
\begin{equation}\label{v-f-def}
	V(\phi) =2 J^2 \exp(\phi) .
\end{equation}

Note that in \eqref{ev-path-integral}, the Influence functional $W$ comes with a prefactor $N_F$, the number of scalar fields. Therefore the effective action, with the scaling given in \eq{nf-scaling} has an overall faction of $N$ and hence the path integral will be saturated by the large $N$ solution of the Liouville mode. The insertion of the operator $\mathcal{O}$ will have no effect on the large $N$ solution of the Liouville mode as long as it is order $1$ in the $N$ scaling, which is the case we consider.

The variation of $S_{SK}$ with respect to the Liouville fields $\phi^+(t)$ and ${\phi^-}(t)$ lead to the following equations of motion for $\phi^+(t)$ and $\phi^-(t)$ respectively:
\begin{align}
	&\frac{\alpha_s}{J} \left[\ddot{\phi}^+(t)+ V'(\phi^+(t)) \right] + i n_f g^2 \left[ F'\left(\phi^+(t)\right) \int_0^T dt' \kappa_{++}^S(t,t') F\left(\phi^+(t')\right) + 2 F'\left(\phi^+(t)\right) \int_0^T dt' \kappa_{+-}(t,t') F\left(\phi^-(t')\right) \right] =0 ,
	\label{phi-eom}\\
	&\frac{\alpha_s}{J} \left[\ddot{\phi}^-(t) + V'(\phi^-(t))\right] -i n_f g^2 \left[F'\left(\phi^-(t)\right) \int_0^T dt' \kappa_{--}^S(t,t') F\left(\phi^-(t')\right) + 2F'\left(\phi^-(t)\right) \int_0^T dt' F\left(\phi^+(t')\right)\kappa_{+-}(t',t) \right] =0 .
	\label{phi-bar-eom}
\end{align}
Notice that only the symmetric kernels
\begin{equation}
	\kappa_{++}^S(t,t') = \kappa_{++}(t,t')+ \kappa_{++}(t',t); \quad  \kappa_{--}^S(t,t') = \kappa_{--}(t,t')+ \kappa_{--}(t',t)
\end{equation}
appear in the equation of motion. %$\kappa_3(t,t')$ is not symmetric in its arguments while the remaining part is inherently symmetric.

%%%%%
\paragraph{Some properties of the kernels}
\begin{align}
	& \text{Re}\left[\kappa_{++}(t,t') + \kappa_{+-}(t,t') \right] =0 , \nonumber \\
	& \text{Im}\left[\kappa^S_{++}(t,t') +2\kappa_{+-}(t,t') \right]= [1+\text{sgn}(t-t')]/2=\theta(t-t') , \nonumber\\
	&2\text{Re}\left[\kappa_{++}(t,t')- \kappa_{+-}(t,t')\right] = K(t,t') , \label{kappa-relations}
\end{align}
where the kernel $K(t,t')$ depends on whether the bath initially is in a pure state or in a thermal state. In case of the pure state,
\begin{align}
	K(t,t') &= -\frac{2}{\pi} \int_{0}^{\infty} \frac{dk}{k} \left[\cos(kt)\cos(kt') \tanh(kL) + \sin(kt)\sin(kt')\coth(kL) \right] \nonumber \\
	&= -\frac{1}{2L}\big(t+t'-|t-t'|\big) - \frac{2}{\pi} \log\left[ \frac{1+e^{-\pi|t+t'|/(2L)}}{1-e^{-\pi|t-t'|/(2L)}} \right] .
	\label{kernel-pure}
\end{align}
It is easy to verify that this kernel is negative definite. Similarly when the bath is initially prepared in thermal state at inverse temperature $\beta_b$,
\begin{align}
	K(t,t') &= -\frac{2}{\pi} \int_{0}^{\infty} \frac{dk}{k} \cos[k(t-t')] \coth\left(\frac{k\beta_b}{2}\right) \nonumber \\
	&= \frac{2}{\beta_b} |t-t'| +\frac{2}{\pi} \log\left[1-e^{-2\pi|t-t'|/\beta_b}\right] \equiv K_{\b_b}(t-t') .
	\label{kernel-thermal}
\end{align}
The kernels have the usual UV divergence for coincident points which goes like $\log|t-t'|$ in 2 dimensions.\footnote{At finite UV cutoff, this divergence is regulated. See appendix \ref{UVkernel} for details.} It is worth noting that the $t-t'$ dependent part of the kernel in the Calabrese-Cardy state is the same as in the thermal state with the identification $\beta_b=4L$.\footnote{This is known from previous studies of correlation functions in (generalized)-Calabrese-Cardy states, see e.g. \cite{Banerjee:2019ilw}.}
\\~\\
Note here that $\log|t-t'|$ behaviour of the above kernel is attained in the large $\beta_b$ limit, whereas in the opposite limit of small $\beta_b$, one obtains the linear term proportional to $|t-t'|$ --- this is the one-dimensional Green's function, as one would expect in a Kaluza-Klein reduction in the presence of a thermal circle.
		
%%%%%%%%%%%%%%%%%%%%
\subsection*{Keldysh rotation}\label{subsec:Keldysh}
We have obtained the effective action \eq{S-sk} on the Schwinger-Keldysh contour Figure \ref{SKc-main} only in terms of Schwarzian mode. We can now study its dynamics more systematically. It is convenient to switch to following variables\footnote{Sometimes these are called classical-quantum variables\cite{kamenev2011field} or the average-difference variables\cite{Haehl:2016pec}. However the name classical-quantum is misleading since $\phi_c$ can also have fluctuations.} 
\begin{equation}
	\phi_c(t) =\frac{1}{2}\left[\phi^+(t) + \phi^-(t)\right] , \qquad \phi_{q}(t) =\frac{1}{2}\left[\phi^+(t) - \phi^-(t)\right] ,
\end{equation}
in terms of which our effective action reads
\begin{equation}
	S_{SK} = N\left\{\frac{\alpha_s}{J}\int_0^T dt \left\{2\dot{\phi}_c \dot{\phi}_q -V(\phi_c+\phi_{q}) +V(\phi_c-\phi_{q}) \right\} -i n_f W[F(\phi_c+\phi_q), F(\phi_c-\phi_q)] \right\} .
	\label{full-kamenev}
\end{equation}
%In these variables, one can verify that
%\begin{align}
%	& \phi_q(t)=0 \nonumber\\
%	& \frac{\alpha_s}{J}\left[\ddot{\phi}_c(t) +V'(\phi_c(t))\right] -n_f g^2 F'(\phi_c(t))\int_0^t dt' F(\phi_c(t'))=0
%\end{align}
We can compute the two equations of motion of this action and verify that
\begin{align}
	& \phi_q(t)=0 , \label{full-solution-q} \\
	& \frac{\alpha_s}{J}\left[\ddot{\phi}_c(t) +V'(\phi_c(t))\right] -n_f g^2 F'(\phi_c(t))\int_0^t dt' F(\phi_c(t'))=0 , \label{full-solution-c}
\end{align}
constitute exact solutions of the above system \eq{full-kamenev} to all orders in $\phi_q$. In terms of the $\pm$ variables, \eqref{full-solution-q} is simply $\phi^+(t) = \phi^-(t)$. With this, the two equations \eq{phi-eom} and \eq{phi-bar-eom} reduce to \eqref{full-solution-c} above with $\phi^+=\phi^-=\phi_c$.

%%%%%%%%%%%%%%%%%%%%%%%%%%%%%%%%%%%%%%%%%%
\subsection{Fluctuations and a Langevin equation}
Let us now look at the fluctuation 
\begin{equation}
	\phi_q(t) = \phi^0_q(t) + \f1{\sqrt{N}}\, \delta \phi_q (t) + O\left(\f{1}{N} \right)
\end{equation}
around the solution $\phi^0_q =0$. In this case, with the use of explicit relations involving the kernels \eqref{kappa-relations}, the action can be rearranged in powers of $N$ as follows
\begin{comment}
\begin{align}\label{kamenev-expansion-tmp}
S_{SK} =& -2 i N g^2 \int_0^T dt \int_0^T dt'\ F(\phi_c(t)) \text{Re}\left[\kappa_{++}(t,t') + \kappa_{+-}(t,t')\right] F(\phi_c(t')) \nonumber \\
&-2 \sqrt{N} \int_0^T dt\ \delta\phi_q(t) \left[\frac{\alpha_s}{J}\left(\ddot{\phi}_c(t) +V'(\phi_c(t))\right) -n_f g^2 F'(\phi_c(t))\int_0^T dt' \text{Im}\left[\kappa^S_{++}(t,t')+\kappa^A_{+-}(t,t')\right] F(\phi_c(t')) \right] \nonumber \\
&- 2 i n_f g^2 \int_0^T dt \int_0^T dt'\ \delta{\phi}_q(t)\left[ F'(\phi_c(t)) \delta{\phi}_q(t) \text{Re}\left[\kappa_{++}(t,t')- \kappa_{+-}(t,t')\right] F'(\phi_c(t'))\right] \delta{\phi}_q(t') + O\left(N^{-3/2}\right).
\end{align}
\end{comment}
\begin{align}\label{kamenev-expansion}
	S_{SK} = &-2 \sqrt{N}\int_0^T dt\, \delta{\phi}_q(t) \left[\frac{\alpha_s}{J}\left(\ddot{\phi}_c(t) +V'(\phi_c(t))\right) -n_f g^2 F'(\phi_c(t))\int_0^T dt' \theta(t-t') F(\phi_c(t'))\right] \nonumber \\
	&+ \f{i}{2} \int_0^T dt \int_0^T dt'\, \delta{\phi}_q(t)\, \tilde K(t,t')\, \delta{\phi}_q(t') + O\left(\frac{1}{\sqrt{N}}\right) ,
\end{align}
where
\begin{equation}
	\tilde{K}(t,t') = -2 n_f g^2 F'(\phi_c(t)) K(t,t') F'(\phi_c(t')) .
\end{equation}
%is a positive definite operator, since $K(t,t')$ for the pure state \eqref{kernel-pure} was a negative definite operator. 
%In the action above, 
Therefore up to this order, we have quadratic action for the fluctuations $\delta{\phi}_q(t')$ and the factors of $N$ are organized such that $\braket{\delta{\phi}_q(t') \delta{\phi}_q(t)} \sim \tilde K^{-1} \sim O(1)$ in $N$ scaling. %Note that due to the structure of the action \eq{kamenev-expansion} $\delta \phi_q(t) =0$ is a solution to all orders in $N$.

We will now replace the quadratic term in $\delta\phi_q$ using an auxiliary variable $\eta$ and the following identity 
\begin{equation}
	\exp\left[{-\frac{1}{2} \int_0^T dt\, dt' \delta{\phi}_q(t) \tilde{K}(t,t') \delta{\phi}_q(t')}\right] = \mathcal{N} \int[D\eta] \exp\left[{-\frac{1}{2} \int_0^T dt \, dt' \eta(t) \tilde{K}^{-1}(t,t') \eta(t') + i\int_0^T dt\, \eta(t) \delta{\phi}_q(t)}\right] .
\end{equation}
With this new variable $\eta$, the action is now linear in $\delta{\phi}_q$,
\begin{align}\label{kamenev-expansion-N2}
	S_{SK} = &-2 \sqrt{N}\int_0^T dt\, \delta{\phi}_q(t) \left[\frac{\alpha_s}{J}\left(\ddot{\phi}_c(t) +V'(\phi_c(t))\right) -g^2 F'(\phi_c(t))\int_0^T dt' \theta(t-t') F(\phi_c(t')) -\frac{1}{2}\f{\eta(t)}{\sqrt{N}} \right] \nonumber \\
	&+\frac{i}{2} \int_0^T dt \, dt' \, \eta(t) \tilde{K}^{-1}(t,t') \eta(t') + O(N^{-1/2}) . 
\end{align}
%and the `noise' correlation $\braket{\eta(t)\eta(t')} \sim \tilde K \sim O(1)$ in $N$ scaling. 
Now variation with respect to $\delta{\phi}_q$ leads us to a Langevin-type equation
\begin{equation}\label{Langevin}
	\frac{\alpha_s}{J}\left(\ddot{\phi}_c(t) +V'(\phi_c(t))\right) - n_f g^2 F'(\phi_c(t))\int_0^t dt' F(\phi_c(t')) = \frac{1}{2} \f{\eta(t)}{\sqrt{N}} ,
\end{equation}
with the following `noise' correlation 
\begin{equation}
	\langle \eta(t) \, \eta(t')\rangle = \tilde{K}(t,t') = -2 n_f g^2 F'(\phi_c(t)) K(t,t') F'(\phi_c(t')) . \label{2pt-noise}
\end{equation}

This Langevin equation is complicated for two reasons. Firstly, the left hand side has an integral which reflects that the equation is non-Markovian and has memory of the dynamics from time $0$ to $t$. Secondly, the Gaussian noise is correlated in a non-local way. It is of great interest to calculate correlation functions of this model and also the time dependence of the probability distribution of the stochastic variable $\phi_c(t)$. The asymptotic behaviour of the correlation functions and the distribution function will not necessarily be thermal like standard Brownian motion (as we will see in Section \ref{sec:2pt-B}) and will carry information about the initial micro-state in terms of the scalar products of the spin of the micro-state $|B_s\ran$ with the spins of the operators $\eqref{om}$ that are coupled to the bath. 

In Section $\ref{sec:fluctuations-2pt}$ we will present a calculation of the equal time correlation function of fluctuations around the two specific solutions of the Langevin equation that correspond, at large times, to a) a black hole at a lower temperature where we verify the Einstein-Smoluchowski relation, and b) to a spacetime without a horizon (complete evaporation) where the large time behaviour is oscillatory (see Figure \ref{fig:2pt-b}, Section \ref{sec:2pt-B}) and characteristic of Brownian motion in a bounded potential.

To tackle the problem of finding an analogue of the Kolmogorov-Fokker-Planck (KFP) equation for the probability distribution it seems best to recast the second order equation $\eqref{Langevin}$ into a local third order equation in time.
\begin{align}\label{third-order-general-noise}
	\frac{\alpha_s}{J} & \left[ F'[\phi_c] \dddot{\phi_c} -F''[\phi_c] \ddot{\phi}_c \dot{\phi}_c + \left\{ F'[\phi_c] V''(\phi_c) - F''[\phi_c] V'(\phi_c) \right\} \dot{\phi}_c \right] - n_f g^2 \left(F'[\phi_c]\right)^2 F[\phi_c] \nonumber \\
	&= \frac{1}{2\sqrt{N}} \left(F'[\phi_c] \,\dot\eta - F''[\phi_c] \,\dot\phi_c \,\eta\right) .
\end{align}
%\begin{align}%\label{third-order-general-noise}
%	&\frac{\alpha_s}{J} \left[ \dddot{\phi}_c + V''(\phi_c) \dot\phi_c - \frac{F''[\phi_c]}{F'[\phi_c]} \left\{ \ddot\phi_c + V'(\phi_c)\right\} \dot{\phi}_c \right] - n_f g^2 F'[\phi_c] F[\phi_c] \nonumber \\
%	&= \frac{1}{2\sqrt{N}} \left( \dot\eta - \frac{F''[\phi_c]}{ F'[\phi_c]} \dot\phi_c \eta \right)
%\end{align}
The proof of the equivalence of the two equations \eqref{Langevin} and \eqref{third-order-general-noise} is done separately for the cases when $F'[\phi_c]\neq 0$ and $F'[\phi_c]=0$. This third order differential equation is supplied by the two initial conditions $\phi_c(0)$, $\dot\phi_c(0)$ and the third initial condition $\ddot\phi_c(0)$ is determined in terms of $\phi_c(0)$ and $\dot\phi_c(0)$ using $\eqref{Langevin}$. Recasting the third order equation into three first order equations for the `position' $x\equiv \phi_c$, velocity $v$, and acceleration $a$, we can write
\begin{align}
	&\dot x = v, \qquad \dot v = a \qquad \text{and} \nonumber \\
	&\dot a - \frac{F''[x]}{F'[x]} a v - \left\{V''(x) - \frac{F''[x]}{F'[x]} V'(x) \right\}v - \frac{J}{\alpha_s} n_f g^2 F'[x] F[x] = \frac{J}{2\alpha_s\sqrt{N}} \left(\dot\eta - \frac{F''[x]}{F'[x]} v \eta\right) .
\end{align}
This can be useful in establishing a KFP type equation for the probability function $P(x,v, a,t)$. Integrating  $P(x,v,a,t)$ over $v$ and $a$ will give the probability distribution of $x=\phi_c$. 

It must be noted that for a generalized KFP equation with greater than two (but finite) number of derivatives, the distribution function ceases to be positive definite. This is known as Pawula's theorem \cite{Pawula1967GeneralizationsAE}.\footnote{We thank Neha Wadia for pointing this out to us.} However as suggested in \cite{Risken1979OnTA}, from a practical viewpoint it is still sometimes useful to work with a finite number of derivatives.

\subsection{Solving the Langevin equation in a perturbative expansion in $1/\sqrt N$}

At large $N$, the RHS of equation \eqref{Langevin} vanishes. The effect of the noise field $\eta(t)$ can only be felt by the fluctuations in $\phi_c(t)$ 
\begin{equation}
	\phi_c(t) = \phi(t) + \frac{1}{\sqrt{N}}\, \delta \phi_c(t) + O\left( \frac{1}{N} \right) .
\end{equation}
Equation \eqref{Langevin} then leads to the following two equations 
\begin{align}\label{ph-cl-eqn}
	& \frac{\alpha_s}{J}\left(\ddot{\phi}(t) +V'(\phi(t))\right) -n_f g^2 F'(\phi(t))\int_0^t dt' F(\phi(t')) =0 , \\
	& \frac{\alpha_s}{J} \left( \delta \ddot{\phi}_c(t) + V''(\phi(t)) \delta \phi_c(t)\right)\nonumber \\ 
	& \hspace{1.65cm} -n_f g^2 \left\{ F''(\phi(t))\, \delta \phi_c(t) \int_0^t dt' F(\phi(t')) + F'(\phi(t)) \int_0^t dt' F'(\phi(t'))\, \delta \phi_c(t') \right\} = \frac{1}{2}\eta(t) . \label{Langevin-separation}
\end{align}
$\phi(t)$ is the solution at strictly large $N$ without any noise and will be studied in detail in Section \ref{sec:results}. The two-point function of $\delta \phi_c(t)$ will be studied in Section \ref{sec:fluctuations-2pt} in the background of the solutions of $\eqref{ph-cl-eqn}$. %with a simple large time behaviour $\phi (t)=-at$.\\

%%%%%%%%%%%%%%%%%%%%%%%%%%%%%%%%%%%%%%%%
\section{Schwarzian + bath dynamics to leading order in large N: description and results} \label{sec:results}
		
In this section, we solve the leading large-$N$ equation \eqref{ph-cl-eqn}, which we reproduce here
%will explore the system with $\phi_q(t)=0$. For convenience, we rewrite \eq{ph-cl-eqn}  with $\phi_c \equiv \phi(t)$:\footnote{Note that for $\phi_q=0$, $\phi_c(t)= \phi(t)= \bar\phi(t)$.}
\begin{align}
	\frac{\alpha_s}{J}\left(\ddot{\phi}(t) +V'(\phi(t))\right) -n_f g^2 F'(\phi(t))\int_0^t dt' F(\phi(t'))=0 .
	\label{ph-cl-eqn-a}
\end{align}
Here $F$ is given by \eq{F-final}; we will treat the two cases of only the relevant coupling \eq{relevant} in Model (a) and the mixed coupling \eq{mixed} in Model (b) separately below. The potential $V$ is given by \eq{v-f-def}. As explained in \cite{Kourkoulou:2017zaj}, (see Appendix \ref{app:init-cond} for more details) we take the initial conditions as
\begin{align}
	\phi(0)= \ln\left(\fr\pi{\b J} \right)^2, \quad \dot\phi(0)=0 .
	\label{init-phi-cl}
\end{align}
We can get rid of the integral in \eqref{ph-cl-eqn-a} by taking one more time derivative, which leads to the third order differential equation at leading order in $N$
\begin{equation}\label{third-order-general}
	F'[\phi(t)] \dddot{\phi} -F''[\phi(t)] \ddot{\phi} \dot{\phi} + \left\{ F'[\phi(t)] V''(\phi(t)) - F''[\phi(t)] V'(\phi(t)) \right\} \dot{\phi} - \frac{J}{\alpha_s} n_f g^2 \left(F'[\phi(t)]\right)^2 F[\phi(t)]=0 .
\end{equation}
We can now solve this differential equation by using the initial conditions \eqref{init-phi-cl} and $\ddot{\phi}(0)=-2J^2 e^{\phi(0)}$. Note that even though this is a third order differential equation, only two sets of initial data are needed for the time evolution of our system and is therefore a causal equation.\footnote{Generically 3rd order equations can have solution which are acausal. The Abraham-Lorentz equation, which describes the non-relativistic motion of an accelerating charge particle, is a well-known third order equation in electrodynamics (see for example \cite{jackson2012classical}). It is known to have pathological solutions in which the particle accelerates before any force is even applied.} This is because $\phi(0)$ and $\dot{\phi}(0)$ are the only two independent initial conditions, while $\ddot{\phi}(0)$ is determined in terms of $\phi(0)$ by putting $t=0$ in \eqref{ph-cl-eqn-a}. The (inverse) temperature of the black hole $\beta^{-1}$ enters through the initial condition.

\subsubsection*{Energy}

The dynamics \eq{ph-cl-eqn-a} is clearly not invariant under time translation, and hence does not have a conserved energy (signifying energy transfer to the bath). As is conventional in treatments of systems coupled to a bath, one can continue to describe the original expression of the  energy (before the coupling) as a `time-dependent energy' of the system, which, in our case, is given by the Schwarzian
\begin{equation}\label{energy-sch}
	E_{sch}= N\frac{\alpha_{s}}{J}\left[ \frac{1}{2}\dot{\phi}^2 + 2 J^2 e^{\phi} \right] .
\end{equation}
Note that the Schwarzian is $SL(2,R)$ gauge-independent quantity. It is also important to note that the Schwarzian equals the ADM mass of the black hole\cite{Maldacena:2016upp}. In the rest of the paper, we will present details of the time dependence of this energy.

A second notion of energy can be obtained as follows:
let us rewrite equation \eqref{third-order-general} as
\begin{equation}\label{energy-equation}
	\frac{d}{d t} E_\Delta(t) = -N_f\, g^2 F[\phi(t)]^2 ,
\end{equation}
where we have defined
\begin{equation}\label{energy-def}
	E_\Delta(t) = N\frac{\alpha_{s}}{J} \left[\frac{1}{2}\dot{\phi}^2 + 2 J^2 e^{\phi} \right] - N_f\, g^2 F[\phi(t)] \int_{0}^{t} dt' F[\phi(t')] .
\end{equation}
The quantity $E_\Delta(t)$ can be interpreted as a time dependent energy of our system. Then the equation \eqref{energy-equation} tells us that this energy can only dissipate with time since the right-hand side is non-positive. A justification for $E_\Delta(t)$ to be a good notion of energy is as follows. We eliminate the non-locality in the equation of motion \eqref{ph-cl-eqn-a} by introducing a new variable $\mathcal{E}(t) = g \int_{0}^{t} dt' F(\phi(t'))$,
\begin{equation}\label{def-epsilon}
	\ddot{\phi}(t) + 2J^2 e^{\phi(t)} -\frac{n_f g J}{\alpha_s} \mathcal{E}(t) F'(\phi(t))=0, \qquad \dot{\mathcal{E}}(t) = g F(\phi(t)) .
\end{equation}
For constant $\mathcal{E}(t) = \epsilon$, we have $E_\Delta = \frac{1}{2}\dot{\phi}^2 + 2J^2 e^{\phi} -\frac{n_f g J}{\alpha_s} \epsilon F(\phi)$. In fact for $F(\phi)=J e^{\phi/2}$, this is exactly the same as the Kourkoulou-Maldacena deformation \cite{Kourkoulou:2017zaj}. For time-dependent $\mathcal{E}(t)$, $E_{\Delta}(t)$ is therefore an appropriate generalization of energy in presence of a bath.

In JT gravity, complete evaporation of black hole has a simple diagnostic -- whether the boundary curve ($e^{\phi}$) hits a zero or not. This is explained in detail in Section \ref{sec:horizon} and Appendix \ref{app:geom}. The absence or presence of horizon is a gauge-invariant statement and therefore this is a smoking gun test of whether the bulk geometry contains a horizon or not.

The solution we will find, up to some strength of coupling, corresponds to cooling of the SYK system, which can be interpreted as black hole losing energy to the bath, in other words, black hole evaporation. This can be qualitatively compared with part of the results of \cite{Almheiri:2019jqq,Maldacena:2019ufo}, except that while they describe the time evolution in terms of the $(G,\Sigma)$ variables, we are able to describe the entire time evolution in terms of the Schwarzian model, thus allowing a gravity interpretation all the way.

Now we turn to solving the equation \eqref{ph-cl-eqn-a} (or equivalently \eqref{third-order-general}) for different types of couplings(\eqref{relevant} and \eqref{mixed}) mentioned in Section \ref{sec:model}.

%%%%%%%%%%%%%%%%%%%%
\subsection{Model (a): Interaction with just the relevant operator ($\Delta=1/2$)}\label{sec:relevant}
%Here we solve \eqref{third-order-general} equation with only the relevant operator 
We start with only the relevant interaction (see \eq{relevant}), i.e. we specialize to
\begin{equation}
	F(\phi) = J e^{\phi/2} . %, \qquad  V(\phi) = 2J^2 e^{\phi}
\end{equation}

%%%%%%%%%%
\subsubsection{Numerical solution}\label{sec:numerical-a}

We simply solve the 3rd order equation \eqref{third-order-general}. For the above interaction it reads (after setting $J=1$ -- this can be regarded as a choice of units, and can be reinserted simply by replacing $\beta \rightarrow \beta J, t\rightarrow t J$, etc.)
%\begin{equation}\label{third-order}% with F=J e^{\Delta \phi} and J neq 1 
%	\dddot{\phi}- \Delta \ddot{\phi} \dot{\phi} +2 J^2 (1-\Delta)e^{\phi} \dot{\phi} -\tilde{g}^2 J^3 e^{2\Delta \phi} =0, \quad \tilde{g}^2= \frac{\Delta n_f g^2}{\alpha_s}
%\end{equation}
\begin{equation}\label{third-order}% with J=1
	\dddot{\phi}-\frac{1}{2} \ddot{\phi} \dot{\phi} + e^{\phi} \dot{\phi} - \tilde{g}^2 e^{\phi} =0, \qquad \tilde{g}^2 \equiv \frac{n_f g^2}{2\alpha_s} .
\end{equation}
Following the comments below \eq{third-order-general}, we impose the initial conditions:
\begin{align}
	\phi(0)=2 \ln\left(\fr\pi{\b} \right), \quad \dot\phi(0)=0, \quad \ddot \phi(0)= -2 \left(\frac{\pi}{\beta}\right)^2
	\label{init-phi-cl-3rd}
\end{align}

\begin{figure}[]
	\centering
	\begin{subfigure}{.5\textwidth}
		\centering
		\includegraphics[width=\linewidth]{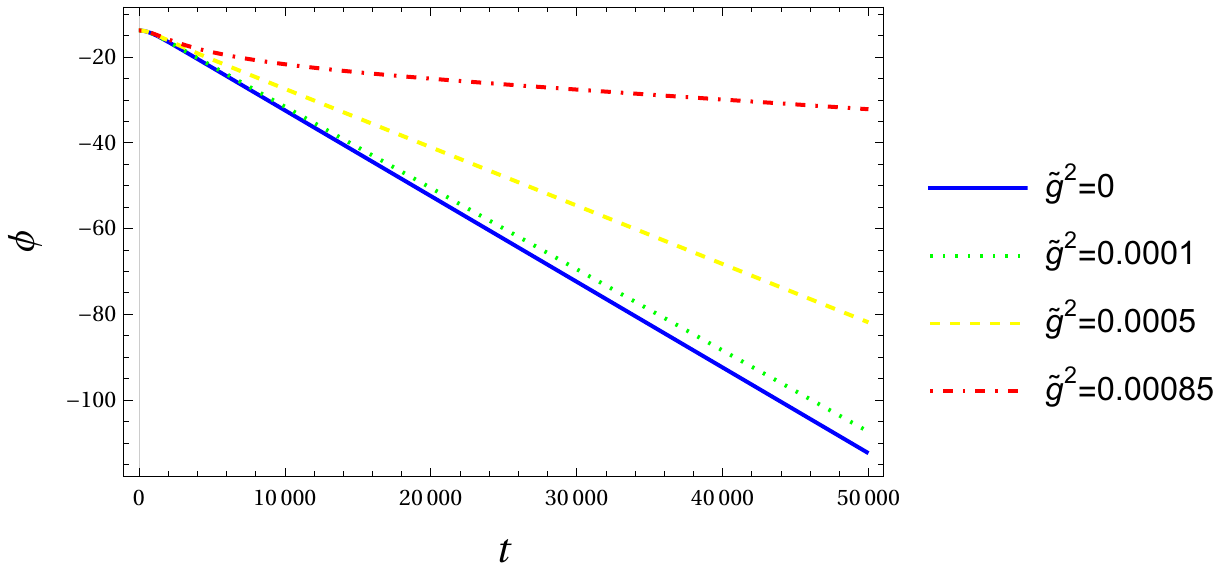}
		\caption{$\phi(t)$}
		\label{fig:phi}
	\end{subfigure}\hfill
	\begin{subfigure}{.5\textwidth}
		\centering
		\includegraphics[width=\linewidth]{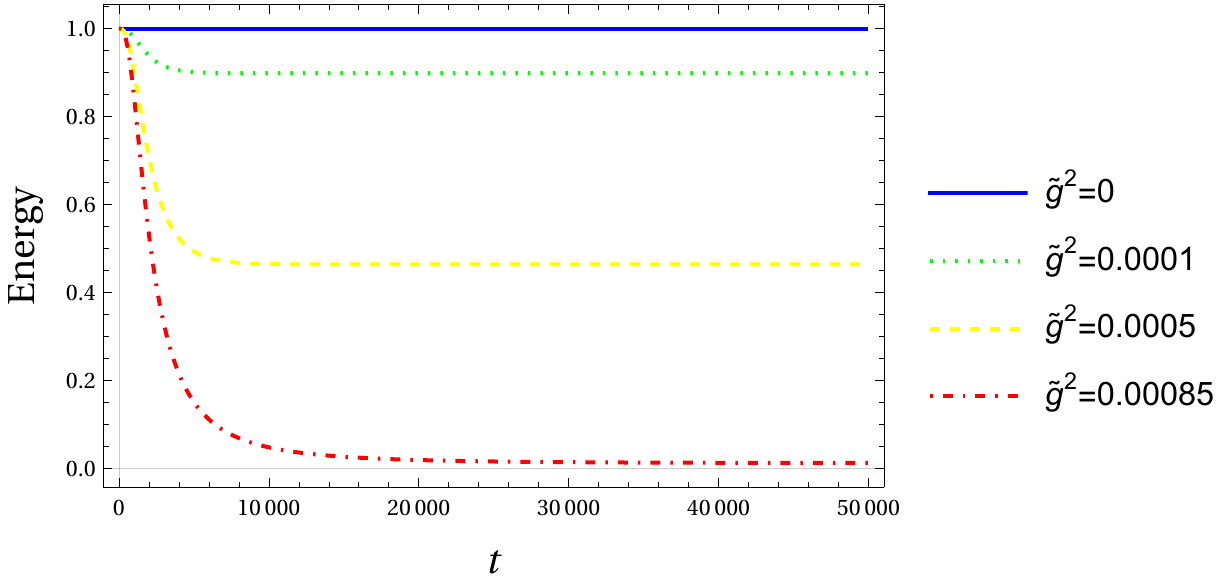}
		\caption{$E_{sch}(t)$}
		\label{fig:energy}
	\end{subfigure}
	\caption{Plot of $\phi(t)$ and energy (estimated by the Schwarzian) for various values of the coupling. Note that the energy loss is greater for stronger coupling. Numerics done for $\beta=1000\pi$ with $J=1$.}
\end{figure}
\begin{figure}[]
	\centering
	\begin{subfigure}{.5\textwidth}
		\centering
		\includegraphics[width=0.9\linewidth]{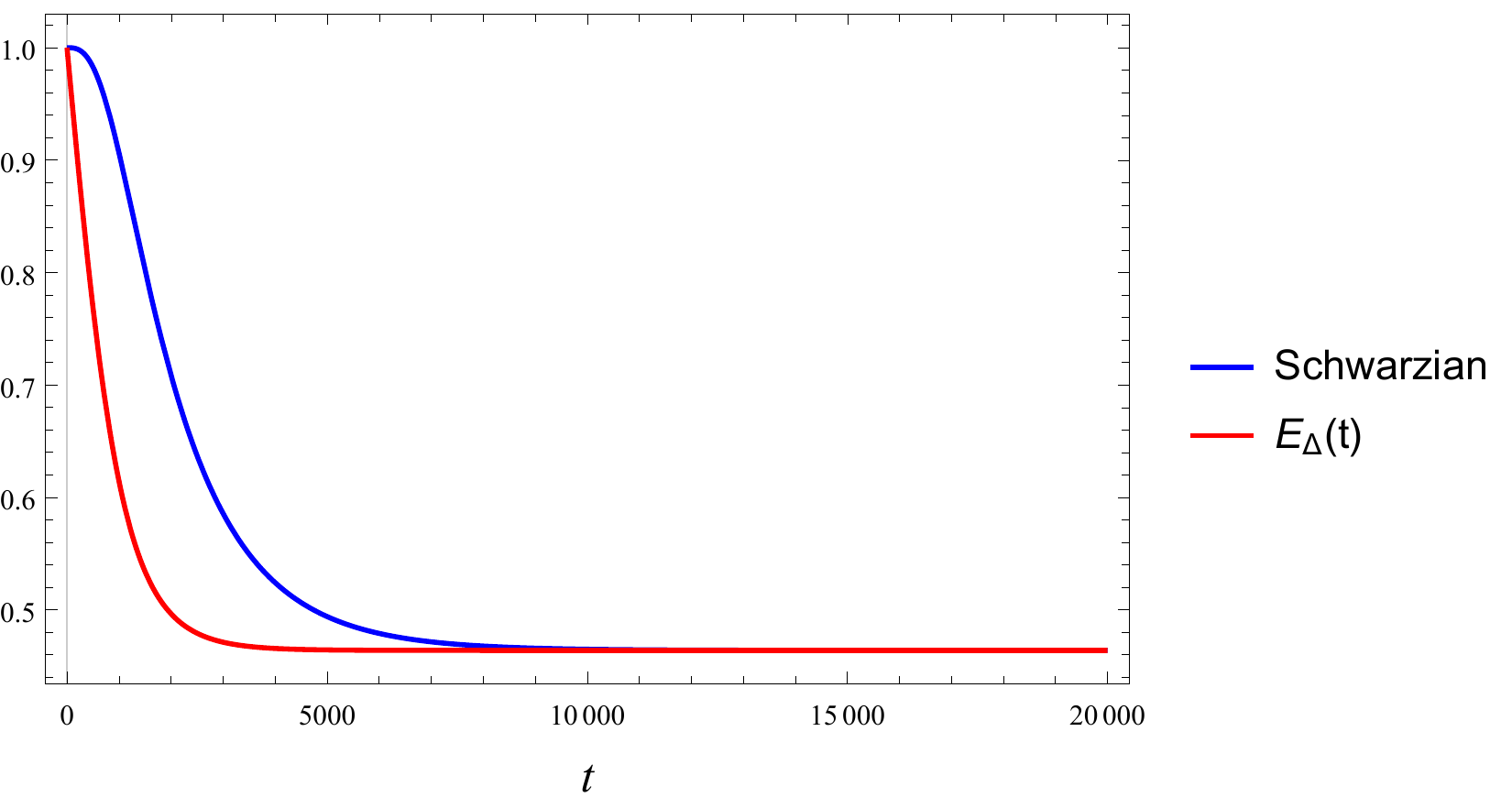}
		\caption{$\tilde{g}^2=0.0005$}
		%\label{fig:sub1}
	\end{subfigure}\hfill
	\begin{subfigure}{.5\textwidth}
		\centering
		\includegraphics[width=0.9\linewidth]{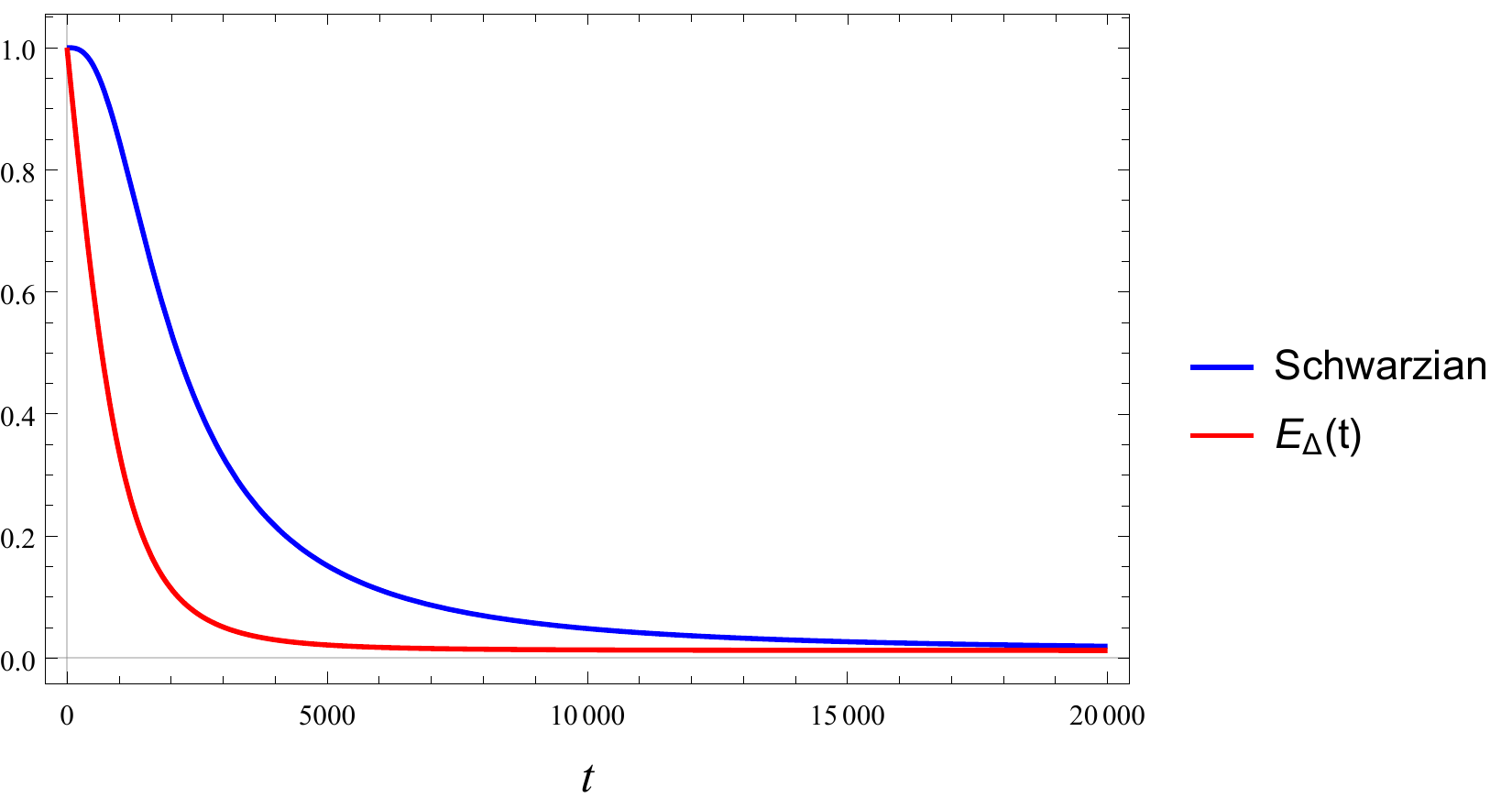}
		\caption{$\tilde{g}^2=0.00085$}
		%\label{fig:sub2}
	\end{subfigure}
	\caption{The energy as a fraction of the initial value at $t=0$. The blue curve is the Schwarzian while the red curve is $E_\Delta$. The two definitions of energy disagree at intermediate times but they agree at late times. Stronger coupling leads to greater energy loss. Numerics done for $\beta=1000\pi$ with $J=1$.}
	\label{fig:energy-loss}
\end{figure}
The solution for $\phi(t)$ as well as the energy (estimated by the Schwarzian) for various values of the coupling $\tilde{g}^2$ is shown in figures \ref{fig:phi} and \ref{fig:energy}. Comparison between the two energies \eqref{energy-sch} and \eqref{energy-def} is also displayed in Figure \ref{fig:energy-loss}. The solutions ($\dot \phi$ and energy) reach an equilibrium and the final energy can be read-off from the slope at late times which can be determined numerically. At late times, $\phi=-a t +\ldots $, the asymptotic energy is 
\begin{equation}
	E_{Sch}=E_{\Delta} = N\frac{\alpha_{s}}{J} \frac{a^2}{2} .
\end{equation}
It may appear surprising at the first sight that the equilibrium configuration of the system is independent of the bath temperature. However it is a familiar phenomenon in usual system-bath models of equilibration. For example, for a pendulum oscillating in a hot viscous medium, the classical motion comes to a stop whose position has nothing to do with the temperature of the medium. However the mean square fluctuations around the classical motion do indeed see the bath temperature. Although our coupling to the bath is more complicated, we find a similar phenomenon. A crucial difference from the pendulum example is that because our coupling is non-linear and non-local in time, the effective friction decreases exponentially in time so that the equilibrium configuration remembers about the coupling strength.

%\textcolor{red}{
Above a certain value of the coupling $\tilde g$, given by
\begin{equation}\label{g-crit-n}
	\tilde g_* \approx 0.65 \sqrt{\frac{2\pi}{\beta J}} ,
\end{equation}
the solution $\phi(t)$, %starts to grow and shows a runaway behaviour
instead of decreasing, grows linearly at late times. This is shown below in Figure \ref{fig:phi-growth}.
\begin{figure}[]
	\centering
	\begin{subfigure}{.4\textwidth}
		\centering
		\includegraphics[width=\linewidth]{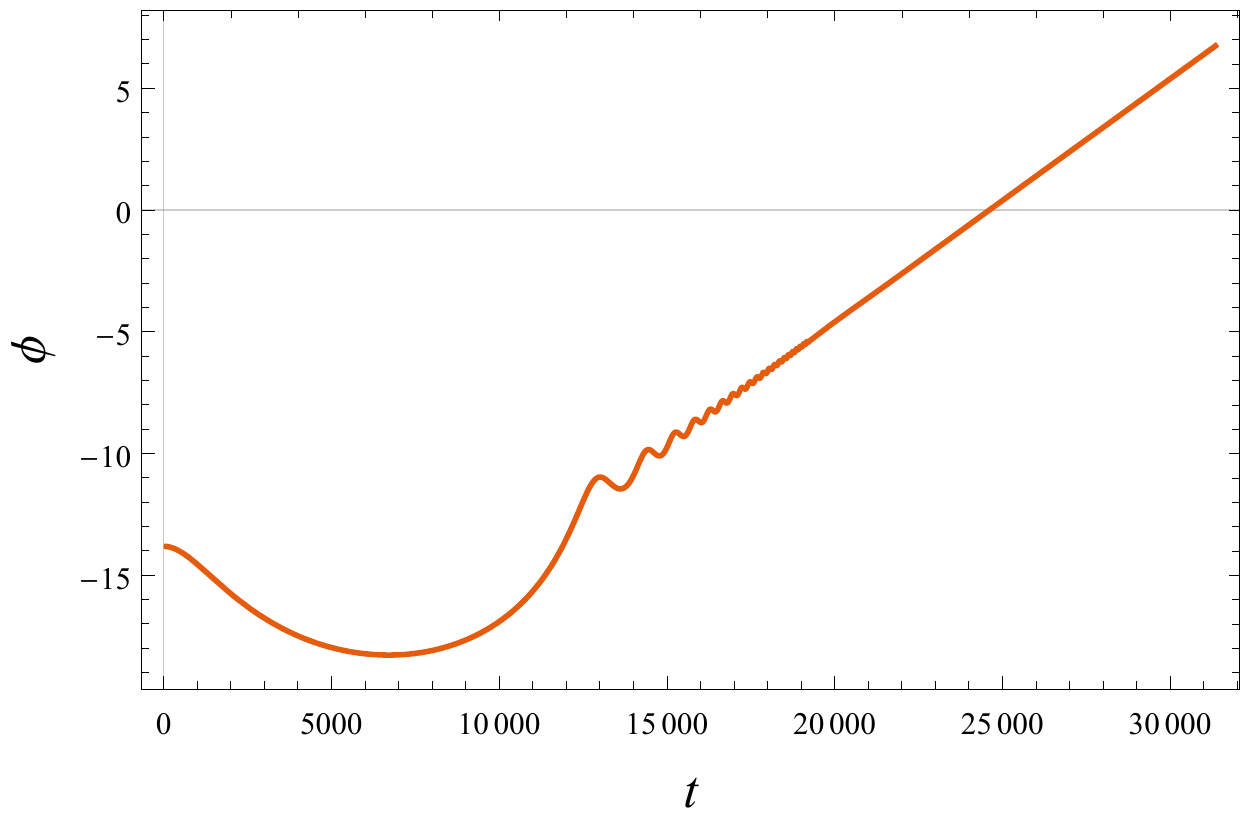}
		\caption{$\phi(t)$}
		\label{fig:phi-growth}
	\end{subfigure}\hfill
	\begin{subfigure}{.52\textwidth}
		\centering
		\includegraphics[width=\linewidth]{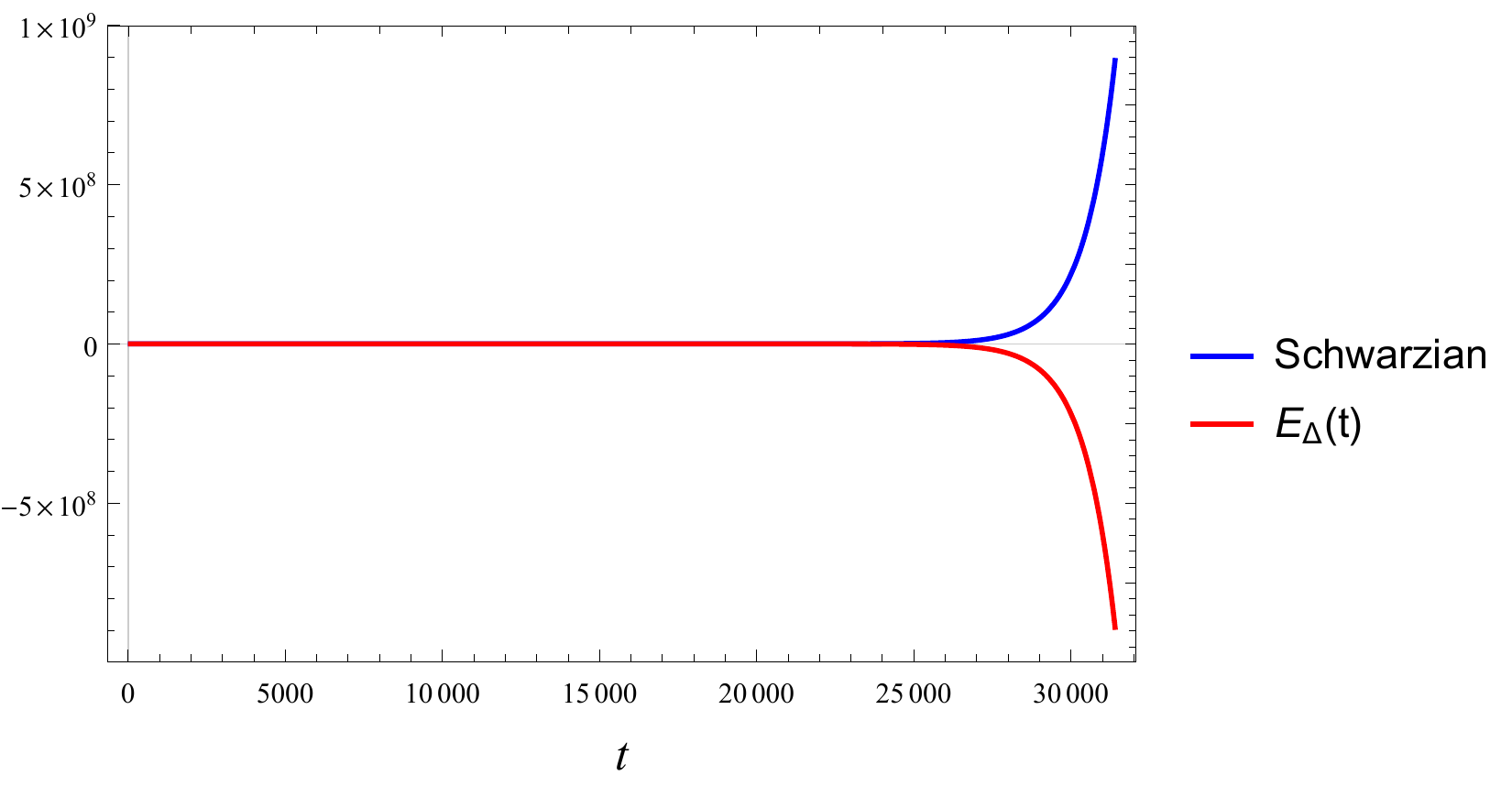}
		\caption{$E_{sch}(t)$}
		\label{fig:energy-growth}
	\end{subfigure}
	\caption{Plot of $\phi(t)$ and energies above critical coupling \protect\eqref{g-crit-n}. Here $\tilde g^2=0.001$, $\beta=1000\pi$ and $J=1$.}
\end{figure}
%%%%%%%%%%
Note that the third order equation \eqref{third-order} possesses an exact, linearly growing, solution
\begin{align}
  \phi(t)=\tilde g^2 t.
  \label{exact-linear-model-a}
\end{align}
Although this solution does not satisfy the initial conditions \eq{init-phi-cl-3rd}, one finds numerically that, the actual solution of \eqref{third-order}, with the correct initial conditions, coincides with \eq{exact-linear-model-a} at late times. Additionally, as shown in Figure \ref{fig:energy-growth}, both the energies $E_{sch}$ and $E_\Delta$, grow unbounded in magnitude. There are a number of ways that one can see that the solution, beyond the critical coupling \eq{g-crit-n}, is unphysical; see Section \ref{sec:unphysical-a}.

\subsubsection{Analytic solution}

After setting $F(\phi)=e^{\phi/2}$, the equation \eq{ph-cl-eqn-a} becomes (after setting $J=1$)
\begin{equation}
	\left(\ddot{\phi}(t) + 2 e^{\phi(t)}\right) -{\tilde g}^2 e^{\phi(t)/2}\int_0^t e^{\phi(t')/2} dt' =0, \qquad {\tilde g}^2= \fr{n_f g^2}{2\alpha_s} .
	\label{ph-cl-eqn-a-half}
\end{equation}
This equation can be solved analytically in a perturbation theory in ${\tilde g}^2$ at least to leading order in this coupling; the answer is
\begin{align}
	&\phi(t)= \phi_0(t) + \gsq \phi_1(t) + O({\tilde g}^4) , \\
	&\phi_0(t)=\log\left(\fr{\pi^2}{\beta^2} \sech^2 \fr{\pi t}{\beta}\right) , \nonumber \\
	&\phi_1(t) = -\frac{1}{\pi} \sech\left( \frac{\pi t}{\beta}\right)  \left\{ \sinh \left(\frac{\pi t}{\beta}\right) \left[ \beta +\beta  \log 2 -\beta \log \left(e^{\frac{2 \pi  t}{\beta }}+1\right)+\pi  t\right] -\beta  \cot ^{-1}\left[ \csch\left( \frac{\pi t}{\beta} \right) \right] \right\} . \label{asym-phi-detailed}
\end{align}
%The expression for $\phi_1(t)$ is very long, and is given in \eq{phi-1} in Appendix \ref{app:gsq-pert}.
At large $t$, the behaviour of the solution can be found from the asymptotic expression
\begin{equation}
	\phi(t) = \left(\tilde{g}^2 -\frac{2\pi}{\beta}\right)t - 2\left[ \log\fr{\beta}{2\pi} + \gsq\fr\beta{2\pi} (1+ \log 2)\right] + O\left(\exp[-\pi t/\beta]\right) . \label{asym-phi-long}
\end{equation}
%\begin{align}
%\phi(t) &=\left( \fr{t}{6\beta}(-12\pi + 6\beta \gsq) \right)
%%  \nonumber\\
%- \left[2\log\fr{\beta}{2\pi} + \gsq\fr\beta{2\pi} (1+ \log 2)\right]
%+ O\left(\exp[-\pi t/\beta]\right) \label{asym-phi}
%\end{align}
At large $t$, the term linear in $t$ dominates, leading to
\begin{equation}
	\phi(t) \to - a t, \quad a= \left(\frac{2\pi}{\beta} - \tilde{g}^2\right), \qquad {\tilde g}^2= \fr{n_f g^2}{2\alpha_s} .
	\label{asym-phi}
\end{equation}
It is easy to see that, for   
\begin{align}
	\tilde g  \ge \tilde g^A_* \equiv \sqrt{\fr{2\pi}{\beta J}} ,
	\label{g-crit-a}
\end{align}
$a <0$ and the solution $\dot f(t)=e^{\phi(t)}$ grows linearly without bound, thus indicating breakdown of perturbation theory in $\gsq$. We found the same phenomenon in the numerical solutions also (see \eq{g-crit-n}) for a slightly different critical value. We compare the analytic and numerical solutions below (see Section \ref{sec:relevant-comparison}).

In the region \eq{g-crit-a}, one can estimate the time $t_*$ beyond which the solution shows a runaway behaviour. In  \eq{asym-phi-long}, for short enough times the constant term dominates, but the linearly diverging term dominates from $t \gtrsim t_*$ where
\[
t_* = \frac{2\left[ \log\fr{\beta}{2\pi} + \gsq\fr\beta{2\pi} (1+ \log 2)\right] }{\left(\tilde{g}^2 -\frac{2\pi}{\beta}\right)} .
\]

\paragraph{Asymptotics:}
A sensible asymptotic solution \eq{asym-phi} is found for couplings in the range
$\tilde g < \tilde g_*$:
\begin{align}
	& \phi(t\to \infty)\to -t ((\tilde g^A_*)^2 - \gsq), ~{\rm hence}~ \dot f(t) \to 0 ,
	\label{asym-a}\\
	& E_{sch}(t\to \infty)= E_\Delta(t\to \infty) \equiv E_\infty \equiv N\alpha_s J \fr12 ((\tilde g^A_*)^2 - \gsq)^2= N\alpha_s J \fr12 \left(\frac{2\pi}{\beta J} -\gsq \right)^2 .
\label{energy-final}
\end{align}
In Section \ref{sec:bh-evap}, we will interpret this classical solution as representing a final black hole solution at a smaller energy compared to the initial black hole, obtained as a result of interaction with the bath.

\begin{figure}[]
	\centering
	\begin{subfigure}{.4\textwidth}
		\centering
		\includegraphics[width=\linewidth]{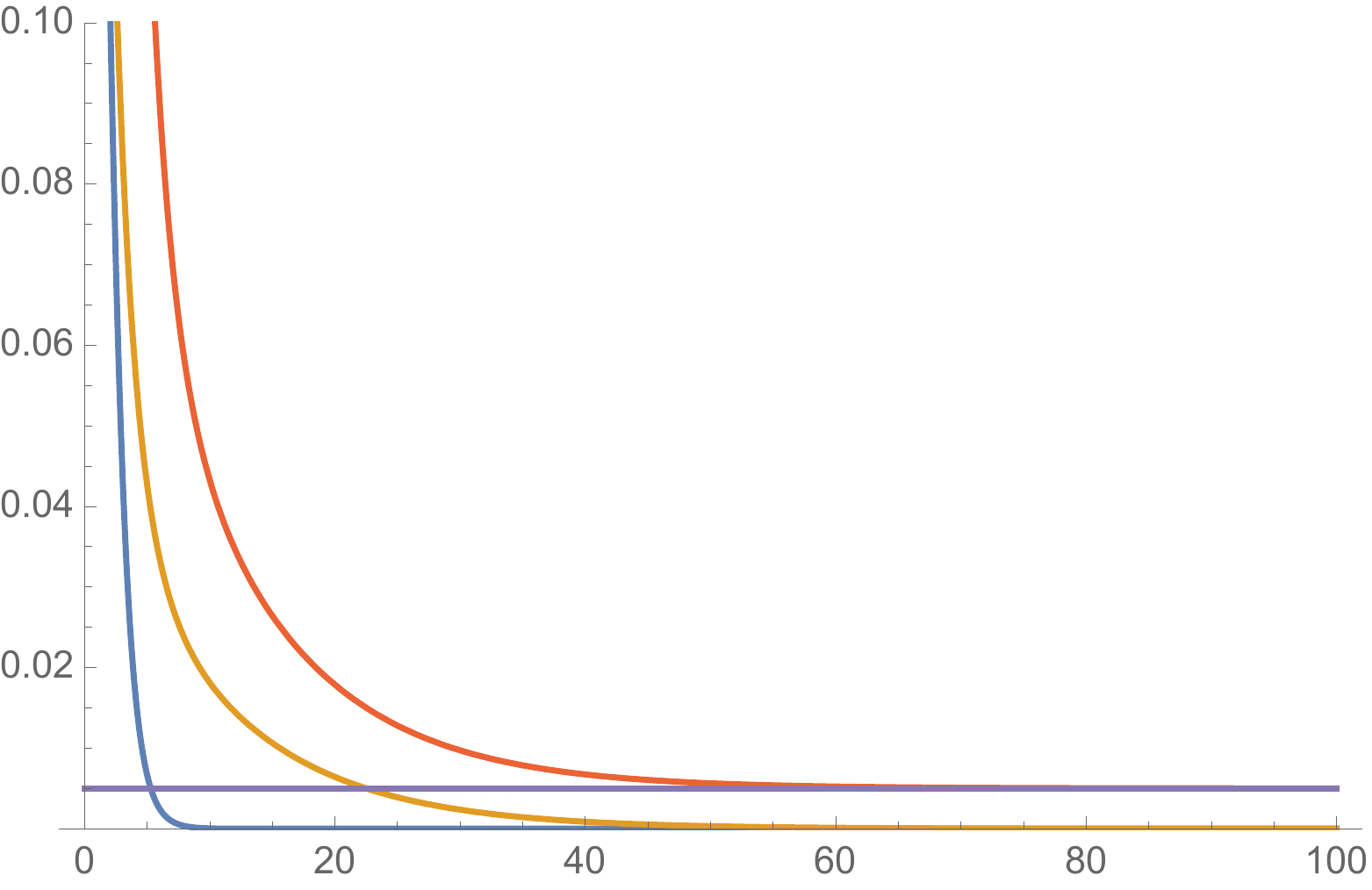}
		\caption{$\tilde{g}^2=0.9$. Red= $E_{sch}$, purple= $E_\Delta$, brown= $E_\infty$, blue=$e^{\phi_0}$, orange=$e^{\phi}$.}
		%\label{fig:sub1}
	\end{subfigure}\hspace{10ex}
	\hspace{5ex}
	\begin{subfigure}{.4\textwidth}
		\centering
		\includegraphics[width=\linewidth]{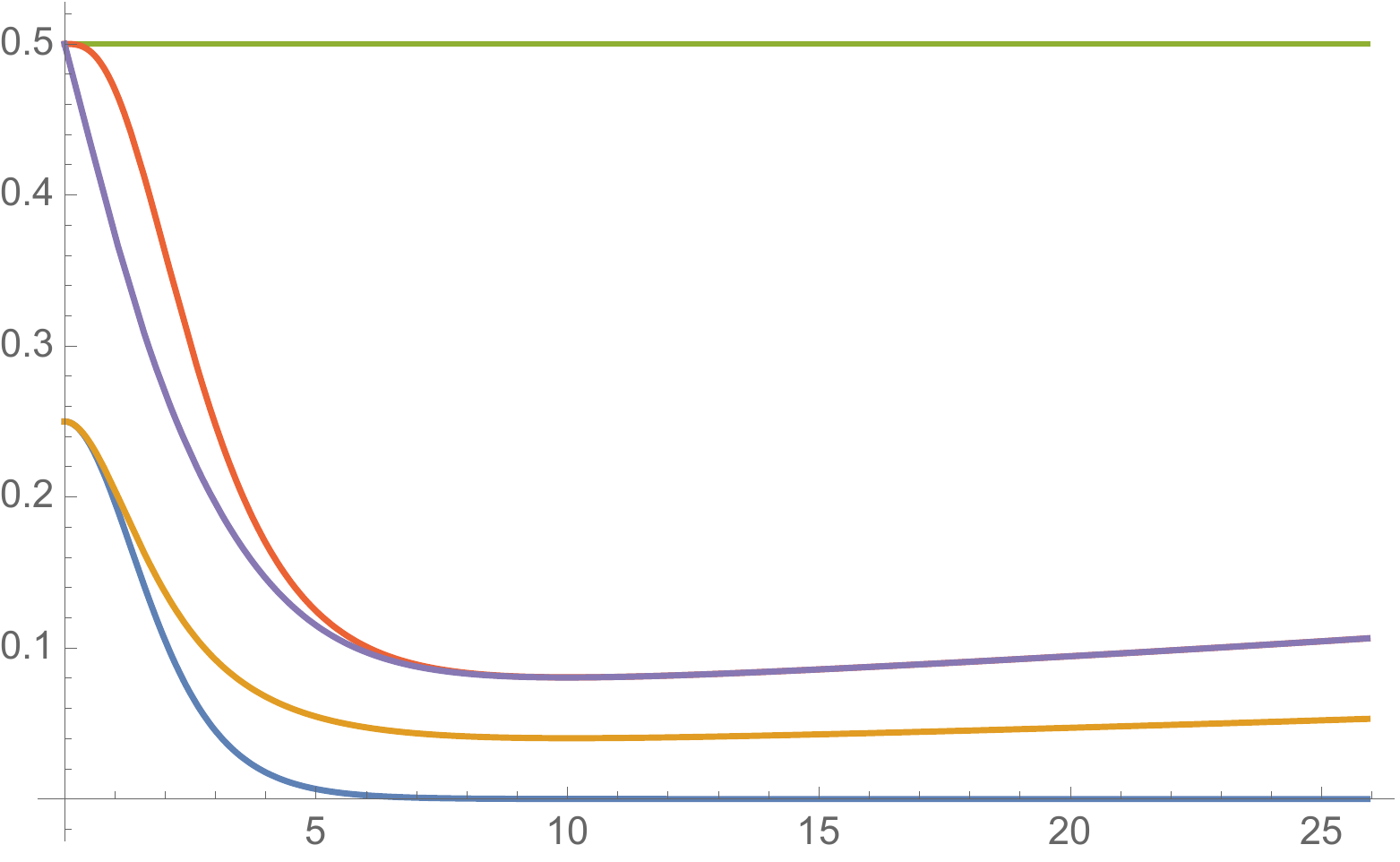}
		\caption{$\tilde{g}^2= 1.02$. Red= $E_{sch}$, purple= $E_\Delta$, green= $E_0$, blue=$e^{\phi_0}$, orange=$e^{\phi}$.}
		%\label{fig:sub2}
	\end{subfigure}
	\caption{In both figures, $\beta J=2\pi$, so that $\tilde g_*^A=1$. The left panel, where $\tilde g < \tilde g_*^A$, shows approach to an asymptotic value for energy as well as $\exp[\phi]$. In the right panel, energy as well as $\exp[\phi]$ show unbounded growth.}
	\label{fig:energy-loss-analytic}
\end{figure}

%%%%%%%%%%
\subsubsection{Comparison of numerical and analytical solutions}\label{sec:relevant-comparison}

We find above that both the numerical and the perturbative analytical methods predict a sensible asymptotic classical solution if the coupling strength $\gsq$ is small enough (with critical coupling dictated by the initial energy of the $|B_s(l)\ran$ state). The small difference between the numerical and analytical values of the critical coupling (see \eq{g-crit-n}, \eq{g-crit-a}) may arise from the fact that \eq{g-crit-a} is obtained from first order perturbation theory in $\gsq$ which may not be valid near criticality.\footnote{This is to be expected if the actual solution is non-analytic in $(g_*^2 - \gsq)$, in which case the radius of convergence in an expansion in $\gsq$ will be $g_*^2$, signaling a breakdown of perturbation theory near criticality. Note that in \cite{Dhar:2018pii} we found a branch cut singularity in the solution as a function of the KM-coupling. Note also that in a first order $\gsq$ expansion, a non-analytic function like $(1 - \fr\gsq{g_*^2})^{^p}$ will behave like $(1 -p \fr\gsq{g_*^2} + ...)$, thus appearing to shift the critical point to $\gsq= p g_*^2$ at first order. \label{ftnt:perturbation-theory}} From the explicit solutions in Figure \ref{fig:rel-N-A-comparison}, one can see that the numerical and analytical solution match well at small couplings but begin to differ at larger couplings as one approaches near criticality.
\begin{figure}[]
	\centering
	\begin{subfigure}{.5\textwidth}
		\centering
		\includegraphics[width=\linewidth]{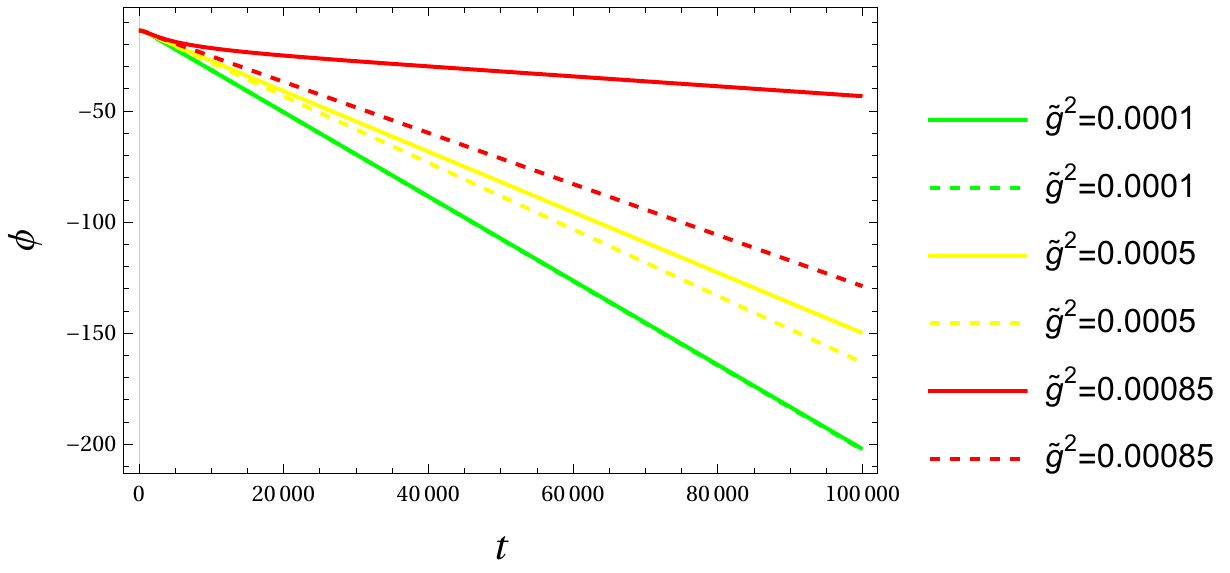}
		\caption{$\phi(t)$}
		%\label{fig:sub1}
	\end{subfigure}\hfill%
	%\hspace{5ex}
	\begin{subfigure}{.5\textwidth}
		\centering
		\includegraphics[width=\linewidth]{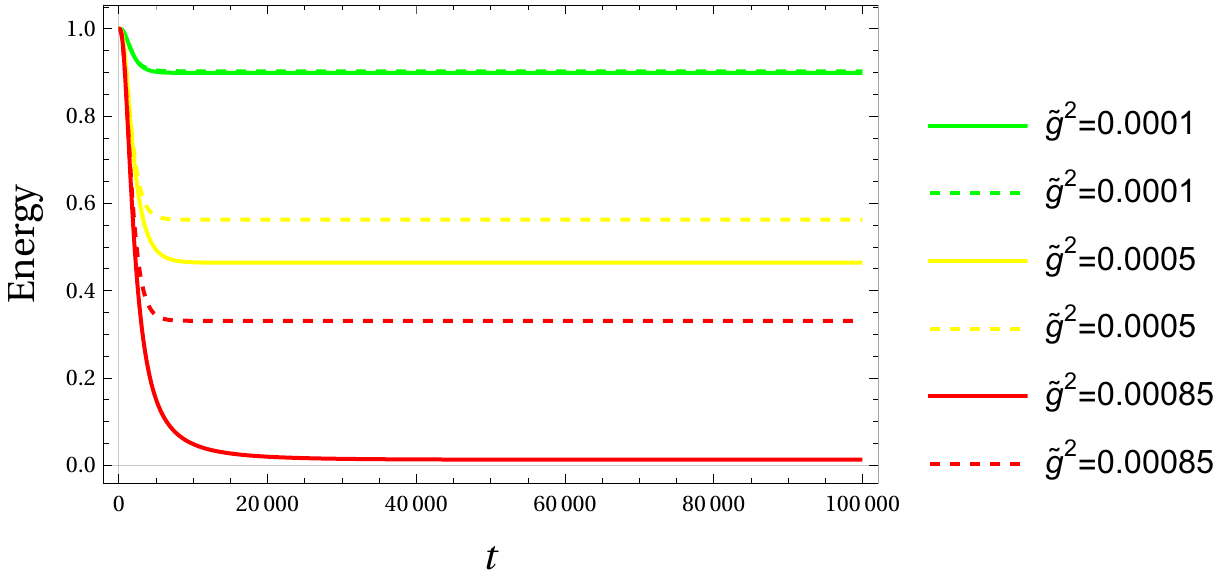}
		\caption{$E_{Sch}(t)$}
		%\label{fig:sub2}
	\end{subfigure}
	\caption{Comparison between numerical (solid lines) and analytical perturbative solutions (dashed lines) for various strengths of the coupling. They match at small couplings but begin to differ at larger couplings.}
	\label{fig:rel-N-A-comparison}
\end{figure}

\subsubsection{Unphysical solutions in Model (a)}\label{sec:unphysical-a}

We found above the existence of diverging solutions beyond the critical coupling \eq{g-crit-n}. There are a number of ways that one can see that these solutions are unphysical:
\begin{enumerate}
\item The unbounded energy $E_{sch}$ soon exceeds the scale $J$, which signals breakdown of the low energy approximation under which the analysis of the model in terms of the Schwarzian is based.
\item %As we will describe below in more detail for model (b), coupling to a bath can be roughly regarded as assigning a time-dependence to the coupling constant(s). The way it works for model (a) is as follows. Consider \eq{eff-pot}, with ${\cal E}(t)$ defined by the first equation of \eq{eq:integral}, both with $g'=0$. With the unbounded solution under discussion here, ${\cal E}(t)$ diverges exponentially in time, thus taking the relevant coupling beyond the regime of conformal perturbation theory around the SYK fixed point.
As we will describe below in more detail for Model (b), coupling to a bath can be roughly regarded as assigning a time-dependence to the coupling constant(s). This can be seen from rewriting the equation of motion as eq. \eqref{def-epsilon} in terms of the time dependent coupling $\mathcal{E}(t)$. With the unbounded solution under discussion here, ${\cal E}(t)$ diverges exponentially in time, thus taking the relevant coupling beyond the regime of conformal perturbation theory around the SYK fixed point.
\item {\bf Anti-dissipation:} The unbounded energy $E_{sch}$ is unphysical simply because it represents absorption of energy by the SYK system from the bath which is effectively colder. This indicates anti-dissipation, or negative friction. In the perturbative solution \eq{asym-phi-long}, one can explicitly see the transition from dissipative behavour to anti-dissipative behaviour ($\dot f(t)$ changes from exponentially decaying to exponentially growing), as the coupling constant crosses the critical value \eq{g-crit-a}.\footnote{The difference of this from \eq{g-crit-n} has been discussed above.} Note that anti-dissipative behaviour is not visible in the standard system-bath models like the Caldeira-Leggett  model \cite{Caldeira:1982iu} which has linear couplings to an Ohmic bath.
\end{enumerate}

It is important to note that the unphysical solutions we find here are not related to the well-known runaway solutions of third order differential equations, e.g. in the radiation reaction problem of electrodynamics.

We will come back to the issue of the unphysical solutions in more details and more generally in Section \ref{domain}.

%%%%%%%%%%%%%%%%%%%%
\subsection{Model (b): Complete Evaporation: Relevant + marginal interaction \label{sec:relevant+marginal}}

First we argue that it is not possible to achieve complete black hole evaporation with a single coupling $F(\phi)= e^{\Delta\phi}$, with $\Delta=1/2$ for relevant and $\Delta=1$ for marginal coupling (we have put $J=1$). It is easy to see that a bounded solution of equation \eqref{ph-cl-eqn-a} is simply not possible! Let us assume that there exists a solution such that
\begin{equation}\label{bounded-phi}
	\phi_{max} > \phi(t\rightarrow\infty) > \phi_{min}
\end{equation}
i.e. it is bounded both above and below. The lower bound is required for disappearance of the horizon while the upper bound is imposed such that the boundary curve does not venture far into the bulk as $\hat z\propto \dot f = \exp(\phi)$ (see Section \ref{sec:horizon} and Appendix \ref{app:geom} for a detailed discussion). The above bound implies that $e^{\Delta\phi_{max}} \,t > \int_0^t e^{\Delta\phi(t')} dt' > e^{\Delta\phi_{min}}\, t $ and therefore the integral grows linearly with time. At sufficiently late times the integral term overtakes the potential term $2 e^\phi$ and the equation of motion \eqref{ph-cl-eqn-a} reads
\[ \ddot{\phi}(t) \approx \frac{n_f g^2 A^2 \Delta}{\alpha_s} t , \]
where $A= \exp(\Delta\phi(t\rightarrow\infty))$ is the asymptotic value that is bounded $\exp(\Delta\phi_{min}) <A< \exp(\Delta\phi_{max})$. The solution 
\[ \phi(t) \approx \frac{n_f g^2 A^2 \Delta}{6\alpha_s} t^3 \]
is unbounded as it keeps growing with time, in contradiction with our assumption \eqref{bounded-phi}. This implies that either $\exp(\phi) \rightarrow 0$ as $t \rightarrow \infty$ and we remain in the black hole phase, or $\exp(\phi) \rightarrow \infty$ as $t \rightarrow \infty$ and the low-energy Schwarzian description breaks down. \\

Now we will overcome this by turning on more than one interaction term simultaneously and achieve complete black hole evaporation. There is another way to achieve complete black hole evaporation if we allow a fixed KM potential term which we briefly mention in appendix \ref{sec:relevant+KM}.

%%%%%%%%%%%%%%%%%%%%

We choose the function $F$ to have both relevant and marginal coupling
\begin{equation}
	g F(\phi)= g J e^{\phi/2} - g' J e^{\phi} .
	\label{relevant+marginal}
\end{equation}
The equation of motion \eqref{ph-cl-eqn-a} in this case reads (after setting $J=1$, which can be reinserted simply by replacing $\beta \rightarrow \beta J, t\rightarrow t J$, etc.)
%\begin{equation}%J neq 1
%\left(\ddot{\phi}(t) + 2J^2 e^{\phi(t)} \right) - \frac{J^2}{\alpha_s} n_f \left(\frac{1}{2} g e^{\phi(t)/2} - g' e^{\phi(t)} \right) \int_0^t J dt' \left(g e^{\phi(t')/2} - g' e^{\phi(t')} \right) =0
%\label{relevant+marginal-EOM}
%\end{equation}
\begin{equation}%J=1
	\left(\ddot{\phi}(t) + 2 e^{\phi(t)} \right) - \frac{n_f}{\alpha_s} \left(\frac{1}{2} g e^{\phi(t)/2} - g' e^{\phi(t)} \right) \int_0^t dt' \left(g e^{\phi(t')/2} - g' e^{\phi(t')} \right) =0 .
	\label{relevant+marginal-EOM}
\end{equation}
It is important to have a relative minus sign between the two couplings. This allows the solution for $\phi$ to approach a finite value. This is achieved if in \eqref{relevant+marginal-EOM} either the integrand vanishes at late times or the pre-factor multiplying the integral effectively goes to zero at late times. Numerical solutions suggest that for generic values of the coupling parameters (see Figure \ref{g-gprime}), it is actually the latter. In Figure \ref{fig:rel+marg} we present the numerical solution where $\phi$ remains bounded at arbitrary large times. This indicates that the bulk geometry is horizonless and therefore the black hole is gone. Without the relative minus sign, the solution $\phi(t)\rightarrow-\infty$ as $t\rightarrow\infty$, which remains a black hole (see Figure \ref{g-gprime}) just as in Model (a).
\begin{figure}[]
	\centering
	\includegraphics[width=0.45\textwidth]{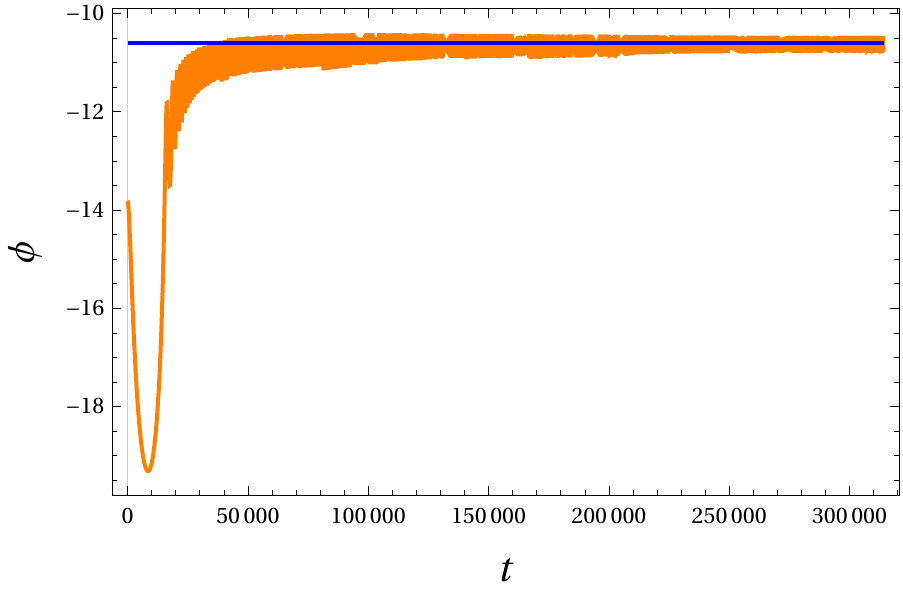}
	\caption{The solution for $\phi$ with both relevant and marginal coupling remains bounded at late times. Numerics done for $\beta=1000\pi,\, J=1,\, n_f=0.01,\,g=0.04$ and $g'=4$. At late times, $\phi$ seems to approach the value $\phi_{eq}=2\log(\frac{g}{2g'}) \approx -10.6$ shown by the blue line.} \label{fig:rel+marg}
\end{figure}

Since $\phi$ approaches $\phi_{eq}\equiv 2\log(\frac{g}{2g'})$ at late times, the integral
\begin{equation}
	\mathcal{E}(t) \equiv \int_0^t dt' g\left(e^{\phi(t')/2} -\frac{g'}{g} e^{\phi(t')} \right) \approx \int_0^t dt' \frac{g^2}{4g'} = \frac{g^2}{4g'} t \label{eq:integral}
\end{equation}
grows linearly with time. We gave also verified this numerically for various values of $g$ and $g'$. It is important that $\phi(t\rightarrow \infty)$ is not exactly $\phi_{eq}=2\log(\frac{g}{2g'})$, but remains very close to it. Otherwise the effect of the coupling would simply vanish. To understand this `attractive' nature of the solution it is useful to consider a time-dependent potential\footnote{We note that this problem is not adiabatic by any means and we will not use the time-dependent potential do calculations. Nevertheless we find it useful in the present context.}
\begin{equation}\label{eff-pot}
	V_{eff}(\phi(t)) = 2 e^{\phi(t)} -\frac{n_f}{\alpha_s} g \left(e^{\phi(t)/2} - \frac{g'}{g} e^{\phi(t)} \right) \mathcal{E}(t) .
\end{equation} 
Since the integral grows linearly, at late times the potential also scales linearly with time
\begin{equation}\label{eff-pot-linear}
	V_{eff}(\phi)\approx \frac{n_f}{4\alpha_s} g^2 \left(e^{\phi} - \frac{g}{g'} e^{\phi/2} \right) t
\end{equation}
and therefore becomes deeper and deeper by stretching along the y-axis linearly with time (see Figure \ref{fig:pot}). Then effectively we have a dynamics of a particle trapped in such a well whose depth increases linearly with time. This also explains why the motion is bounded.
\begin{figure}[]
	\centering
	\includegraphics[width=0.6\textwidth]{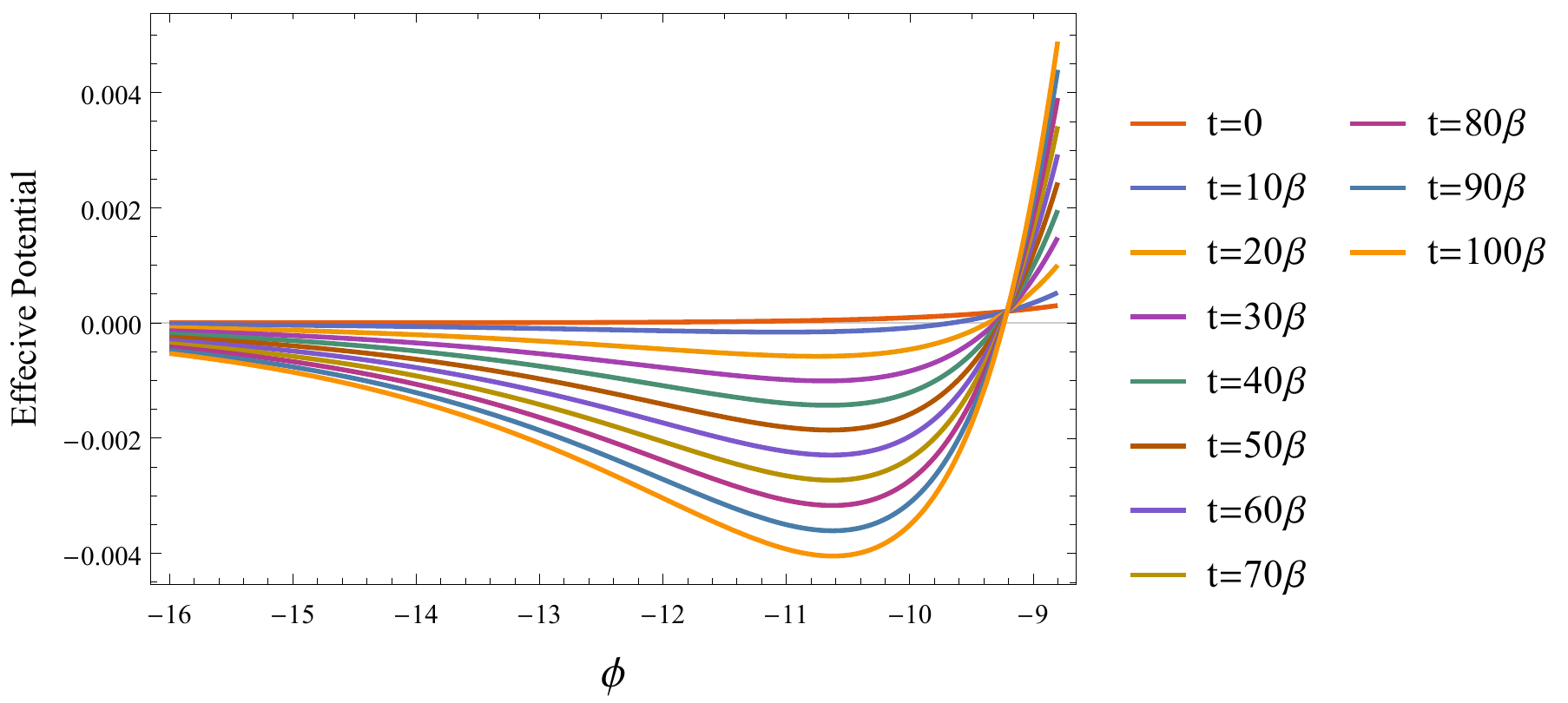}
	\caption{The effective potential \protect\eqref{eff-pot} becomes deeper and deeper as time increases. This is because the integral increases linearly with time. Here $\beta=1000\pi$ is the initial inverse temperature of the black hole. Further we chose $n_f=0.01,g=0.04$ and $g'=4$.} \label{fig:pot}
\end{figure}

\begin{figure}[]
	\centering
	\includegraphics[width=0.5\textwidth]{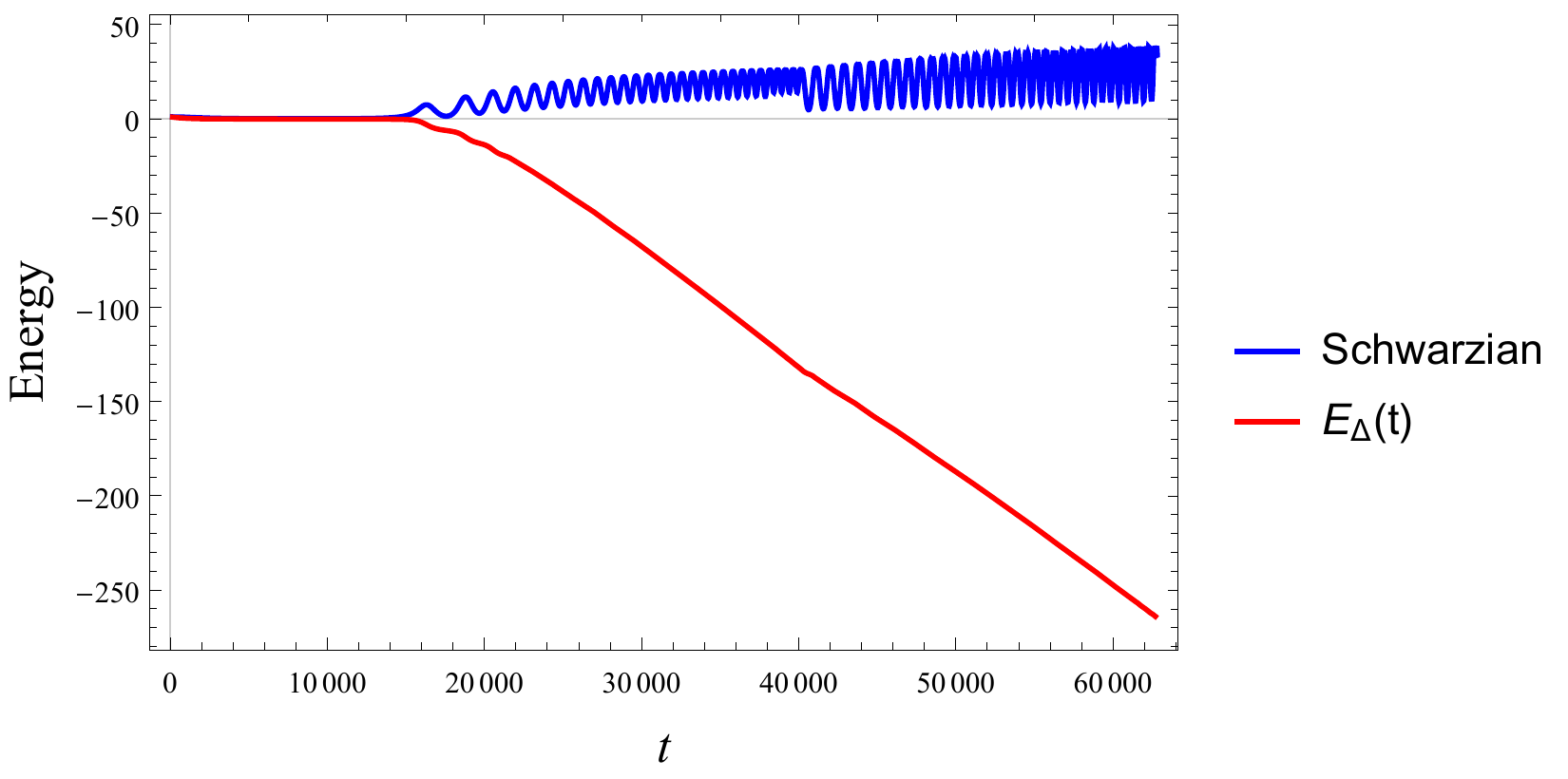}
	\caption{Energies as a function of time. Here $\beta=1000\pi, n_f=0.01, g=0.04$ and $g'=4$.} \label{fig:ecomp}
\end{figure}
The energies (\eqref{energy-sch} and \eqref{energy-def}) for the bounded solution of \eqref{relevant+marginal-EOM} however also keep increasing with time, see Figure \ref{fig:ecomp}.

%%%%%%%%%%
\subsubsection{Analytic solution at large times in terms of Airy functions}\label{sec:Airy}
From the numerical solution we saw that at late times $\phi$ approaches a fixed value $\phi_{eq}= 2\log(\frac{g}{2g'})$. This suggests that we separate %\doubt {\color{blue}Change $\delta \phi$ to something else} \doubt
\begin{equation}
	\phi(t) = \phi_{eq} + \psi(t) .
\end{equation}
Note that this is outside the validity of perturbation theory in small coupling. The equation of motion to leading order in small $\psi$ reads (setting $J=1$)
%\begin{equation}%J neq 1
%	\frac{\alpha_s}{J} \left[\delta \ddot{\phi}(t) + 2J^2 e^{\phi_{eq}} (1+\psi(t)) \right] +\frac{n_f g^2 J^2}{32} \left(\frac{g}{g'}\right)^2 t\, \delta\phi(t) =0
%\end{equation}
\begin{equation}\label{preAiry} %J=1
	\ddot{\psi}(t) + 2 e^{\phi_{eq}} (1+\psi(t)) +\frac{n_f g^2}{32\alpha_s} \left(\frac{g}{g'}\right)^2 t\, \psi(t) =0 .
\end{equation}
The last term comes from the integral in \eqref{relevant+marginal-EOM} and grows linearly as we saw in \eqref{eq:integral}. At very late times it will overtake the potential $e^{\phi_{eq}}$, which is just a constant.\footnote{One can also solve the equation with the constant term present but we omit it here as it is not particularly illuminating.} The equation then becomes of the Airy form
%\begin{equation}\label{airyEQ}%J neq 1
%	\frac{\alpha_s}{J} \ddot{\delta\phi}(t) +\frac{n_f g^2 J^2}{32} \left(\frac{g}{g'}\right)^2 t\, \delta\phi(t) =0
%\end{equation}
\begin{equation}\label{airyEQ}%J=1
	\ddot{\psi}(t) +\frac{n_f g^2}{32\alpha_s} \left(\frac{g}{g'}\right)^2 t\, \psi(t) =0 ,
\end{equation}
with the solution
\begin{align}%J neq 1
	\psi(t) =& c_1 Ai(-ct) + c_2 Bi(-ct) \label{airy} , \\
	c=&\frac{c'}{2\cdot 2^{2/3}} (-c')^{-2/3}, \quad c'=\frac{n_f}{\alpha_s}\frac{g^4}{g'^2} .\nonumber
\end{align}
The asymptotic form reads
\begin{equation}%J neq 1
	\phi(t)= \phi_{eq} +  c_1 \frac{\sin \left(\frac{2}{3}(ct)^{3/2} + \frac{\pi}{4} \right)}{\sqrt{\pi}(ct)^{1/4}} + c_2 \frac{\cos \left(\frac{2}{3}(ct)^{3/2}+ \frac{\pi}{4} \right)}{\sqrt{\pi}(ct)^{1/4}} + O(t^{-7/4}) . \label{airyAsymp}
\end{equation}
Thus the solution approaches the constant value $\phi_{eq}$ slowly as a power law modulated by oscillations whose frequency increases with time.

Note that the above solution is not valid for arbitrarily small values of the coupling as discussed above (see Figure \ref{g-gprime}).

%%%%%%%%%%
\subsubsection{Quench and stabilizing the effective potential}\label{sec:quench}
For the solution of \eqref{relevant+marginal-EOM}, as we saw above both the energies and the effective potential kept increasing without bound. This means that sooner or later the Schwarzian approximation we are working with will break down. More precisely, the low-energy approximation that we have used to locate the IR fixed point is $\omega \ll J$. Applying the same principle to the linearized Airy equation \eqref{airyEQ}, the time-dependent frequency should satisfy %(we have reinstated $J$)
%\begin{equation}%J neq 1
%	\omega(t) = \left(\frac{n_f}{32\alpha_s} \frac{g^4 J^2}{g'^2} Jt\right)^{1/2} \ll J=1
%\end{equation}
\begin{equation}%J neq 1
	\omega(t) = \left(\frac{n_f}{32\alpha_s} \frac{g^4}{g'^2} t\right)^{1/2} \ll 1 .
\end{equation}
This gives us a time-scale $t_*$
\begin{equation}\label{tbreakdown}
	t_* = \frac{32\alpha_s}{n_f} \frac{g'^2}{g^4}
\end{equation}
at which our solution breaks down. In other words we can only trust our solution for times $t\ll t_*$.
 
Now we quench the coupling parameters in such a way that the potential and energies stabilize at late times. We are allowed to this because the couplings are by themselves not dynamical variables of the system and thus can be tuned. By quench we mean that both $g$ and $g'$ are made explicit functions of time. Note that it is really the UV couplings $\hat g_1, \hat g_2$ that are being quenched and the effect carries over to $g,g'$ as they are linearly related \eqref{mixed}. We choose the following quench protocol. %Do we quench $\hat g_1, \hat g_2$ or $g,g'$? UV and IR parameters should be quenched in the same way. In fact the UV quench should imply the IR quench
\begin{align}\label{quench-g}
	g(t)=& g \left[\Theta(t)-\Theta(t-t_{max})\right] + g\,\Theta(t-t_{max}) \sqrt{\frac{t_{max}}{t}} ,\\
	g'(t)=& g' \left[\Theta(t)-\Theta(t-t_{max})\right] + g'\,\Theta(t-t_{max}) \sqrt{\frac{t_{max}}{t}} , \nonumber
\end{align}
i.e. after $t=t_{max}$, the couplings go down in strength as $t^{-1/2}$. This time must satisfy $t_{max}\ll t_*$. Note that the ratio $g'(t)/g(t)$ remains constant. For time-dependent parameters the equation of motion \eqref{relevant+marginal-EOM} is modified to
%\begin{equation}%J neq 1
%	\left(\ddot{\phi}(t) + 2J^2 e^{\phi(t)} \right) - \frac{J^2}{\alpha_s} n_f g(t) \left(\frac{1}{2} e^{\phi(t)/2} - \frac{g'}{g} e^{\phi(t)} \right) \int_0^t Jdt' g(t')\left(e^{\phi(t')/2} - \frac{g'}{g} e^{\phi(t')} \right) =0
%	\label{time-dep-EOM}
%\end{equation}
\begin{equation}%J=1
	\left(\ddot{\phi}(t) + 2 e^{\phi(t)} \right) - \frac{n_f}{\alpha_s} g(t) \left(\frac{1}{2} e^{\phi(t)/2} - \frac{g'}{g} e^{\phi(t)} \right) \int_0^t dt' g(t')\left(e^{\phi(t')/2} - \frac{g'}{g} e^{\phi(t')} \right) =0 .
	\label{time-dep-EOM}
\end{equation}
\begin{figure}[]
	\centering
	\includegraphics[width=0.45\linewidth]{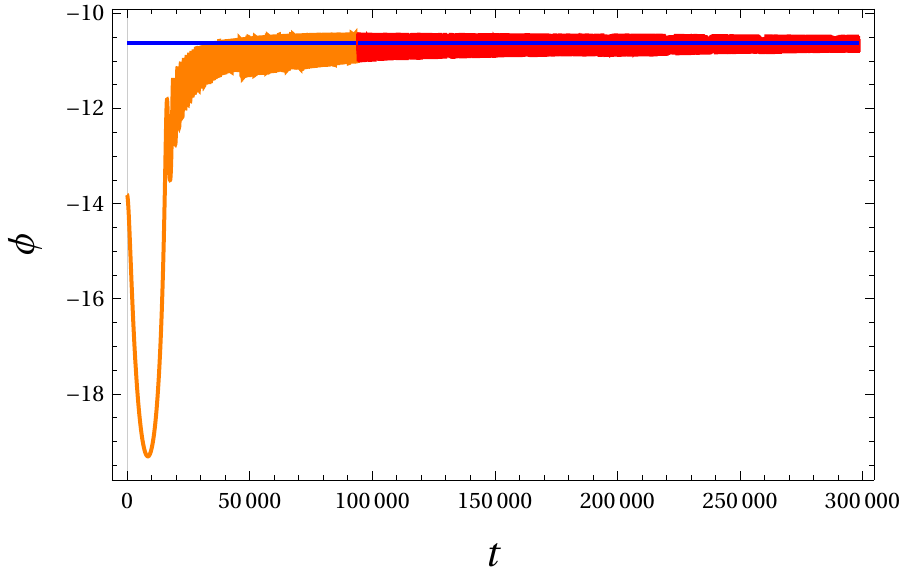}
	\caption{The solution for $\phi$ with both relevant and marginal coupling remains bounded even while quenching the parameters as in \protect\eqref{quench-g}. The orange curve is before the quench, which is initiated at $t=t_{max}=30\beta$. After that the solution shown in red, remains finite and bounded at all times. In fact $\phi$ approaches the value $2\log(\frac{g}{2g'})-\frac{8\alpha_s}{8\alpha_s + n_f t_{max} g^2} \approx -10.63$ shown by the blue line. Here $J=1, \beta=1000\pi, n_f=0.01, g=0.04, g'=4$.}\label{fig:phiquench}
\end{figure}
\begin{figure}[h!]
	\centering
	\includegraphics[width=0.6\linewidth]{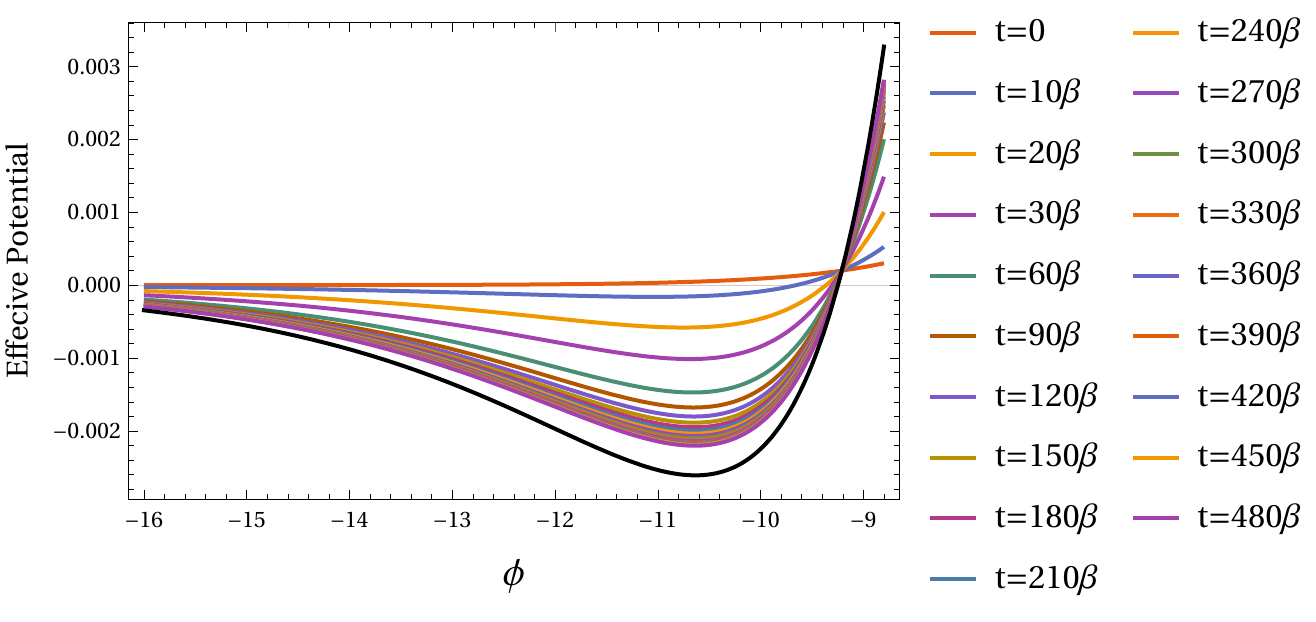}
	\caption{We plot the effective potential \protect\eqref{eff-pot} while quenching the parameters $g$ and $g'$ as in equation \protect\eqref{quench-g}. The black curve is the asymptotic value ($ t\rightarrow \infty $) of the effective potential \protect\eqref{eff-pot-const}. The post-quench solution stabilizes and seems to slowly approach the black curve asymptotically. Also compare with the previous case in Figure \protect\ref{fig:pot}, where the potential kept on increasing. We quench both the coupling parameters $g$ and $g'$ at $t=t_{max}=30\beta$. Here $J=1, \beta=1000\pi, n_f=0.01, g=0.04, g'=4$.} \label{fig:effPquench}
\end{figure}
We solve this modified equation with the quench protocol \eqref{quench-g} and find that even now $\phi$ approaches $\phi_{eq}=2\log \frac{g}{2g'}$ and therefore is still finite and bounded as shown in Figure \ref{fig:phiquench}. In Figure \ref{fig:effPquench} and Figure \ref{fig:quenchEn} we plot the effective potential and the energies (both \eqref{energy-sch} and \eqref{energy-def}) respectively. After $t=t_{max}$ we see that all of them are bounded unlike without the quench. For the specific values of parameters, namely $n_f=0.01,g=0.04, g'=4$ and $\alpha_s\approx 0.007$, for which we present the plots, we choose $t_{max}$, such that $t_{max} \approx 10^5 \ll t_*\approx 10^8$.
\begin{figure}[]
	\centering
	\begin{subfigure}{.45\textwidth}
		\centering
		\includegraphics[width=\linewidth]{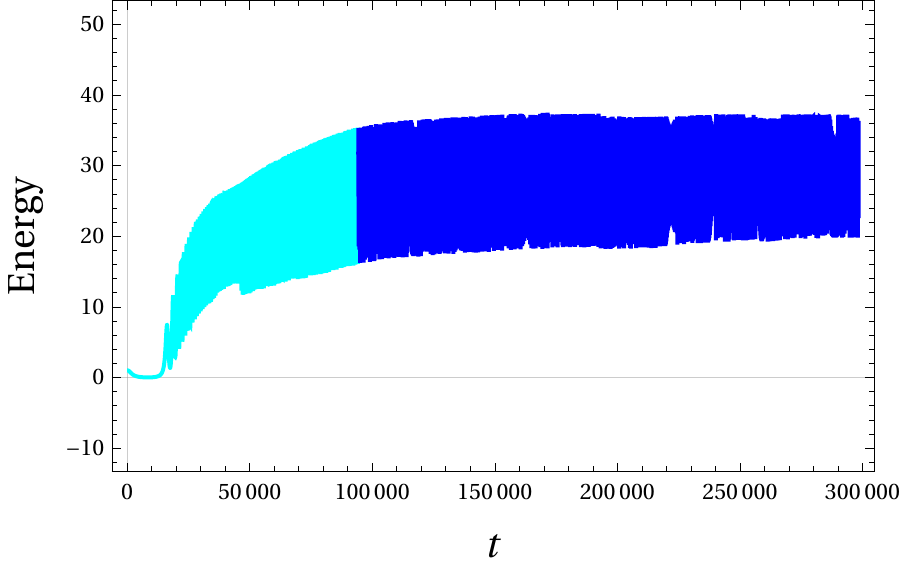}
		\caption{Energy $E_{sch}$}
		\label{fig:sch-quench}
	\end{subfigure}\hfill
	\begin{subfigure}{.45\textwidth}
		\centering
		\includegraphics[width=\linewidth]{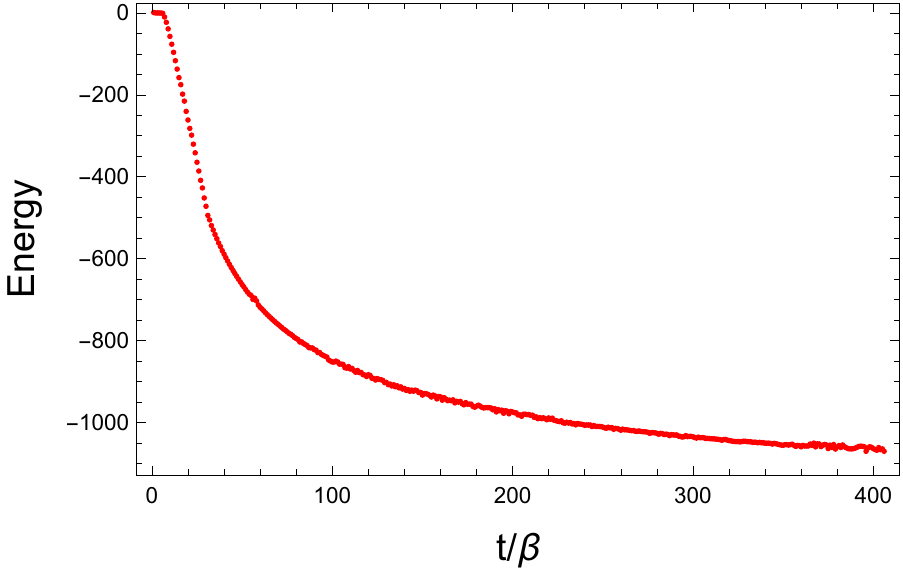}
		\caption{Energy $E_{\Delta}$}
		\label{fig:energy-quench}
	\end{subfigure}
	\caption{We plot the energies, namely Schwarzian \protect\eqref{energy-sch} (left) and $E_\Delta$ \protect\eqref{energy-def} (right). We quench the parameters $g$ and $g'$ as in eq. \protect\eqref{quench-g} with $t_{max}=30\beta\approx 94000$. Before $t_{max}$ (shown in light blue) the amplitude of the Schwarzian keeps on increasing, however after $t_{max}$ (shown in dark blue) the amplitude growth stops, but it keeps on oscillating. For $E_\Delta$, the linear growth stops at $t= t_{max}=30\beta$. Both the energies are shown in units of the initial energy. Here $J=1, \beta=1000\pi, n_f=0.01, g=0.04, g'=4$.} \label{fig:quenchEn}
\end{figure}
%%%%%
\paragraph{Long times after quench:} Since the solution remains finite and bounded at all times, we can do a similar analysis that led us to the Airy equation before. Writing
\begin{equation}
	\phi(t) = \phi_{eq} + \psi(t) ,
	\label{asym}
\end{equation}
the equation of motion to leading order in small $\psi$ at long times after the quench ($t\gg t_{max}$) reads
\begin{equation}%J=1
	\ddot{\psi}(t) + 2 e^{\phi_{eq}} \left( 1+\psi(t) \right) +\frac{n_f t_{max} g^2}{16\alpha_s} \left(\frac{g}{g'}\right)^2 \psi(t) =0 .
\end{equation}
We have set $J=1$. Compared to equation \eqref{preAiry}, the last term does not grow linearly with time. We can solve this equation, the solution is simple
\begin{equation}\label{sin-cos}
	\psi(t) = -\frac{8\alpha_s}{8\alpha_s + n_f t_{max} g^2} + c_1 \sin(\Omega t) + c_2 \cos(\Omega t), \qquad \Omega = \left(\frac{8\alpha_s g^2+n_f t_{max} g^4}{16\alpha_s g'^2}\right)^{1/2} .
\end{equation}
Therefore we find that at late times the solution oscillates with constant frequency $\Omega$ around the value 
\begin{equation} \label{phiTeq}
	\tilde\phi_{eq} = \phi_{eq} -\frac{8\alpha_s}{8\alpha_s + n_f t_{max} g^2} = 2\log(\frac{g}{2g'})-\frac{8\alpha_s}{8\alpha_s + n_f t_{max} g^2} \,.
\end{equation}
For the parameter values $n_f=0.01,g=0.04, g'=4, \alpha_s\approx 0.007$ used in Figure \ref{fig:phiquench}, $\tilde\phi_{eq} \approx-10.63$ and the frequency $\Omega \approx 0.037$. There is a simple way to understand \eq{sin-cos}. At long times ($t\gg t_{max}$), we find that the integral in equation of motion \eqref{time-dep-EOM} stabilizes, and we can rewrite the equation simply as %\doubt check!!!
\begin{equation}\label{eq-asymp}
	\ddot{\phi} + V'_{eff}(\phi) =0 ,
\end{equation}
with the asymptotic value of the effective potential (we have reinstated $J$)
\begin{equation}\label{eff-pot-const}% J neq 1
	V_{eff}(\phi)= 2 J^2 e^{\phi}  + \frac{J t_{max}}{2\alpha_s} n_f g^2 J^2 \left(e^{\phi} - \frac{g}{g'} e^{\phi/2} \right) ,
\end{equation}
%\begin{equation}\label{eff-pot-const}% J=1
%	V_{eff}(\phi)= 2 e^{\phi}  + \frac{n_f}{2\alpha_s} t_{max}\, g^2 \left(e^{\phi} - \frac{g}{g'} e^{\phi/2} \right)
%\end{equation}
which is time-independent.\footnote{In arriving at this expression we have dropped terms that are power law suppressed as $1/\sqrt{t}$.} This is also shown in Figure \ref{fig:effPquench} for comparison. In a similar spirit, one can show that the energy $E_\Delta$ \eqref{energy-def}, asymptotically corresponds to particle motion in the above asymptotic potential \eqref{eff-pot-const} (we have reinstated $J$):
\begin{equation}\label{e-d-const}
	E_\Delta(t\rightarrow \infty) = N\frac{\alpha_{s}}{J} \left[\frac{1}{2}\dot{\phi}^2 + 2 J^2 e^{\phi} + \frac{J t_{max}}{2\alpha_s} n_f g^2 J^2 \left(e^{\phi} - \frac{g}{g'} e^{\phi/2} \right) \right]= N\frac{\alpha_{s}}{J} \left[\frac{1}{2}\dot{\phi}^2 +V_{eff}(\phi)  \right] .
\end{equation}
Note that  $E_{sch}$, which has only the term $2J^2 e^\phi$ of the potential, misses coupling term. Thus, $E_\Delta$ is the more appropriate definition of energy here than $E_{sch}$.\footnote{It is important to realize that although it appears that the coupling $g$ goes to zero as a power law, it is not enough to decouple the system from the bath; as a result the two notions of energy remain different even asymptotically.}

%%%%%%%%%%
\subsection{Nature of solutions and admissible domain of the $g$-$g'$ plane}\label{domain}

%\textcolor{red}{
Note first of all that Model (a) can be regarded as a special case of Model (b), i.e. Model (a) is obtained by putting $g'=0$ in Model (b). As we saw above in this section, the nature of the solutions strongly depends on the couplings $g$ and $g'$. From our numerical analysis, we were able to chart out various solutions in the $g$-$g'$ plane which is shown in Figure \ref{fig:grid}. 
\begin{figure}[]
	\centering
	\includegraphics[width=0.5\linewidth]{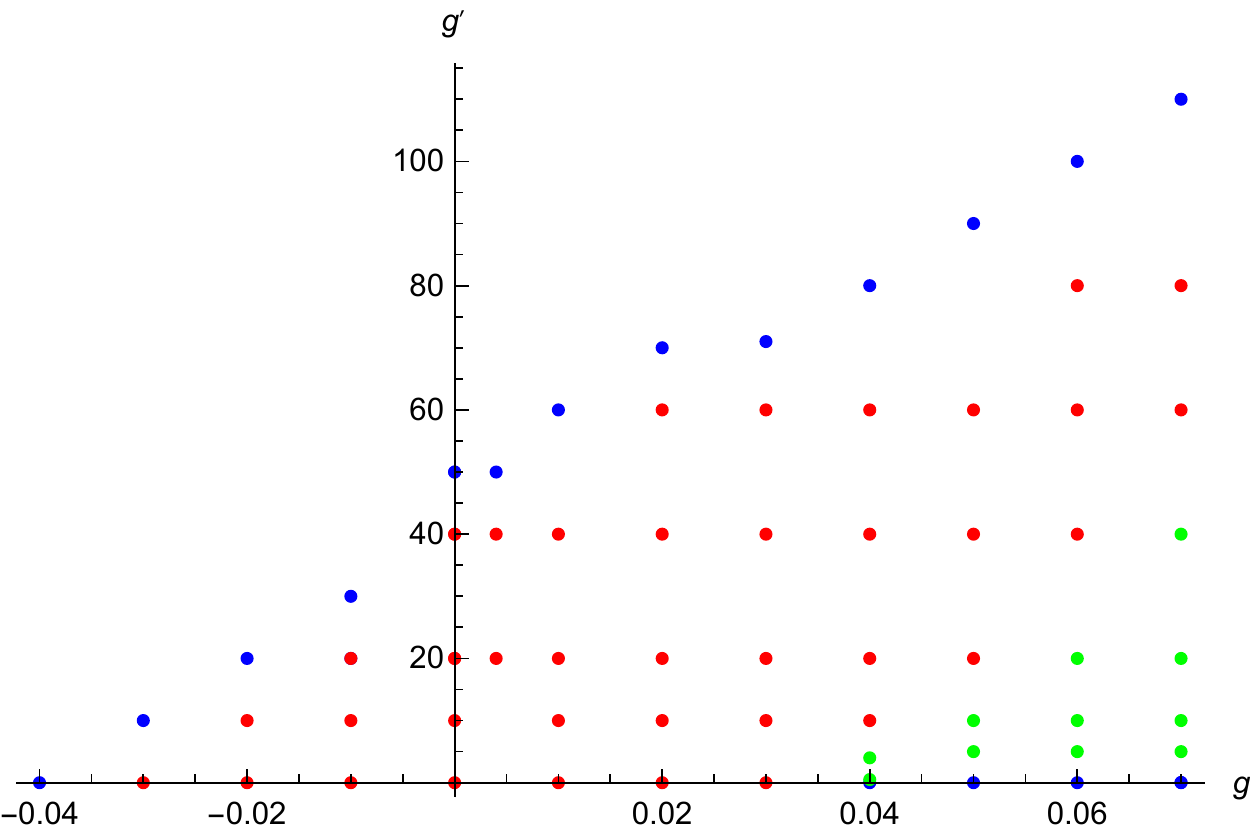}
	\caption{The various solutions in the $g$-$g'$ plane: (i) well-behaved solutions, $\phi(t\to \infty) \to -\infty$, that correspond to black hole geometries (shown in red), (ii) bounded solutions that oscillates around a fixed value $\phi_{eq}$ with a diminishing amplitude (shown in green), and (iii) runaway solutions $\phi(t\to \infty) \to \infty$, which we disregard (shown in blue). Models represented by green dots lead to horizonless geometries after one turns on a quench protocol \protect\eqref{quench-g}. }
	\label{fig:grid}
\end{figure}

Broadly we found three types of solutions, (i) well-behaved solutions, $\phi(t\to \infty) \to -\infty$, that correspond to black hole geometries (shown in red), (ii) bounded solutions that oscillates around a fixed value $\phi_{eq}$ with a diminishing amplitude (shown in green), and (iii) runaway solutions $\phi(t\to \infty) \to \infty$, which we disregarded (shown in blue). Based on this, a schematic cartoon of the solution space is constructed and is shown in Figure \ref{g-gprime}. While we have not marked the full plane, the entire region `outside' corresponds to runaway (blue) solutions.

We note that near the origin, the solutions are always (smaller) black holes. However there is a finite region in $g$-$g'$ plane for which the solutions as well as energies at late times remain finite and bounded, after using the quench protocol \eqref{quench-g}. They correspond to horizonless geometries, as shown in Figure \ref{fig:rel+marg} and are shown as green dots and regions in Figures \ref{fig:grid} and \ref{g-gprime}. This means that horizonless geometries cannot be obtained in perturbation theory in couplings, as we already found.
%}
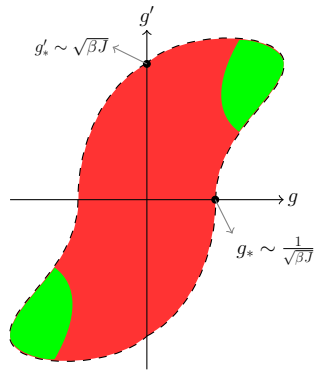
\begin{figure}[h!]
	\centering
	\begin{tikzpicture}[scale=0.45, every node/.style={transform shape}]
		%\draw[help lines, very thin, gray] (-5,-5) grid (5,5);
		\draw[thick, dashed] (0,4) .. controls (-1,3.5) and (-2,2) .. (-2,0);
		\fill[red!80] (0,4) .. controls (-1,3.5) and (-2,2) .. (-2,0) -- (0,0) -- (0,4);
		\draw[thick, dashed] (0,-4) .. controls (1,-3.5) and (2,-2) .. (2,0);
		\fill[red!80] (0,-4) .. controls (1,-3.5) and (2,-2) .. (2,0) -- (0,0) -- (0,-4);
		\draw[thick, dashed] (2,0) .. controls (2,2) and (4,3) .. (4,4) .. controls (4,5) and (1,5) .. (0,4);
		\fill[red!80] (2,0) .. controls (2,2) and (4,3) .. (4,4) .. controls (4,5) and (1,5) .. (0,4) -- (0,0) -- (2,0);
		\draw[thick, dashed] (-2,0) .. controls (-2,-2) and (-4,-3) .. (-4,-4) .. controls (-4,-5) and (-1,-5) .. (0,-4);
		\fill[red!80] (-2,0) .. controls (-2,-2) and (-4,-3) .. (-4,-4) .. controls (-4,-5) and (-1,-5) .. (0,-4) -- (0,0) -- (-2,0);
		\fill[green] (2.7,2) .. controls (3,2.5) and (4,3.4) .. (4,4) .. controls (4,4.5) and (3,4.75) .. (2.7,4.7) .. controls (2,3.5) and (2,2.5) .. (2.7,2);
		\fill[green] (-2.7,-2) .. controls (-3,-2.5) and (-4,-3.4) .. (-4,-4) .. controls (-4,-4.5) and (-3,-4.75) .. (-2.7,-4.7) .. controls (-2,-3.5) and (-2,-2.5) .. (-2.7,-2);
		\draw[->] (-4,0) -- (4,0);
		\node[right] at (4,0) {\LARGE $g$};
		\draw[->] (0,-5) -- (0,5);
		\node[above] at (0,5) {\LARGE $g'$};
		\draw[fill=black] (2,0) circle [radius=0.1]; 
		\draw[gray, ->] (2,0) -- (2.5,-1); \node[right] at (2.5,-1.5) {\LARGE $g_*\sim \frac{1}{\sqrt{\beta J}}$};
		\draw[fill=black] (0,4) circle [radius=0.1]; 
		\draw[gray, ->] (0,4) -- (-1,4.5); \node[left] at (-1,4.5) {\Large $g'_*\sim \sqrt{\beta J}$};
		%\draw[thick, yellow] (0,0) --(1.5,4.7);
		%\draw[thick, blue] (0,0) --(3.5,4.6);
	\end{tikzpicture}
	\caption{The `phase-diagram' in the $g$-$g'$ plane. This is a cartoon based on our numerical investigation (see Figure \ref{fig:grid}) and is not drawn to scale. This diagram has used the fact that the allowed domain has a reflection symmetry in the origin of the $g-g'$ plane. The color coding is the same as Figure \ref{fig:grid}, except now the blue dots populate the white region outside the black dashed curve. We note that the place where this curve intersects the $g$ axis is $g_*\sim O(1/\sqrt{\beta J})$, and the place where it intersects the $g'$ axis is $g'_* \sim O(\sqrt{\beta J})$ which is very large. In principle some part of the green region can extend to infinity. %Inside the dashed black curve, the low-energy approximation remains valid and the solutions are well-defined. Outside, the solutions show a runaway behaviour. In particular, the place where this curve intersects the $g$ axis is $g_*\sim O(1/\sqrt{\beta J})$, and the place where it intersects the $g'$ axis is $g'_* \sim O(\sqrt{\beta J})$ which is very large. In the red region, solutions remain black holes while green regions correspond to non black hole solutions. The blue line represents $g/g'=10^{-2}$ which crosses into green region at $(g,g')\approx (0.038,3.8)$ and then crosses the dashed curve at $(g,g')\approx (0.058,5.8)$. The yellow line represents $g/g'=10^{-4}$ remains in the red region till it directly crosses the dashed line at $(g,g')\approx (0.0045,45)$. Note that the allowed domain has a reflection symmetry in the origin of the $g-g'$ plane. Since green regions are finite distance away from the origin, they are outside the applicability of perturbation theory in the coupling. %\doubt{\color{red} MAKE THE OUTER PERIPHERY DOTTED/WIGGLY? AND ELONGATE FIGURE ALONG Y-AXIS - DONE}
	\label{g-gprime}}	
\end{figure}

%\textcolor{red}{
Whenever the solution $\phi(t)$ grows unbounded, the Schwarzian $E_{sch}$ also grows without bound (due to the exponential potential $2e^{\phi(t)}$). Therefore outside the red region, there is anti-dissipation and the system takes in energy from the (colder) bath. This kind of behaviour is unphysical, and is a generalization of such behaviour in the special case of Model (a) ($g'=0$) as elaborated in Section \ref{sec:unphysical-a}.  In the green region too, the energy increases, but it does so in a much milder, power law, fashion. With an appropriate time-dependent coupling as in \eqref{quench-g}, one obtains in this case finite energy solutions that remain bounded and correspond to horizonless geometry.
%}

Lastly, we note that in order to generate Figure \ref{fig:grid}, we solved the third order equation \eqref{third-order-general} and not the second order equation \eqref{ph-cl-eqn-a}. One might worry that the runaway solutions are pathologies of the 3rd order equation, similar to the radiation reaction problem in electrodynamics.\footnote{ We are grateful to R. Loganayagam and Suvrat Raju for raising this point.} However we numerically verify that the solutions we find satisfy the original 2nd order equation, so that the runaway solutions are indeed pathologies of the 2nd order equation.

%Just as in Model (a), the solutions of Model (b) described in Section \ref{sec:relevant+marginal} above show a runaway behavior for large values of $g$ and $g'$. From our numerics we are able to infer that there is a curve in the $g$-$g'$ plane which separates the well-behaved and the runaway solutions, see Figure \ref{fig:grid}. This is also shown as the dashed black curve in the schematic cartoon Figure \ref{g-gprime}. However there is a finite region in $g$-$g'$ plane for which at late times the solutions are bounded corresponding to horizonless geometries, as shown in Figure \ref{fig:rel+marg}. These are shown as green regions in Figure \ref{g-gprime}, whereas the red region corresponds to solutions that remain a (smaller) black hole at late times. 
%Note that it is $\hat g$ and $\hat g'$, which are tunable parameters, not $g$ and $g'$ since they depend on $s\cdot s' \propto \cos\theta$. However this angle can be recovered from the asymptotic geometry. This is true whether the asymptotic solution is black hole or not. Therefore this forms the basis of the recovery of the initial state as elaborated in section \ref{sec:recovery}.

\section{Information recovery \label{sec:recovery}}

We would like to determine our black hole microstate from the asymptotic measurements. In other words, we would like to determine the spin vector $s$ of the SYK pure state $\ket{B_s(l)}$.  

To do this we imagine doing the following ``experiment". At time $t=0$ we couple the SYK system  to a bath using a certain coupling of our choice. In the previous section, in both models (a) and (b), we found that in a certain range of the coupling parameters there exist asymptotic solutions and that they carry an imprint of the spin vector $s$. In the following we will investigate whether this imprint is enough to determine the spin vector $s$ entirely. We will consider models (a) and (b) in turn.

\subsection{Model (a): partial evaporation\label{sec:recovery-a}}

In this case the SYK system is coupled to the bath through the operator  $\hat g_1 \mathcal{O}_1^{\{s' \}}$. We will consider $\hat g_1 $  and spin $s'$ as the coupling parameters, which we imagine being able to choose freely.\footnote{We will come back to this point shortly. See Section \ref{sec:s'}.}

We first show that the existence of a well defined asymptotic solution is ensured if we  fix $\hat g_1$ as follows
\begin{align}
	\hat g_1  = c_1  \frac{1}{\sqrt{\beta J}},
	\label{range-a}
\end{align}
where $c_1$ is any fixed number satisfying
\begin{align}
  c_1 \ll 4\pi \sqrt{\frac{\alpha_s}{n_f}}.
  \label{c1-range}
  \end{align}

Proof: To show this, note that the effective IR coupling, $g$ or $\tilde g$, is given in terms of the coupling parameters $\hat g_1$ and $s'$ by (see \eqref{cos-alpha}, \eqref{relevant} and \eqref{third-order})
\begin{align}
	\tilde g= g \sqrt{\frac{n_f}{2\alpha_{s}}}, \quad g = \frac{\hat g_1}{\sqrt{4\pi}}  \cos\theta, 
	\quad \cos\theta = s'.s \equiv \frac{1}{N/2} \sum_{i=1}^{N/2} s'_i  s_i .
	\label{maps}
\end{align}
\eq{range-a} implies that 
\begin{align}
  \frac{{\tilde g}^2}{2\pi/(\beta J)}= \left(\frac{c_1}{4\pi\sqrt{\alpha_{s}/n_f}}\right)^2 \cos^2\theta \ll \cos^2\theta \ll 1 ,
  \label{bound-a}
\end{align}
where in the penultimate step we have used \eq{c1-range}. The final
inequality ensures that bounds such as \eq{g-crit-n} or \eq{g-crit-a} are satisfied which ensures the existence of a well-defined asymptotic solution. Note that it is enough to fix $\hat g_1$ as above; the bound \eq{bound-a} is satisfied irrespective of what $s'$ is. [QED]

Once the existence of a well-defined asymptotic solution is ensured in this way, we would like to ask whether the asymptotic solution retains some information about the original black hole microstate $\ket{B_s(l)}$. It is easy to see that the asymptotic energy \eq{energy-final}, given by 
\begin{align}
	E_\infty  = N\alpha_s J \fr12 \left(\frac{2\pi}{\beta J} -\gsq \right)^2
	\label{energy-final-a}
\end{align}
does retain such information, since by measuring it, we can determine $(\tilde g)^2$ and running through the maps in \eq{maps} and using the fixed value of $\hat g_1$ in \eq{range-a} and \eq{c1-range}, we can determine square of the inner product $(s'\cdot s)^2$. We have imagined here that the spin $s'$ is part of the preparation of the experiment, which we know and can tune (see Section \ref{sec:s'}). For any given choice of $s'$, the quantity $(s'\cdot s)^2$ carries a rather small amount of information about the spin vector $s$, which we are interested in. However by repeating the experiment for $\binom{N/2}{2}$ distinct choices of the spin vector $s'$, say, $s'= s^{(\alpha)}, \alpha=1,2,..., \binom{N/2}{2}$, then one can determine the spin vector $s$ up to a sign. This is explained in detail in Appendix \ref{app:Stirling}.

We now turn to Model (b). We will see here that we can determine the spin vector $s$ uniquely. 

\subsection{Model (b): complete evaporation\label{recovery-b}} 

Recall that in this model we use the mixed coupling
\begin{equation}
	\hat g_1 \mathcal{O}_1^{\{s' \}}+ \hat g_2 \mathcal{O}_2^{\{s',s' \}},
\end{equation}
which is characterized by the data $\{ \hat g_1, \hat g_2,s' \}$. The relation to the IR data is given by
\begin{equation}\label{UV-IR}
		g=\hat g_1\frac{\cos\theta}{\sqrt{4\pi}}, \quad g' = -\hat g_2 \left(\frac{\cos \theta}{\sqrt{4 \pi}}\right)^2, \quad \cos\theta=s'.s
\end{equation}
As in the previous subsection, we begin by examining the constraint on the couplings coming from the requirement of existence of a well-defined asymptotic solution. In model (a), it was given by an upper bound on $g$ or $\tilde g$. In model (b), it is given by the coloured region of the $g-g'$ plane, as explained in Figure \ref{g-gprime}. The algorithm here is divided into two steps: 1) finding the spin $s$ up to overall sign and 2) then fixing the sign.

Keeping in mind the different scaling of the two axes with respect to $\sqrt{\beta J}$ we fix the UV parameters as
\begin{equation}\label{fix-UV}
	\hat g_1 = \frac{c_1}{\sqrt{\beta J}}, \qquad \hat g_2 = c_2\sqrt{\beta J}
\end{equation}
where $c_1,c_2$ are $O(1)$ numbers ensuring that $g,g'$ always remain within the red region near the origin in Figure \ref{g-gprime}. The solution in the red region remains a black hole and the asymptotic energy (Schwarzian) is a function of $g$ and $g'$. From the energy one can recover the inner product $(s\cdot s')^2$ numerically or analytically in perturbation theory using\footnote{This expression for $a$ can be easily derived by combining the results of equations \eqref{asym-phi} and \eqref{asym-phi-2}.}
\begin{equation}\label{read-a}
	E_{sch}= N\frac{\alpha_{s}}{J} \frac{a^2}{2}, \quad	a=\left(\frac{2\pi}{\beta} - \tilde{g}^2 -\frac{\pi^2}{3\beta^2} \tilde{g}'^2 \right).
\end{equation}
Repeating the experiment for $\binom{N/2}{2}$ distinct choices of the spin vector $s'$, just as in Model (a), we can recover the initial spin $s$ up to a sign.\footnote{Note that we did not really need Model (b), for this Model (a) was sufficient.} This completes the first step.

In the 2nd step, we fix $c_1$ and $c_2$ such that if $\cos\theta=1$, then the point $(g,g')$ would lie in the green region. Here we will choose the values
\begin{equation}
	g=0.04, g'=4 \implies c_1 =0.04\sqrt{4\pi \beta J}, c_2 =-\frac{16\pi}{\sqrt{\beta J}} 
\end{equation}
which we know lie in the green region. Now we take either of 2 possible signs ($s'=\pm s$) and run our experiment with the above values of the UV parameters. If the late time solution oscillates around $\phi_{eq}=2\log(\frac{g}{2g'})$, then we know that it is the correct sign. If on the other hand we find some other behaviour, then we know that the other sign is the correct one. Thus we have determined the spin $s$ completely. Therefore we can determine the spin $s$ completely by doing $O(N^2)$ experiments. The polynomial dependence in $N$ has already been remarked on in point 3 of Section \ref{sec:intro}.

\subsection{Choice of $s'$ \label{sec:s'}}

As we explained above,  the low energy experiments performed at asymptotic times can measure the IR couplings $g$ and $g'$. In the protocol described above, we have assumed that we can freely choose the UV data, in particular we can choose the spin $s'$ characterizing the SYK operators which couple to the bath at $t=0$. It is important to note that this requires access to microscopic physics which goes beyond the low energy approximation. This should be contrasted with the asymptotic measurement of the energy (for Model (a)) or the frequency of the stable oscillation (for Model (b)) which can be expressed in terms of the Schwarzian mode. The important point is that the distinction between different $\ket{B_s(l)}$ states cannot be made by operators built out of the Schwarzian mode, {\it i.e.} the ``boundary graviton''. In our construction of the SYK-bath coupling, we have crucially used the KM operator ${\cal O}_1^{s'}$ which can make such a distinction among the various black hole microstates. As has been discussed in the literature, such operators cannot be described purely in terms of the boundary graviton, but possibly requires non-local constructions in the bulk (see, e.g. \cite{Gao:2016bin, Maldacena:2017axo},\cite{Kourkoulou:2017zaj, Almheiri:2018ijj, Maldacena:2018lmt}). Our ability to choose such couplings to the bath is to be understood in terms of some such bulk construction; we hope to come back to this important issue in detail in the future.

\section{Two-point function of fluctuations} \label{sec:fluctuations-2pt}

Until now, we have only focused on the leading large-N solution. Now we also look at the two-point function $\langle \delta\phi_c(t)\, \delta\phi_c(t') \rangle$, which is suppressed in $N$. Formally one can rewrite \eqref{Langevin-separation} as
\[	\mathcal{D}_t \cdot \, \delta\phi_c = \frac{1}{2} \eta(t) , \]
%\[
%	\mathcal{D}_t\cdot(\ldots) = \left[\frac{\alpha}{J} \left( \partial_t^2 + V''(\phi_0(t)) \right) -n_f g^2  F''(\phi_0(t)) \xi(t) \int_0^t dt' F(\phi_0(t')) \right] (\ldots) -n_f g^2 F'(\phi_0(t)) \int_0^t dt' F'(\phi_0(t')) (\ldots) 
%\]
%\doubt Define $\xi$, $\tilde \eta$ --- ANURAG \doubt
where $\mathcal{D}$ is the integro-differential operator in \eqref{Langevin-separation}. To find the two-point function of $\delta\phi_c$, one needs to invert this operator
\begin{equation}
	\braket{\delta\phi_c(t) \, \delta\phi_c(t')} = \frac{1}{4} \mathcal{D}_t^{-1} \mathcal{D}_{t'}^{-1} \braket{\eta(t) \, \eta(t')} .
\end{equation}
In general this is hard because of two complications -- firstly it involves an integral and secondly it is a function of both $t-t'$ and $t+t'$. Since the classical solution in the two models (a) and (b) are starkly different, we compute the 2-point functions in these models separately.

%%%%%%%%%%%%%%%%%%%%
\subsection{Model (a)}\label{sec:2pt-A}
In model (a) \ref{sec:relevant}, the classical solution reaches another equilibrium corresponding to a smaller black hole. At late times the solution behaves as $\phi(t) =-at+\ldots$, which allows us to self-consistently drop the integrals in the Langevin equation \eqref{Langevin-separation}. Therefore at late times, equation \eqref{Langevin-separation} essentially becomes
\[ \frac{\alpha_s}{J}\, \delta\ddot{\phi_c}(t) = \frac{1}{2} \eta(t) , \]
or equivalently the differential operator simply reads
\[ \mathcal{D}_t = \frac{\alpha_s}{J} \partial_t^2 . \]
Now inversion is achieved trivially by integrating twice with respect to $t$ and $t'$. The two-point function at late times ($t\gg a^{-1}$) is given in terms of an integral
\begin{align}\label{2ptF}
	\langle \delta\phi_c(t) \, \delta\phi_c(t')\rangle=& e^{-a(t+t')/2} \Bigg\{-\frac{2C}{\pi} \int_{-\Lambda}^{\Lambda} \frac{dk\, e^{ik(t-t')}}{k \left( k^{2}+ a^{2}/4 \right)^2} \coth(2kL) \\
	&+\frac{C}{4\pi} \int_{-\Lambda}^{\Lambda} \frac{dk\, e^{ik(t+t')}}{k \left( k^{2}+ a^{2}/4 \right)^4}\left[\frac{a^4}{2}-12 a^2k^2+8k^4 -i \left(16a k^3 -4a^3 k \right) \right] \csch(2kL) \Bigg\} , \nonumber
\end{align}
where $C=2 n_f g^{2}\left(\frac{J^2}{2 \alpha_{s}}\right)^{2}$. 
%\begin{equation}\label{eqc}
%	C=2 n_f g^{2}\left(\frac{J^2}{2 \alpha_{s}}\right)^{2}
%\end{equation}
This is UV finite unlike the noise correlator \eqref{2pt-noise} and therefore we can take the cutoff $\Lambda\rightarrow \infty$. Then one can evaluate this integral and the  full form of the 2-point function is given by \eqref{eq:2pt-full} in the appendix \ref{app:2pt}. In particular the equal-time correlator ($t=t'\gg a^{-1}$) reads
\begin{equation}
	\langle \delta\phi_c(t) \, \delta\phi_c(t)\rangle= e^{-a t} \left[\frac{16 C}{a^4 L} t + \text{constant} + O(e^{-\pi t/L}) + O(e^{-at}) \right] .
\end{equation}
The diffusion constant appears in the Einstein-Smoluchowski form (see \cite{pathria2011statistical} for example) in terms of the effective\footnote{We use the term ``effective'' temperature because the bath is in a pure state in which the late time correlators behave as thermal correlators at this temperature.} bath temperature $(4L)^{-1}$.  In a similar fashion one can also evaluate the 2-point function for the energy $E_\Delta$ defined in \eqref{energy-def}.

%%%%%%%%%%%%%%%%%%%%
\subsection{Model (b)}\label{sec:2pt-B}

\begin{figure}[h]
	\centering
	\begin{subfigure}{.45\textwidth}
		\centering
		\includegraphics[width=\linewidth]{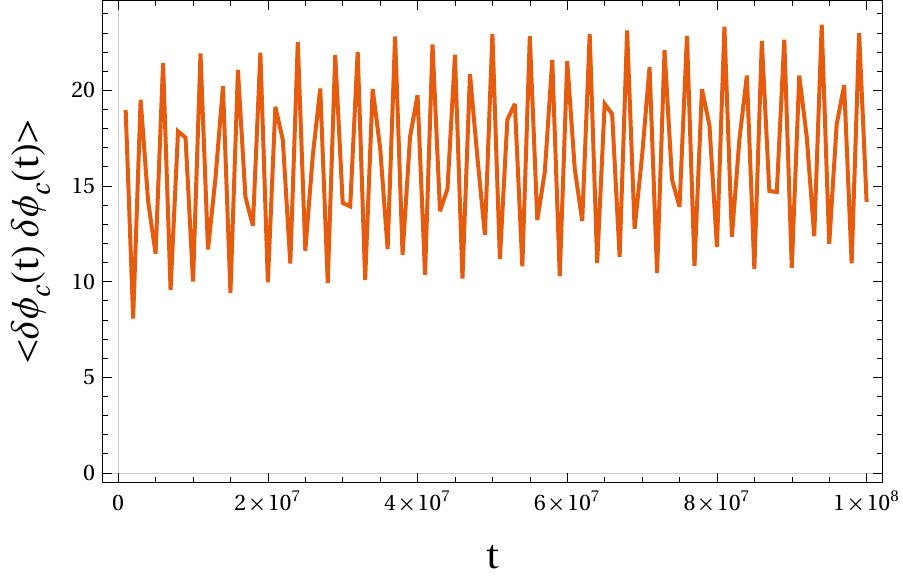}
		\caption{Fixed bath temperature $T_b=2.5\times 10^{-6}$}
		\label{fig:fixed-L}
	\end{subfigure}\hfill
	\begin{subfigure}{.45\textwidth}
		\centering
		\includegraphics[width=\linewidth]{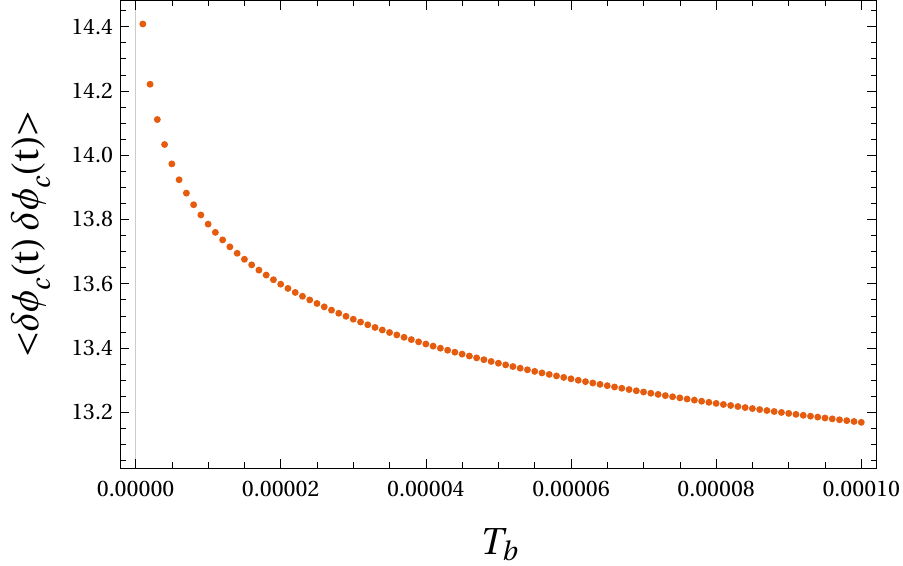}
		\caption{Fixed time $t=10^8$}
		\label{fig:fixed-t}
	\end{subfigure}
	\caption{We plot the equal-time 2-point function $\langle\delta\phi_c(t) \delta\phi_c(t)\rangle $ by evaluating \protect\eqref{eq:integral2ptb} numerically. In the left panel we plot it as a function of time $t$ at fixed effective bath temperature $T_b=(4L)^{-1}=2.5\times 10^{-6}$. In the right panel we plot it as a function of effective bath temperature $T_b$ at time $t=10^8\gg t_{max}\approx 10^5$. The temperature dependence is clearly not linear as in the Einstein-Smoluchowski relation. Here $J=1, \beta=1000\pi, n_f=0.01, g=0.04, g'=4$ and $t_{max}=30\beta\approx 10^5$.} \label{fig:2pt-b}
\end{figure}

In model (b) \ref{sec:relevant+marginal}, the final geometry is horizonless. To get rid of the integral involving fluctuation $\delta \phi_c$ in \eqref{Langevin-separation}, we take one time more derivative. The third order equation with time-dependent coupling reads
\begin{align}
	&\dddot{\delta \phi_c}(t) -\ddot{\delta \phi_c}(t) \left\{\frac{F''(\phi(t))}{F'(\phi(t))} \dot{\phi}(t) +\frac{\dot{g}(t)}{g(t)}\right\} + \left\{V''(\phi(t))- \frac{J n_f}{\alpha_s} g(t) F''(\phi(t)) \int_0^t dt' g(t') F\left(\phi (t')\right)\right\} \dot{\delta \phi_c}(t) \nonumber \\
	+& \Bigg[V'''(\phi(t)) \dot{\phi}(t) - \left\{ \frac{F''(\phi(t))}{F'(\phi(t))} \dot{\phi}(t) + \frac{\dot{g}(t)}{g(t)}\right\} V''(\phi(t)) - \frac{J n_f}{\alpha_s} g(t) \bigg\{ F'''(\phi(t)) \dot{\phi}(t) \int_0^t d t' g(t^{\prime}) F\left(\phi (t^{\prime}) \right) \nonumber \\
	+& F''(\phi(t)) g(t) F(\phi(t)) + g(t)\left(F^{\prime}(\phi(t))\right)^2 -\frac{\left(F''(\phi(t))\right)^2}{F'(\phi(t))} \dot{\phi}(t) \int_0^t dt' g(t^{\prime}) F\left(\phi(t')\right) \bigg\} \Bigg] \delta \phi_c(t) \nonumber \\
	=& \frac{J}{2\alpha_s} \dot{\eta}(t) -\frac{J}{2\alpha_s} \left\{\frac{F''(\phi(t))}{F'(\phi(t))} \dot{\phi}(t) +\frac{\dot{g}(t)}{g(t)}\right\} \eta(t) . %\equiv \sigma(t)
\end{align}
Now we make a technical simplification, namely that the classical solution at late times after quench sits at the bottom of the potential \eqref{eff-pot-const}, which we call $\phi_m$. This is equivalent to putting $c_1=c_2=0$ in \eqref{sin-cos} and therefore $\dot\phi_m = 0$. Plugging in this solution at late times after quench \eqref{quench-g}, the 3rd order equation reads
\begin{align}
	&\dddot{\delta \phi_c}(t) +\frac{1}{2t} \ddot{\delta \phi_c}(t) +\left\{V''(\phi_m)-\frac{2}{\alpha_s} \eta_f g^2 J t_{max} F''(\phi_m) F(\phi_m)\right\} \dot{\delta \phi_c}(t) \nonumber \\
	&-\frac{1}{t}\left[\frac{n_f}{\alpha_s} g^2 J t_{max} \left\{F''(\phi_m) F(\phi_m) +\left(F'(\phi_m)\right)^2 \right\} -\frac{1}{2} V''(\phi(t)) \right] \delta \phi_c(t)=\frac{J}{\alpha_s}\left\{\frac{1}{2} \dot{\eta}(t)+\frac{1}{4 t} \eta(t)\right\} .
\end{align}
At very late times we can further drop the $1/t$ terms, then we are only left with
\begin{align}
	\dddot{\delta \phi_c} + \omega^2 \delta\dot\phi_c &= \frac{J}{2\alpha_s} \dot\eta ~~~ \xrightarrow{\text{\small integrate}} ~~~ \ddot{\delta \phi_c} + \omega^2 \delta \phi_c = \frac{J}{2\alpha_s} \eta , \\
	\omega^2 &= V''(\phi_m)-\frac{2}{\alpha_s} \eta_f g^2 J t_{max} F''(\phi_m) F(\phi_m) .
\end{align}
The fluctuation correlation can now be calculated in terms of the noise correlation as
\begin{align}\label{xi2ptB}
	\left\langle\delta \phi_c(t) \delta \phi_c\left(t^{\prime}\right)\right\rangle &=\left(\frac{J}{2 \alpha_s}\right)^2 \int_0^{\infty} d t_1 \int_0^{\infty} d t_2\, G_R\left(t,t_1\right) G_R\left(t^{\prime}, t_2\right) \langle\eta (t_1) \, \eta (t_2)\rangle \\
	&=\left(\frac{J}{2 \alpha_s \omega}\right)^2 \int_0^t d t_1 \int_0^{t^{\prime}} d t_2\, \sin \left[\omega (t-t_1)\right] \sin \left[\omega(t'-t_2)\right] \langle\eta (t_1) \, \eta (t_2)\rangle , \nonumber
\end{align}
where we have used the retarded Green's function
\begin{equation}\label{GreenHO}
	G_R(t-t') = \theta(t-t') \frac{\sin[\omega(t-t')]}{\omega} .
\end{equation}
We already know the noise 2-point function \eqref{2pt-noise}
\begin{equation}
	\langle\eta (t_1)\, \eta(t_2)\rangle =-2\, n_f\, g(t_1) F'\left(\phi(t_1)\right) K(t_1,t_2)\, g(t_2) F'\left(\phi(t_2)\right) \approx -2 \frac{n_f\, g^2\, t_{max}}{\sqrt{t_1} \sqrt{t_2}} F'(\phi_{m})^2 \, K(t_1,t_2) ,
\end{equation}
where in the right most expression we have put the classical solution $\phi=\phi_{m}$ at late times. Using the integral form of the kernel \eqref{kernel-pure}, this allows us to express the 2-point function as the following integral
\begin{equation}\label{eq:integral2ptb}
	\langle\delta \phi_c(t)\, \delta \phi_c(t')\rangle =\frac{n_f t_{max}}{\pi} \left(\frac{J g F'(\phi_m)}{\alpha_s \omega}\right)^2 \int_0^{\Lambda} \frac{dk}{k} \Big\{f_c(k,t) f_c(k,t') \tanh(k L) + f_s(k,t) f_s(k,t') \coth(k L)\Big\} ,
\end{equation}
where
\begin{align}
	f_s(k,t)= \int_0^t dt' \sin[\omega(t-t')] \frac{\sin(k t')}{\sqrt{t'}} =& \sqrt{\frac{\pi}{2}} \frac{\sin(\omega t) S\left(\sqrt{\frac{2}{\pi}} \sqrt{t(k+\omega)}\right) +\cos(\omega t) C\left(\sqrt{\frac{2}{\pi}} \sqrt{t(k+\omega)}\right)}{\sqrt{k+\omega}} \, + \nonumber\\
	&\sqrt{\frac{\pi}{2}} \frac{\sin(\omega t) S\left(\sqrt{\frac{2}{\pi}} \sqrt{t(k-\omega)}\right) -\cos(\omega t) C\left(\sqrt{\frac{2}{\pi} \sqrt{t(k-\omega)}}\right)}{\sqrt{k-\omega}} , \\
	f_c\left(k,t\right) = \int_0^t d t' \sin[\omega(t-t')] \frac{\cos(k t')}{\sqrt{t'}} =& \sqrt{\frac{\pi}{2}} \frac{\sin (\omega t) C\left(\sqrt{\frac{2}{\pi}} \sqrt{t(k+\omega)}\right)-\cos (\omega t) S\left(\sqrt{\frac{2}{\pi}} \sqrt{t(k+\omega)}\right)}{\sqrt{k+\omega}} \, + \nonumber \\
	&\sqrt{\frac{\pi}{2}} \frac{\sin(\omega t) C\left(\sqrt{\frac{2}{\pi}} \sqrt{t(k-\omega)}\right) +\cos(\omega t) S\left(\sqrt{\frac{2}{\pi}} \sqrt{t(k-\omega)}\right)}{\sqrt{k-\omega}} ,
\end{align}
with the Fresnel integrals defined as
\begin{equation}
	S(z) = \int_{0}^{z} \sin(\pi x^2/2) dx , \qquad  C(z) = \int_{0}^{z} \cos(\pi x^2/2) dx .
\end{equation}
This is UV finite unlike the noise correlator \eqref{2pt-noise} and therefore we can take the cutoff $\Lambda\rightarrow \infty$ in \eqref{eq:integral2ptb}. 

In Figure \ref{fig:2pt-b} we plot the equal-time correlation function $\langle \delta\phi_c(t) \delta\phi_c(t)\rangle $, both as a function of time at fixed $L$ and also a function of $L$ at fixed time. For this we have to evaluate the integral \eqref{eq:integral2ptb} numerically. As is evident from the graph \ref{fig:fixed-t}, the temperature dependence here is not linear as in the Einstein-Smoluchowski relation. Further the time-dependence has oscillatory behaviour as is characteristic of Brownian motion in a bounded potential (see Figure \ref{fig:effPquench}). For Brownian motion in a harmonic potential, which is exactly soluble, the time dependence of equal time 2-point function is oscillatory \cite{HOBM}.

%%%%%%%%%%%%%%%%%%%%%%%%%%%%%%%%%%%%%%%%
\section{Diagnostics of black hole evaporation\label{sec:bh-evap}}

Since our discussion of the SYK model coupled to the bath is always in a regime described by the Schwarzian mode, there is a direct correspondence between the SYK dynamics and black hole dynamics. In Model (a) (Section \ref{sec:relevant}), the solutions which at late times behave as $\phi(t)= -a t+ \ldots$, correspond to black hole geometries with a smaller mass. Model (b) (Section \ref{sec:relevant+marginal}) consists of solution in which $\dot f(t)=e^{\phi(t)}$ remains finite and bounded at all times and they correspond to non-black hole geometries. In this section we make these statements a bit more precise. We can explicitly see the phenomenon of black hole evaporation, in various ways:

%%%%%%%%%%%%%%%%%%%%
\subsection{Energy loss}\label{sec:energy-loss}

In Model (a) we are able to explicitly calculate the Schwarzian \eqref{energy-sch} and $E_\Delta$ \eqref{energy-def}. In Figure \ref{fig:energy-loss} we plotted both Schwarzian and $E_\Delta$ as a function of time and found that they decrease with time. While the two energies disagree at intermediate times they agree at late times, when equilibrium is reached. The total loss of energy depends on the coupling $g^2$. Note that by tuning $g^2$ to be arbitrarily close to its critical value, we can have an asymptotic black hole solution which is arbitrarily small; thus, using our set-up we can potentially explore questions related to information loss, while always remaining within weak coupling.

In Model (b), we found for the fixed coupling parameters $g$ and $g'$, the energies (\eqref{energy-sch} or \eqref{energy-def}) keep growing in magnitude (see Figure \ref{fig:ecomp}), and our solution breaks down at a timescale $t_*$ given by \eqref{tbreakdown}. For this reason we did a quench \eqref{quench-g}, after which both the energies are bounded (see Figure \ref{fig:quenchEn}). In particular $E_\Delta$, which is the more appropriate energy in this scenario, has a well-defined asymptotic form \eqref{e-d-const} at late times which is smaller than the initial energy, as is evident from  Figure \ref{fig:energy-quench}. %. However in this scenario, we argue below that the horizon disappears and therefore there is complete black hole evaporation.

%%%%%%%%%%%%%%%%%%%%                    
\subsection{Existence and location of the horizon}\label{sec:horizon}

In this subsection, we determine whether the solutions we found in Models (a) and (b) correspond to geometries with a horizon or not. The details of how to do it are discussed in Appendix \ref{app:horizon}, where we find that for a (future) horizon to exist,  in the limit $t\to \infty$, we must have that $\dot f(t) \to 0$ and that $f(t)$ reaches a finite value $\hat t_H$, which demarcates the point where the horizon meets the boundary $\hat z=0$.

As noted in that Appendix, the above condition is invariant under an $SL(2,R)$ transformation of $f(t)$. We can therefore determine the above condition in any given SL(2,R) gauge, in particular, the one we already used in fixing one of the initial conditions $\dot\phi(0)=0=\ddot f(0)$ for solving the classical equation \eqref{ph-cl-eqn-a} and choice of the Lagrange multiplier (see footnote \ref{ftnt:lagrange}) in arriving to the Liouville action \eqref{liouville}, together with $f(0)=0$. We determine $f(t)$ from a solution for $\phi(t)$ as follows
\begin{equation}
	\dot f(t) =  e^{\phi(t)}, \; f(t)=  \int_{0}^{t}dt'e^{\phi(t')} .
	\label{f-phi}
\end{equation}

\paragraph{Model (a)}

\underbar{Analytic solution} 
In \eqref{asym-phi-detailed}, \eqref{asym-phi}, an analytic solution is presented in a perturbation theory in the coupling $g$. It is easy to see from there that the condition for existence of a future horizon is satisfied since
\begin{align}
& \dot f= \exp[\phi(t)] \stackrel{t\to \infty}{\to} \exp[-a t] \to 0,  \; a= \left(\frac{2\pi}{\beta J} - \tilde{g}^2\right)  \nonumber\\
& \hat t_H= f(t)  \stackrel{t\to \infty}{\to}  \int_{0}^{\infty}dt'e^{-a t}= \frac1{\frac{2\pi}{\beta J}  - \tilde{g}^2} < \infty .
\label{analytic-a}
\end{align}
Note that the second line evaluates the coordinate of the endpoint of the horizon $\hat t_H= f(\infty)$ (which is the same as the endpoint of the boundary curve, see the detailed discussion in Appendix \ref{app:horizon}). Note that as $\tilde g$ increases, the location of the endpoint of the horizon rises, that is, the horizon shifts to the future. The perturbative result indicates that $\hat t_H \to \infty$, i.e. the horizon disappears when $\tilde g^2 = \frac{2\pi}{\beta J}$.

\underbar{Numerical solution} 
The analytic solutions are perturbative. In Section \ref{sec:numerical-a}, we discussed numerical solutions which do not take recourse to perturbation theory.  Based on the results there, we present in Figure \ref{fig:tbulk}, the result for $f(t)$. We find that the results (shown in from Figure \ref{fig:tbulk} ) are in agreement with the analytic findings, namely that $\hat t_H=f(t)$ asymptotes to a finite value, and  that as we increase the coupling $\tilde{g}^2$, the horizon meets the boundary at larger and larger times $\hat t$, i.e. the horizon shifts to the future. We stress again that this comparison makes sense only in a given fixed gauge.

The spacetime diagram for Model (a) is presented on the right panel of Figure \ref{fig:tbulk}, which falls in the ($t>0$ portion of) class (B) of the diagrams Figure \ref{fig:topology} and Figure \ref{fig:cutout}.

This model, therefore, represents a final black hole of lower energy \eq{energy-final} than the initial energy of the  microstate. We will say more on the final state in Section \ref{model-a}.

\paragraph{Model (b)}

\underbar{Analytic solution} The late time (post-quench) analytic solution in this case is given by \eq{asym}, \eq{sin-cos}. Clearly $\dot f(t) \not \to 0$, as $t \to \infty$. Hence the solution corresponds to a geometry without a horizon. The same result is supported by the numerical calculations (see Section \ref{sec:quench}). 

The spacetime diagram for Model (b) is presented on the right panel of Figure \ref{fig:tbulk}, which falls in the ($t>0$ portion of) class (A) of the diagrams Figure \ref{fig:topology} and Figure \ref{fig:cutout}.

The final geometry \eqref{AAdS}, for both Models (a) and (b), is presented in Section \ref{sec:final-state}.

\begin{figure}[]
	\centering
	\begin{subfigure}{.5\textwidth}
		\centering
		\includegraphics[width=\textwidth]{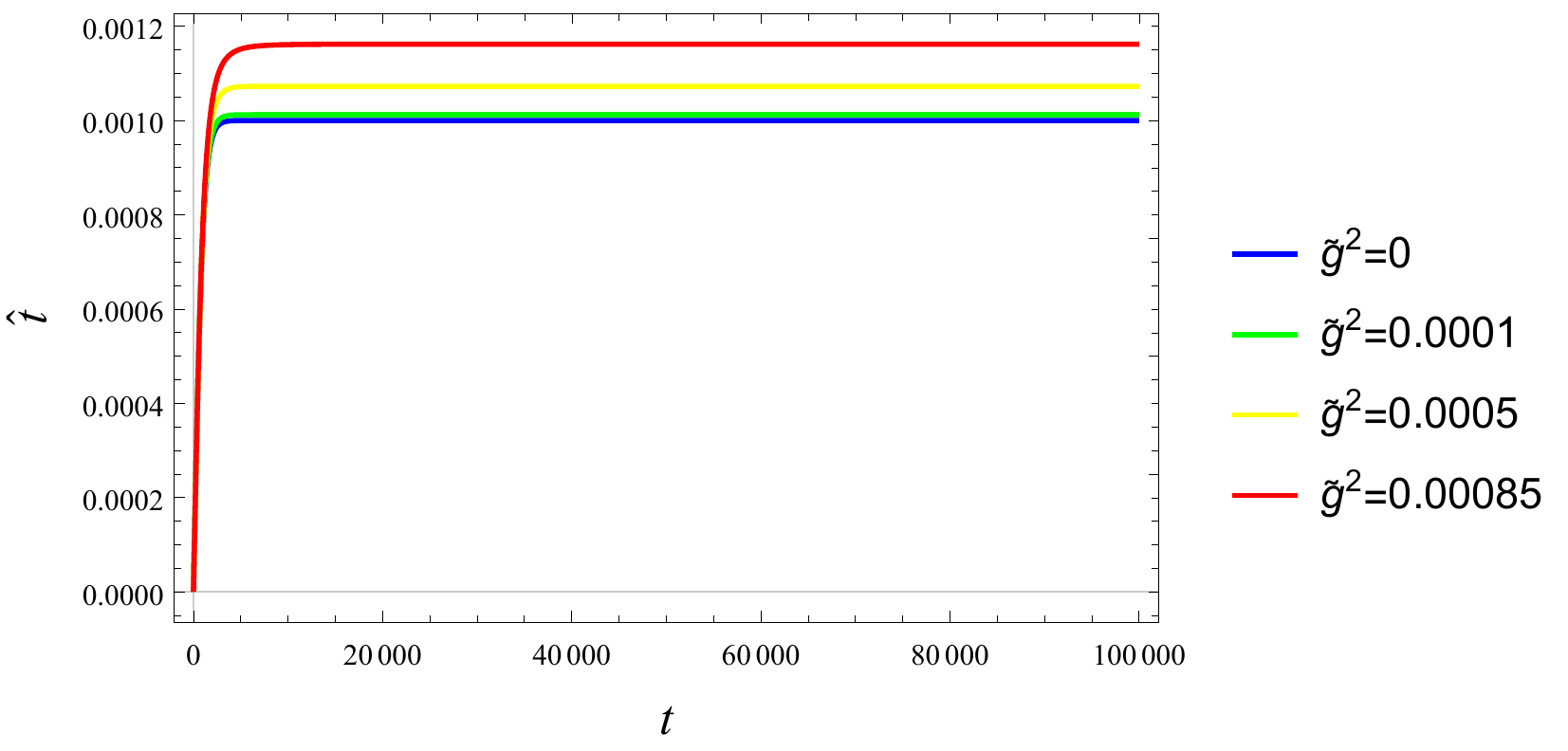}
		%\caption{}
		%\label{fig:tbulk}
	\end{subfigure}\hfill
	\begin{subfigure}{.5\textwidth}
		\centering
		\begin{tikzpicture}[scale=0.7, every node/.style={transform shape}]
			%\draw[help lines, very thin, gray] (-4,0) grid (4,4);
			\draw (-2,0) -- (2,0) -- (2,4);
			\draw[very thick] (2,4) -- (-2,0);
			\fill[blue!50] (-2,0) -- (1.7,0) .. controls (1.7,1) .. (2,4) -- (-2,0);
			\fill[red!50] (0,0) -- (1.7,0) .. controls (1.8,1) .. (2,2) -- (0,0);
			\draw[red, thick] (1.7,0) .. controls (1.8,1) .. (2,2) -- (0,0);
			\draw [->, thick, decorate, decoration={snake}]  (0,1) -- (1,2);
			\draw[blue, thick] (1.7,0) .. controls (1.7,1) .. (2,4);
			\draw (-2,0) arc (180:360:2);
			\node at (0,-1) {Euclidean};
			\node[red, right] at (2,2) {(a) $\hat z=\delta. \dot f=0$}; %\draw[->] (3.1,2) -- (4,2);
			\node[blue, right] at (2,3.9) {(b) $\hat z =\delta.\dot f \neq 0$}; %\draw[->] (3.1,3.9) -- (4,3.9);
		\end{tikzpicture}
	%\caption{}
	%\label{fig:AdS2}
	\end{subfigure}
	\caption{(Left) Bulk time $\hat t=f(t)$ for various strengths of single relevant interaction in model (a). Note that this comparison makes sense only in a fixed gauge. Numerics done for $\beta=1000\pi, J=1$. (Right) The top half portrays Poincar\'e patch for $\hat t\ge 0$. In model (a), the red curve hits $\hat z=0$ at a finite bulk time, so any signal after that will not be received by the boundary observer. On the other hand in model (b) the blue curve never hits $\hat z=0$ and therefore covers the full Poincar\'e patch. This corresponds to no horizon.} \label{fig:tbulk}
\end{figure}
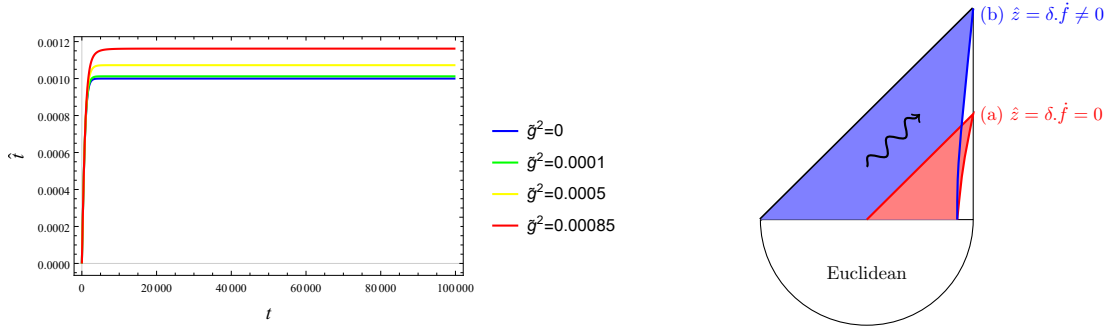

%The notion of the horizon `area' in two spacetime dimensions is replaced by the value of the dilaton at the horizon. The value of the dilaton decreases away from the boundary; hence when the horizon shifts away from the boundary, the horizon `area' decreases. \doubt ???? 

%\iffalse{
%***ENTROPY 
%At late times the solution is still a black hole with a smaller temperature. Thus one would expect that the (near-extremal) entropy should be given by \cite{Maldacena:2016upp}
%\begin{equation}\label{entropy}
%	S= N \frac{4\pi^2\alpha_s}{J} \mathcal{T}_{final} 
%\end{equation}
%Since we started with a pure state, the entanglement entropy of our Schwarzian system with the bath should be bounded by this quantity. (WHICH WE CAN ESTIMATE BY THE SUBLEADING TERM??).
%}\fi		

%%%%%%%%%%%%%%%%%%%%
\subsection{Radiation into the bath}\label{sec:energy-gain}

In the first scenario, as shown above in Section \ref{sec:energy-loss}, the system energy decreases. Since the total energy of the system and bath is constant, it can be inferred that the energy has gone into the bath. In our description of the dynamics (see \eq{ev-path-integral}, \eq{S-sk}) since the bath degrees of freedom have been integrated, the above inference is indirect. In order to directly compute the radiation into the bath, one can compute the expectation value of the energy flux $T^{\rm bath}_{xt}(x,t)=\sum_{i=1}^{N_f} \fr12 \del_x \Phi_i \del_t \Phi_i(x,t)$ at an interior point $x>0$ of the bath at time $t$. This can be done by inserting the operator $T^{\rm bath}_{xt}$, instead of the system operator $O(t)$, in \eq{ev}. It is easy to see that this quantity is zero in the decoupled case; once the coupling is switched on, it can be seen from perturbative calculations, that the energy flux turns non-zero and it moves towards $x\to \infty$ with the speed of light; in other words, the system emits a radiation of energy into the bath. More generally, it would be interesting to understand the time dependence of bath correlation functions and entanglement properties by treating the bath as the open system obtained after tracing over the SYK.

%\doubt Merge with para above \doubt
%In this paper we have traced over the bath degrees of freedom and studied SYK dynamics as an open quantum system. One may ask about the possibility of tracing over the SYK degrees of freedom and studying the bath as on open quantum system. It would be interesting to understand the time dependence of bath correlation functions and entanglement properties from this viewpoint.

%{\color{blue}In Model (a), the interaction terms goes to zero when the system reaches a new equilibrium as $t\rightarrow\infty$. But the total energy of SYK+bath is conserved
%\begin{equation}\label{eq:conservation}
%	E_{SYK}(0^-) + E_{bath}(0^+) = E_{SYK}(\infty) + E_{bath}(\infty)
%\end{equation}
%Therefore energy lost by the black hole is the same as gained by the bath as $t\rightarrow\infty$.}

%%%%%%%%%%%%%%%%%%%%%%%%%%%%%%%%%%%%%%%%
\section{Discussion}\label{sec:discussion}

%%%%%%%%%%%%%%%%%%%%
\subsection{Comments on thermalization and comparison with other works}\label{sec:G-Sigma}

As mentioned earlier, the issue of thermalization in the SYK + bath model has been discussed before, in particular in  \cite{Almheiri:2019jqq} and \cite{Maldacena:2019ufo}. In these papers, the analysis is at large $N$, but not restricted to large $J$, which necessitates the use of the bilocal variable $G(t_1, t_2)$ (the description in terms of the Schwarzian or the Liouville mode $\phi(t)$ is available only at large $J$) It is argued in these papers that the SYK + bath system asymptotically thermalizes to the bath temperature. The argument is that the bilocal variable $G$ is essentially the two-point of the fermions; thus, the classical evolution of $G$ represents the large $N$ time-evolution of the two-point function.  Assuming that the two-point function satisfies the fluctuation-dissipation theorem, applying it to its asymptotic form of $G$ one can determine the asymptotic temperature, which is found equal to the bath temperature. 

In our paper, we find something ostensibly different; the classical equation of motion of the Liouville field $\phi(t)$ is independent of the bath `temperature'\footnote{see comments below \eqref{product-state}} $T_b = \frac{1}{4L}$, so the asymptotic solution for $\phi(t)$ does not see the bath temperature at all (this is true in both of our Models (a) and (b)). The difference is not unwarranted, however. As already mentioned, the classical quantity of interest in \cite{Almheiri:2019jqq, Maldacena:2019ufo}, is itself a {\it two-point} function. It is well-known, e.g. from the Caldeira-Leggett model of an oscillator (the ``system") + bath (see \cite{Caldeira:1982iu,kamenev2011field}), that the classical motion of the system, in particular its equilibrium position, is independent of the bath temperature. It is the equilibrium {\it two-point function} of the system which gives the bath temperature. This turns out to be true in our work as well. Both in Models (a) and (b), although the classical motion of the Liouville field, given by $\phi_c(t)$, does not see the bath temperature, two point function described in Section \ref{sec:fluctuations-2pt}, are sensitive to the bath temperature. In Model (a) the two-point function actually satisfies a version of the Einstein-Smoluchowski relation,  which can be used to identify the equilibrium temperature with the bath temperature. Model (b) is more non-trivial, however. Here the two-point function is not proportional to the bath temperature. This can perhaps be attributed to the fact that the solution investigated in this model is non-perturbative in the coupling (see Figure \ref{g-gprime} and Section \ref{domain}), and is outside of the linear response regime, which is necessary for the naive version of the fluctuation-dissipation theorem.\footnote{We do not rule out some modified FDT relation using which one can still read off the equilibrium temperature from the two-point function; this is an interesting issue which we wish to come back to.\label{ftnt:FDT}}

%%%%%%%%%%%%%%%%%%%%
\subsection{Some comments on black hole entropy}\label{sec:purity}
%\doubt Our model as a dynamical version of PSSY model \cite{Penington:2019kki}??? If YES, then entanglement entropy would show phase transition and island should appear (follows from their calc.) \\
%\doubt Calculation of purity -- existence of 2nd saddle and its relation to islands??? 

We might wonder what the  dynamical solutions computed in this paper imply for the time evolution of entanglement entropy (EE). This question  can be addressed by considering the so-called purity of the time-evolved state. Let us start from the factorized state 
\begin{equation}\label{t0state-norm}
	\ket{\mathbf\chi_0}=\ket{\widetilde{B_s(l)}} \otimes \ket{\widetilde{\Psi_0(L)}}
\end{equation}
which is the normalized version of \eq{eq:t0state}.\footnote{Note that the norm of $\ket{B_s(l)}$ is
	$\sqrt{Z_{SYK}(2l)}$. Similarly the norm of $\ket{\Psi_0(L)}$ is
	$\sqrt{Z_{bath}(4L)}$.}
The Hamiltonian of the full system is of the form
\begin{equation}
	H= H_{SYK} + H_{bath} + g H_{int}
\end{equation}
which can be read off from the total action \eq{main-model}.  The time-evolved state, at time $T$, is given by
\begin{equation}\label{t0state-time}
	\ket{\mathbf\chi_T}=\exp[-i H T] \ket{\mathbf\chi_0}	.
\end{equation}
Whether the SYK system is entangled with the bath in this state can be found by defining the reduced density matrix\footnote{One could alternatively define a reduced density matrix for the bath $\rho_{bath}(t)$, as in \cite{Penington:2019kki}, which gives an equivalent measure of entanglement (e.g. the same Renyi entropies) as long as the total state of the system and bath is a pure state.}
\begin{equation}\label{RDM-norm}
	\rho_{SYK}(T)= {\rm Tr}_{bath} \ket{\mathbf\chi_T} \langle{\mathbf\chi_T}| ,
\end{equation}
which is the same as \eq{RDM} except that \eq{RDM-norm} is correctly normalized with unit trace. As is well known, if the final state of the full system, \eq{t0state-time}, is factorizable between the SYK system and the bath, then \eq{RDM-norm} is a pure state density matrix (with only one nonzero eigenvalue $\lambda_1=1$ and the rest are zero), which implies that  the purity $\gamma \equiv {\rm Tr} (\rho_{SYK}(T)^2)=1$. On the other hand, it is a straightforward exercise in time-dependent perturbation theory to show that, with $H_{int}$ of the form $\mathcal{O}_{SYK} \otimes \mathcal{O}_{bath}$ the factorizable initial state turns non-factorizable. The interaction Hamiltonian causes simultaneous  transition to orthogonal states both in the SYK Hilbert space as well as in the bath Hilbert space. Thus, e.g., to first order in $g$, and for sufficiently short time,  the final state is a linear combination of orthogonal factorized states, which itself is not factorizable.\footnote{One can verify these statements in a simple toy model consisting of two qubits, each with energy levels $\pm E_1, \pm E_2$. Start with a product state e.g. $\ket{++}$ and some interaction Hamiltonian of the form $H_{int}= g \sigma_{X,1} \sigma_{X,2} $. For small $g$, the state immediately becomes mixed, although after a sufficiently long time $T$, depending on the value of $g$, it comes back to a product state and the process repeats in a periodic fashion. \label{ftnt:toy}} One can generalize to any order of perturbation theory to show that  the final state is not factorizable. This is equivalent to saying that $\rho_{SYK}(T)$ must have  multiple non-zero eigenvalues  $\lambda_i <1$, so that the purity is
\begin{align}
	  \gamma \equiv  {\rm Tr} \left(\rho_{Sch}(T)^2\right) <1 .
	\label{purity}
\end{align}
Note that this strict inequality is valid at large $N$ since the transition amplitudes discussed above are caused by $H_{int}$ which is proportional to $N$.
 
The computation of purity corresponds to computing  a path integral along a contour  described in Figure \ref{fig:purity}.
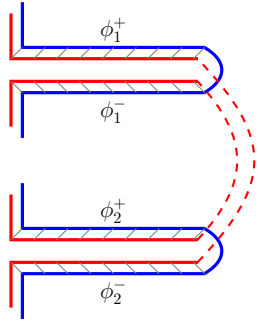
\begin{figure}[]
	\centering
	\begin{tikzpicture}[scale=0.6, every node/.style={transform shape}]
	%\draw[help lines, very thin, gray] (-2,-2) grid (6,6);
	\draw[blue, very thick] (0,1) -- (0,0) -- (4,0);
	\draw[blue, very thick] (0,-2) -- (0,-1) -- (4,-1);
	\draw[blue, very thick] (3.99,0) .. controls (4.5,-0.3) and (4.5,-0.7) .. (3.99,-1);
	\draw[red, very thick]  (-0.25,0.75) -- (-0.25,-0.25) -- (3.85,-0.25);
	\draw[red, very thick]  (-0.25,-1.75) -- (-0.25,-0.75) -- (3.85,-0.75);
	%now the 2nd copy
	\draw[blue, very thick] (0,5) -- (0,4) -- (4,4);
	\draw[blue, very thick] (0,2) -- (0,3) -- (4,3);
	\draw[blue, very thick] (3.99,4) .. controls (4.5,3.7) and (4.5,3.3) .. (3.99,3);
	\draw[red, very thick]  (-0.25,4.75) -- (-0.25,3.75) -- (3.85,3.75);
	\draw[red, very thick]  (-0.25,2.25) -- (-0.25,3.25) -- (3.85,3.25);
	%joining the contours
	\draw[red, thick, dashed] (3.84,-0.25).. controls (5,1) and (5,2) .. (3.84,3.25);
	\draw[red, thick, dashed] (3.84,-0.75).. controls (5.5,1) and (5.5,2) .. (3.84,3.75);
	%labelling the fields
	\node[above, font=\Large] at (2,0) {$\phi_2^+$};
	\node[below, font=\Large] at (2,-1) {$\phi_2^-$};
	\node[above, font=\Large] at (2,4) {$\phi_1^+$};
	\node[below, font=\Large] at (2,3) {$\phi_1^-$};
	%interactions show in gray
	\draw[gray] (-0.25,-0.25) -- (0,0);
	\draw[gray] (0.25,-0.25) -- (0.5,0);
	\draw[gray] (0.75,-0.25) -- (1,0);
	\draw[gray] (1.25,-0.25) -- (1.5,0);
	\draw[gray] (1.75,-0.25) -- (2,0);
	\draw[gray] (2.25,-0.25) -- (2.5,0);
	\draw[gray] (2.75,-0.25) -- (3,0);
	\draw[gray] (3.25,-0.25) -- (3.5,0);
	\draw[gray] (3.75,-0.25) -- (4,0);
	\draw[gray] (-0.25,-0.75) -- (0,-1);
	\draw[gray] (0.25,-0.75) -- (0.5,-1);
	\draw[gray] (0.75,-0.75) -- (1,-1);
	\draw[gray] (1.25,-0.75) -- (1.5,-1);
	\draw[gray] (1.75,-0.75) -- (2,-1);
	\draw[gray] (2.25,-0.75) -- (2.5,-1);
	\draw[gray] (2.75,-0.75) -- (3,-1);
	\draw[gray] (3.25,-0.75) -- (3.5,-1);
	\draw[gray] (3.75,-0.75) -- (4,-1);
	\draw[gray] (-0.25,3.75) -- (0,4);
	\draw[gray] (0.25,3.75) -- (0.5,4);
	\draw[gray] (0.75,3.75) -- (1,4);
	\draw[gray] (1.25,3.75) -- (1.5,4);
	\draw[gray] (1.75,3.75) -- (2,4);
	\draw[gray] (2.25,3.75) -- (2.5,4);
	\draw[gray] (2.75,3.75) -- (3,4);
	\draw[gray] (3.25,3.75) -- (3.5,4);
	\draw[gray] (3.75,3.75) -- (4,4);
	\draw[gray] (-0.25,3.25) -- (0,3);
	\draw[gray] (0.25,3.25) -- (0.5,3);
	\draw[gray] (0.75,3.25) -- (1,3);
	\draw[gray] (1.25,3.25) -- (1.5,3);
	\draw[gray] (1.75,3.25) -- (2,3);
	\draw[gray] (2.25,3.25) -- (2.5,3);
	\draw[gray] (2.75,3.25) -- (3,3);
	\draw[gray] (3.25,3.25) -- (3.5,3);
	\draw[gray] (3.75,3.25) -- (4,3);
	\end{tikzpicture}
	\caption{The path integral contour to calculate purity \eq{purity}. There are two copies of the Schwinger-Keldysh contour (we omit the space direction), denoted by  the upper and lower parts of the figure, each of which corresponds to SYK+bath system as in Figure \ref{SKc-main}. The blue contours refer to the bath which are joined within each copy, representing the partial trace over the bath Hilbert space. The two replicas are joined by the identification  \eq{identification} shown by the dashed red lines. The fields $\phi_r^\pm$, $r=1,2$, denote the Schwarzian path integral variables over the two replica copies.}   \label{fig:purity}
\end{figure}
%Consider for simplicity the calculation of purity (2nd Renyi entropy) within our model. This would involve two replica copies of the SK contour as shown in figure \ref{fig:purity}. The fields $\phi_r^\pm$ now have an extra replica index $r=1,2$. First one integrates out the bath (indicated by joining the blue contours in fig. \ref{fig:purity}) to get the reduced density matrix $\rho_{SYK}(T)$ of the system at time $T$. The purity is then obtained by
%\begin{equation}\label{2renyi}
	%S_2 = -\log\text{Tr} \left(\rho_{SYK}^2 (T)\right)
%\end{equation}
%The action on this 2-replica contour reads
%\begin{align}\label{action-r2}
%	S=& N \left\{\frac{\alpha_s}{J} \int dt \left[\frac{1}{2} \dot{\phi_1^+}^2 - V(\phi_1^+)\right] - \frac{\alpha_s}{J} \int dt \left[\frac{1}{2} \dot{\phi_1^-}^2 - V(\phi_1^-)\right] -i n_f W[F(\phi_1^+),F(\phi_1^-)] \right\} + \\
%	& N \left\{\frac{\alpha_s}{J} \int dt \left[\frac{1}{2} \dot{\phi_2^+}^2 - V(\phi_2^+)\right] - \frac{\alpha_s}{J} \int dt \left[\frac{1}{2} \dot{\phi_2^-}^2 - V(\phi_2^-)\right] -i n_f W[F(\phi_2^+),F(\phi_2^-)] \right\} \nonumber
%\end{align}
At leading order in $N$, the path integral in the above figure can be computing by solving the saddle-point equations (cf. \eq{phi-eom}, \eq{phi-bar-eom})
\begin{align}\label{saddle-r2}
	&\frac{\alpha_s}{J} \left[\ddot{\phi}_r^+(t)+ V'(\phi_r^+(t)) \right] + i n_f g^2 \left[ F'\left(\phi_r^+(t)\right) \int_0^T dt' \kappa_{++}^S(t,t') F\left(\phi_r^+(t')\right) + 2 F'\left(\phi_r^+(t)\right) \int_0^T dt' \kappa_{+-}(t,t') F\left(\phi_r^-(t')\right) \right] = 0 , \\
	&\frac{\alpha_s}{J} \left[\ddot{\phi}_r^-(t) + V'(\phi_r^-(t))\right] -i n_f g^2 \left[F'\left(\phi_r^-(t)\right) \int_0^T dt' \kappa_{--}^S(t,t') F\left(\phi_r^-(t')\right) + 2F'\left(\phi_r^-(t)\right) \int_0^T dt' F\left(\phi_r^+(t')\right)\kappa_{+-}(t',t) \right] =0,
\end{align}
which are to be solved subject to the identification  
\begin{align}
	\phi_1^+(T)= \phi_2^-(T), \qquad \phi_1^-(T)= \phi_2^+(T) .
	\label{identification}
\end{align}
{\bf Trivial saddle point:} Note that there is a trivial saddle point solution 
\begin{align}
	\phi_1^+(t) = \phi_1^-(t) = \phi_2^+(t) = \phi_2^-(t) ,
	\label{trivial}
\end{align}
all equal to the solution $\phi_c(t)$ which satisfies \eq{full-solution-c}. It is easy to see that this solution  corresponds to the large $N$ path integral for the disconnected diagram $\left(\text{Tr}\, \rho_{SYK}(T)\right)^2 $ which is $1^2=1$,\footnote{Note that besides the twisted identification \eq{identification}, the solution \eq{trivial} also satisfies $\phi_1^+(T)= \phi_1^-(T)$, $\phi_2^+(T)= \phi_2^-(T)$ appropriate to tracing over the system Hilbert space in each replica copy.} Thus, if the above saddle  point were the only one, we would get the purity $\gamma =1$.  

\gap{1}
{\noindent {\bf Nontrivial saddle point:} Since we found in \eq{purity} that the strict inequality $\gamma <1$ must be valid at large $N$, there must be at least one non-trivial saddle point solution besides \eq{trivial}. Such a solution must connect the two replica copies non-trivially, making  use  of the twisted identifications \eq{identification}. In the Keldysh notation}
$\phi_{r,\pm}(t)= \phi_{r,c}(t) \pm \phi_{r,q}(t)$, \eq{identification} reads 
\begin{align}
	\phi_{1,c}(T) + \phi_{1,q}(T) = \phi_{2,c}(T) -  \phi_{2,q}(T), \qquad \phi_{1,c}(T) - \phi_{1,q}(T) = \phi_{2,c}(T) +  \phi_{2,q}(T) .
	\label{identification-cq}
\end{align}
If we choose $\phi_{r,q}=0$, we are immediately forced to the trivial solution \eq{trivial}. For a non-trivial solution, we must have non-zero $\phi_{r,q}$, which cannot be obtained from \eq{full-kamenev} perturbatively in $\phi_{r,q}$; hence it has to be a non-perturbative solution. Contribution from the non-trivial saddle point to the path integral in Fig \ref{fig:purity} will have to be negative, to reduce the purity from 1 which is obtained from the trivial saddle point \eq{trivial}. We hope to come back to this issue in the future.
%It must also become more dominant at late times.

%%%%%%%%%%%%%%%%%%%%
\subsection{The final state\label{sec:final-state}}

%\doubt Gautam : Move model (a) before model (b) \doubt 

\subsubsection{Model (b): the evaporation model}\label{model-b}

We will first consider model (b) which exhibits complete evaporation.

\begin{comment}
	We argued above (see \eq{purity}) that early in  the time development, the purity must decrease from 1, or equivalently, the entanglement entropy between the system and the bath must increase from 0. This argument, based on time-dependent perturbation theory, was valid, however, for short times. We gave a simple toy example in footnote \ref{ftnt:toy} where two systems start with a product state and under an interaction immediately becomes a mixed state, but after a sufficiently large time, it comes back to a product state. Since the model discussed in this paper represents black hole evaporation, one expects the entanglement entropy to follow a Page curve, which implies that the SYK system + bath asymptotically comes back to a product state, of the form $\ket{\rm final_{SYK}} \otimes \ket{\rm final_{bath}}$. If we accept this statement, we can derive the nature of the state of the SYK system  $\ket{\rm final_{SYK}} $, as follows.
\end{comment}

By the final state, we will imply here the RDM (reduced density matrix) \eqref{RDM} as $T\to \infty$. In case we are interested in computing expectation values of low energy SYK operators  ${\mathcal O}$ which are diagonal in the $\phi$-representation, such as $\phi$ itself, the expectation value \eqref{rdmev} involves only diagonal elements of $\rho$, namely $\lan \phi^+| \rho | \phi^-\ran $, with $\phi^+ = \phi^-$. Such diagonal elements are obtained by saddle point solutions $\phi_c$ of \eqref{full-solution-c}, which are described in detail in  Section \ref{sec:results}. By using large $N$ factorization, expectation values of any function $F[\phi]$ are also given by $F[\phi_c]$, e.g. we can find the asymptotic value of $\dot f(t) = \exp[\phi(t)]$ using this argument.

Using the above arguments, the asymptotic form of the Schwarzian mode $f(t)$ for Model (b) is given by \eq{asym} and \eq{sin-cos}: 
\begin{equation}
	\dot f= \exp(\phi), \quad \phi = {\rm constant} + c_1 \sin \Omega t + c_2 \cos \Omega t, \quad  \Omega = \left(\frac{8\alpha_s g^2+n_f t_{max} g^4}{16\alpha_s g'^2}\right)^{1/2} .
	\label{f-asym}
\end{equation}

In the boundary dual, $f(t)$ represents a Diff/SL(2,R) transformation $U_f$ of the ground state of the approximately conformal SYK theory.\footnote{These statements are the two-dimensional counterparts of a Brown-Henneaux diffeomorphism of AdS$_3$ which correspond to a conformal transformation of the vacuum of the dual CFT$_2$. The difference in the lower dimensional case  is that the IR fixed point at $1/J =0$ is singular, hence one works with small but non-zero $1/J$.} Hence $\ket{\rm final_{SYK}}= U_f \ket{0}$.\footnote{Note that the initial SYK state, in a low energy projection, is given by, $\ket{B_s(l)} \approx   U_{f_0}\ket{0}$, where $f_0(t) = \pi/\beta \tanh(\pi t/\beta)$ is the Diff element corresponding to a thermal transformation.} It is possible to write the unitary transformation $U_f$  explicitly, however, it is more useful to characterize the state  $U_f\ket{0}$ in terms of expectation values of various operators. Thus, the fermion two-point function in such a state will be given by
\begin{align}
	G(t_1, t_2)=\lan 0| U_f^\dagger \psi_i(t_1) \psi_i(t_2) U_f |0\ran =
	C \frac{\left( f'(t_1) f'(t_2) \right)^\Delta}{\left(f(t_1)- f(t_2)\right)^{2\Delta}
	}
\label{asym-2pt}
\end{align} 
where the last equality follows by applying the conformal transformation $f(t)$ to
$\lan 0 | \psi_i(t_1) \psi_i(t_2) | 0 \ran= C/(t_1 - t_2)^{2\Delta}$. \eq{asym-2pt} can be explicitly evaluated by using  the expression for $f(t)$ from \eq{f-asym}.

\subsubsection*{The final geometry after evaporation}

As we discussed in Section \ref{sec:horizon}, the geometry in this case does not have a horizon. The explicit geometry \eq{AAdS}, at large times, can be obtained from \eq{f-asym} which gives $\dot f(t)=e^{\phi(t)}$:
\begin{align}
	ds^2 &= \frac{dz^2}{z^2} -\frac{d t^2}{z^2} \left(1 + \frac{z^2}{2} g(t)\right)^2 , \nonumber \\
	g(t) &=\left\{f(t),t\right\} =-\frac{1}{2} ({c_1} \Omega \cos(\Omega t)-{c_2}\Omega \sin(\Omega t))^2 -{c_1}\Omega^2 \sin(\Omega t) -{c_2} \Omega^2 \cos(\Omega t) .
	\label{AAdSb}
\end{align}

Note that if  the assumption about the final state being a product state is not true, then the final state of the SYK will be described as an RDM as in the above discussion and the asymptotic form of $f(t)$ should be interpreted as in Model (a) (see below).

\subsubsection{Model (a): incomplete evaporation}\label{model-a}

As described in Section \ref{sec:horizon}, in this model the final geometry has a horizon, whose location (in a fixed gauge) is given by \eqref{analytic-a}. It represents a black hole with a lower mass. 

In the quantum theory, the thermal microstate $\ket{B_s(l)}$, after coupling to the bath, becomes entangled with the bath. Since the final configuration is a black hole, the state is expected to be an entangled state between the bath and the system; for system observables, the state can  be represented as a RDM (reduced density matrix) $\rho_{_{\rm SYK}}(\rm final)$ with nontrivial von Neumann entropy. The final  energy  \eq{energy-final} as well as the asymptotic form of $f(t)=\int_0^t \exp[\phi(t)]$ obtainable from \eq{asym-a} should be interpreted as the expected values of the corresponding operators, in this RDM $\rho_{_{\rm SYK}}(\rm final)$.

The fermion two-point function is given by the conformal transformation from \eq{asym-2pt} where $f(t)$ for Model (a) is to be read off from \eqref{analytic-a}. The final geometry \eq{AAdS} is given by 
\begin{align}
	ds^2 &= \frac{dz^2}{z^2} -\frac{d t^2}{z^2} \left(1- \frac{z^2 a^2}{4} \right)^2 .
	\label{AAdSa}
\end{align}
The horizon is the null surface $z= 2/a$ which translates to $\hat t + \hat z = f(\infty) =0$ (in accordance with Appendix \ref{app:horizon}).

\section*{Acknowledgements}
We would like to thank Soumyadeep Chaudhuri, R Loganayagam, Juan Maldacena, Shiraz Minwalla, Kyriakos Papadodimas, Onkar Parrikar, Suvrat Raju, Subir Sachdev, Ashoke Sen, Ritam Sinha, Nilakash Sorokhaibam, Sandip Trivedi and Neha Wadia for discussions and comments during the course of this work. S.R.W. would like to thank the support of the Infosys Foundation Homi Bhabha Chair at ICTS-TIFR. A.K. and G.M. acknowledge support from the Quantum Space-Time Endowment of the Infosys Science Foundation.

%%%%%%%%%%%%%%%%%%%%%%%%%%%%%%%%%%%%%%%%
\begin{appendices}
	
%%%%%%%%%%%%%%%%%%%%%%%%%%%%%%%%%%%%%%%%
\section{SYK Operators in the IR}\label{app:IRoperators}
In this section, we will compute the expectation value of the operator
\begin{equation}
	\mathcal{O}_m(t) = (-1)^{m+1} J \prod_{j=1}^{m}\left(\f{i}{N} \sum_{k=1}^{N/2} s^{(j)}_k \psi_{2k-1}(t) \psi_{2k}(t)\right),
\end{equation}
in a thermal microstate $\ket{B_s(l)}$.
\begin{equation}
	\frac{1}{Z} \braket{B_s(l)| \mathcal{O}_m(t)|B_s(l)} = (-1)^{m+1} \left(\f{i}{N}\right)^m J \left(\frac{1}{Z} \braket{B_s(l)|\prod_{j=1}^{m} s^{(j)}_k\, \psi_{2k-1}(t) \psi_{2k}(t)|B_s(l)}\right) .
\end{equation}
Here index $k$ is summed over $k=1,...,N/2$.
Let us define $\hat{S}_k = 2i \psi_{2k-1}\psi_{2k}$, such that $\hat{S}_k\ket{B_s} = s_k \ket{B_s}$ where $s_k$ is the $k$th component of the spin vector $s$. It is also clear that $\hat{S}_k^2 = \mathbb{I}$. With this
\begin{align}
	\frac{1}{Z} \braket{B_s(l)|\prod_{j=1}^{m} s^{(j)}_k\, \psi_{2k-1}(t) \psi_{2k}(t)|B_s(l)} &= \frac{1}{Z} \braket{B_s| e^{-lH_0}\prod_{j=1}^{m}\left( \sum_{k=1}^{N/2}s^{(j)}_k\,  \psi_{2k-1}(t)\psi_{2k}(t) e^{-lH_0} \hat{S}_k^{2}\right) |B_s} \nonumber \\
	&= \frac{1}{Z} \braket{B_s(l)| \prod_{j=1}^{m}\left(\sum_{k=1}^{N/2} s^{(j)}_k s_k\, \psi_{2k-1}(t) \psi_{2k}(t) \hat{S}_k(il)\right)|B_s(l)} \nonumber \\
	&= (2i)^m \frac{1}{Z} \braket{B_s(l)| \prod_{j=1}^{m}\left(\sum_{k=1}^{N/2} s^{(j)}_k s_k\, \psi_{2k-1}(t)\psi_{2k}(t) \psi_{2k-1}(il) \psi_{2k}(il)\right)|B_s(l)} \nonumber \\
	&=\prod_{j=1}^{m}\left(2i\sum_{k=1}^{N/2} s^{(j)}_k s_k\right) G_\beta(t-il)^2.
\end{align}
The last expression is obtained at large $N$ after the disorder average. With the above result we get
\begin{align}
	\frac{1}{Z} \braket{B_s(l)| \mathcal{O}_m(t)|B_s(l)} &= -J G_\beta(t-il)^{2m} \prod_{j=1}^{m}\left(\f{2}{N}\sum_{k=1}^{N/2} s^{(j)}_k s_k \right) \nonumber\\
	&= - J \left(\prod_{j=1}^{m} \cos(\theta^{(j)})\right) G_\beta(t-il)^{2m} ,
\end{align}
where we have defined $\cos(\theta^{(j)}) = \f{2}{N}\sum_{k=1}^{N/2} s^{(j)}_k s_k$.

%%%%%%%%%%%%%%%%%%%%%%%%%%%%%%%%%%%%%%%%
\section{Effective Action and Feynman-Vernon phase%(Integrating out bath field)
}\label{app:bath_details}
In this section, we will derive the effective action along the real time piece of the Schwinger-Keldysh contour given in Figure \ref{fig:SK-contour}, starting from \eq{main-model} and integrating out the bath fields. The interaction term in the \eq{main-model} is such that each bath field can be dealt with separately. The effective action for $N_f$ fields can be obtained by multiplying the effective action of one field (obtained by integrating out that one field), by a factor of $N_f$.

Our model is described by the action \eq{main-model} on the contour in the Figure \ref{SKc}. The interaction term, coupling the two theories is non-vanishing only along the real time part of the contour. In the euclidean part of the contour, both the theories evolve with their free Hamiltonians. Let us describe the Euclidean part of the contour given in Figure \ref{SKc}, and we will only show the construction of a ket, the construction of the bra follows similarly. In App \ref{App:B1} we will construct the state at $t=0$ with Euclidean time evolution from $\ket{B_s}\otimes \ket{Bd}$, in App \ref{app:coupling} we will evolve this state with interactions in real time and in App \ref{app:effective_action}, we will construct the density matrix $e^{-iHT}\hat{\rho}_{sch}\otimes \hat{\rho}_{bath}e^{iHT}$ and integrate out the bath fields. In App\ref{UVkernel} we will comment on the UV nature of the effective Feynman-Vernon influence functional and in App \ref{app:thermal} we will discuss the effective Feynman-Vernon influence functional with the bath in a thermal state. 
\begin{figure}[]
	\centering
	\begin{tikzpicture}
	\node[above right] at (3.1,2) {$t = -iL, \ket{Bd}$};
	\node[below right] at (3.1,-2) {$t = iL, \bra{Bd}$};
	\node[above left] at (2.7,1.5) {$t = -il, \ket{B_s}$};
	\node[below left] at (2.7,-1.5) {$t = il, \bra{B_s}$};
	
	\node[above right] at (3,-.05) {$\hat{\rho}_{sch}\otimes \hat{\rho}_{bath}$};
	%\node[below right] at (3,.05) {$\bra{B_s(l)}\otimes\bra{\Psi_0(L)}$};
	
	%\node[above right] at (12,1) {$e^{-iHT}\ket{B_s(l)}\otimes\ket{\Psi_0(L)}$};
	\node[above] at (6,0.7) {+ contour};
	\node[below] at (6,-0.7) {- contour};
	\draw[decoration={brace,raise=5pt},decorate] (13,.7) -- node[right=8pt] {$\hat{\rho}(T)$} (13,-.7);
	\draw [<-](3,-2.5)--(3,2.5);
	\draw [->,blue,thick] (3,2)--(3,2);
	%\draw [very thick,green] (2.8,2.9)--(3.2,2.9);
	\draw [blue,thick](3,2)--(3,0.5);
	\draw [->,blue,thick] (3,-.5)--(3,-1);
	\draw [blue,thick] (3,-1)--(3,-2);
	\draw [->](0,0)--(13,0);
	\draw [->,thick,blue](3,.5)--(6,.5);
	\draw [thick,blue](3,.5)--(12.5,.5);
	\draw [->,thick,blue](12.5,-.5)--(6,-.5);
	\draw [thick,blue](6,-.5)--(3,-.5);
	% \draw [dashed] (3,1)--(3,-1);
	%\draw [dashed] (5,1)--(5,-1);
	%\draw[very thick,green] (5,.3)--(5,.7);
	%\draw[very thick,green] (8,.3)--(8,.7);
	\draw[fill,blue,thick] (3,2) circle [radius=0.08];
	\draw[fill,blue,thick] (3,-2) circle [radius=0.08];
	%\draw[fill,green,thick] (3,2.9) circle [radius=0.08];
	%\draw[fill,green,thick] (5,.5) circle [radius=0.08];
	
	\draw [red](2.8,-1.8)--(2.8,-.7);
	\draw[red] (2.8,-.7)--(12.5,-.7);
	\draw [red](2.8,1.8)--(2.8,.7);
	\draw[red] (2.8,.7)--(12.5,.7);
	%\draw[dashed,red] (12.5,.7) to [out=315,in=45] (12.5,-.7);
	%\draw[green] (12.5,.7) -- (12.5,-.7);
	\draw[fill,red,thick] (2.8,-1.8) circle [radius=0.08];
	\draw[fill,red,thick] (2.8,1.8) circle [radius=0.08];
	\draw[gray,dashed,ultra thick] (3,.5)--(2.8,.7);
	\draw[gray,dashed,ultra thick] (4,.5)--(3.8,.7);
	\draw[gray,dashed,ultra thick] (5,.5)--(4.8,.7);
	\draw[gray,dashed,ultra thick] (6,.5)--(5.8,.7);
	\draw[gray,dashed,ultra thick] (7,.5)--(6.8,.7);
	\draw[gray,dashed,ultra thick] (8,.5)--(7.8,.7);
	\draw[gray,dashed,ultra thick] (9,.5)--(8.8,.7);
	\draw[gray,dashed,ultra thick] (10,.5)--(9.8,.7);
	\draw[gray,dashed,ultra thick] (11,.5)--(10.8,.7);
	\draw[gray,dashed,ultra thick] (12,.5)--(11.8,.7);
	%\draw[gray,dashed,ultra thick] (12.5,.5)--(12.3,.7);
	
	\draw[gray,dashed,ultra thick] (3,-.5)--(2.8,-.7);
	\draw[gray,dashed,ultra thick] (4,-.5)--(3.8,-.7);
	\draw[gray,dashed,ultra thick] (5,-.5)--(4.8,-.7);
	\draw[gray,dashed,ultra thick] (6,-.5)--(5.8,-.7);
	\draw[gray,dashed,ultra thick] (7,-.5)--(6.8,-.7);
	\draw[gray,dashed,ultra thick] (8,-.5)--(7.8,-.7);
	\draw[gray,dashed,ultra thick] (9,-.5)--(8.8,-.7);
	\draw[gray,dashed,ultra thick] (10,-.5)--(9.8,-.7);
	\draw[gray,dashed,ultra thick] (11,-.5)--(10.8,-.7);
	\draw[gray,dashed,ultra thick] (12,-.5)--(11.8,-.7);
	%\draw[gray,dashed,ultra thick] (12.5,-.5)--(12.3,-.7); 
	\node [below left] at (3,0) {$ t =0$}; 
	%\draw [dashed] (8,1)--(8,-1);
	%\node [below right] at (8.1,0) {$t = t$};
	%\node [below right] at (5,0) {$t = t'$};
	\node [below right] at (11.5,0) {$t = T $};
	\draw[blue,dashed] (12.5,.5) to [out=300,in=60] (12.5,-.5);
	%\draw[blue,very thick,dashed] (0,2)--(10,2);
	%\draw[blue,very thick,dashed] (0,-2)--(10,-2);
	%\draw[fill,green,thick] (8,.5) circle [radius=0.08];
	% \draw[decoration={brace,mirror,raise=5pt},decorate] (3,.6) -- node[right=8pt] {$l'$} (3,3);
	%\draw[decoration={brace,raise=5pt},decorate] (3,-.6) -- node[right=8pt] {$l$} (3,-2);
	\end{tikzpicture}
	\caption{This is the same diagram as Figure \ref{fig:SK-contour} with the $x$ direction suppressed. The bath contour is depicted in blue and the SYK path is depicted by red. The blue and red dots on the top depict the kets $\ket{Bd}$ and $\ket{B_s}$ respectively and on the bottom depicts the bras $\bra{Bd}$ and $\bra{B_s}$ respectively. The state at $t=0$ is $\hat{\rho}_{sch}\otimes \hat{\rho}_{bath}$ and the state at $t=T$ is the reduced density matrix defined as $\hat{\rho}(T)=\text{Tr}_{bath}\left[e^{-iHT}\hat{\rho}_{sch}\otimes \hat{\rho}_{bath}e^{iHT}\right]$. The tracing out of the bath fields is depicted by the blue dashed line connected the upper and lower bath contour. The `+' (upper) contour and the `-' (lower) contour are exactly at $Im(t)=0$, the gap in the above figure is only shown for better presentation, the value of real time along both contours is also the same. Hence, to distinguish fields on the upper contour from the fields on the lower contour we add a $\pm$ superscript.}
	\label{SKc}
\end{figure}
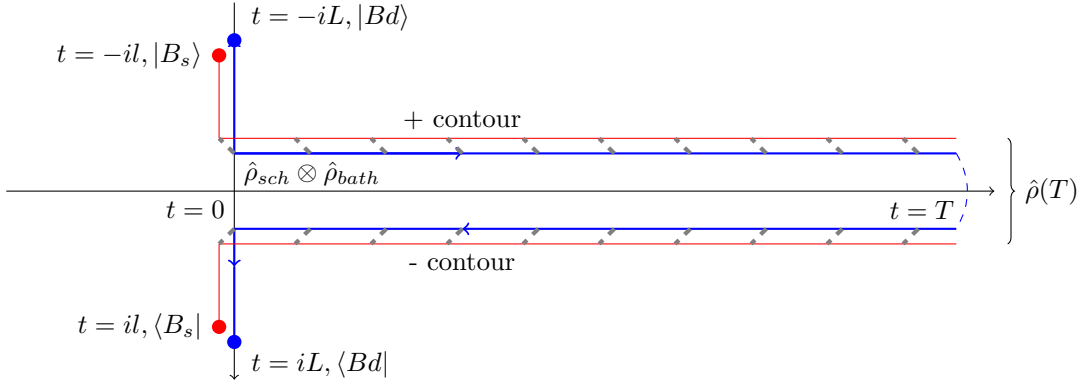

%%%%%%%%%%%%%%%%%%%%
\subsection{Euclidean evolution and the initial bath wavefunctional}\label{App:B1}
As stated in the text, the initial bath state is the CC state \cite{Calabrese:2005in,Calabrese:2007rg} which is obtained by Euclidean evolution of a pure boundary state
\begin{equation}\label{CC}
	|\Psi_0 (L)\rangle = e^{-L H_{bath}} |Bd\rangle , 
\end{equation}
with the boundary state defined by $\Phi |Bd\rangle =0$. We will use the convention that the boundary state $|Bd\rangle$ is at the time $t= i\tau= -iL$. The bath state at time $t=0$ can be formally given by
\begin{equation}
	\ket{\Psi_0(L)} = \int D\Phi\ \Psi_0[\Phi(x)] \ket{\Phi(x)} ,
\end{equation}
with the wavefunctional $\Psi_0[\Phi(x)]$ defined as follows
\begin{equation}
	\Psi_0[\Phi(x)]= \int_{\Phi(\tau=-L,x)=0}^{\Phi(\tau=0,x)=\Phi(x)} D\Phi \, e^{-\int_{-L}^0 d\tau \int_0^\infty dx\  L_{bath}^E [\Phi(\tau,x)]} = \braket{\Phi(x)|\Psi_0(L)}.
\end{equation}
We can evaluate this path-integral by a saddle point approximation (which is exact for a quadratic theory). The bath field lives on a semi-infinite line. Variation of the action leads to the equation of motion $\partial^2 \Phi=0$ with a boundary term $\int d\tau\  \partial_{x}\Phi\ \delta \Phi(\tau,x)|_{x=0}$ at $x=0$, where we impose Neumann boundary condition $\partial_{x}\Phi(\tau,x)|_{x=0}=0$. The expansion of the bath field in momentum modes is
\begin{equation}
	\Phi(\tau,x) =\frac{2}{\pi} \int_{0}^{\infty} dk\,\alpha_k(t)\cos(kx) .
\end{equation}
The equation of motion reads
\begin{equation}
	(\partial_{\tau}^2 + \partial_{x}^2)\Phi(x,\tau) = \frac{2}{\pi} \int_{0}^{\infty} dk\, \cos(kx) \left[\ddot{\alpha}_k(\tau) -k^2 \alpha_k(\tau)\right] =0 ,
\end{equation}
with the solution $\alpha_k(\tau) = a(k) \cosh(k\tau) + b(k) \sinh(k\tau)
$. Imposing the boundary condition $\Phi(-L,x)=0$ gives $b(k) = a(k) \coth(k L)$. The on-shell action simply reduces to
%\begin{equation}
%	\kern-10pt	S=\frac{1}{2}\int d^2x \left(\partial \Phi\right)^2 = \frac{1}{2}\int d^2x\left[\partial\cdot(\Phi \partial \Phi)- \Phi \partial^2 \Phi\right] = \frac{1}{2} \int_{0}^{\infty}dx \left[\Phi(0,x)\dot{\Phi}(0,x)- \Phi(-L,x)\dot{\Phi}(-L,x)\right]
%\end{equation}
%Here we have already used $\del_x \Phi(\tau,x=0)=0$. Noting further the bc $\Phi(-L,x)=0$, we find that $S$ contains only the following term
\begin{equation}
	S= \frac{1}{2} \int_{0}^{\infty}dx\, \Phi(0,x)\dot{\Phi}(0,x) = \frac{1}{\pi}\int_{0}^{\infty} k\,dk \frac{a(k)^2}{\tanh(kL)} .
\end{equation}
%where in going to the second line we have invoked the product to sum formula for cosines and used
%\begin{equation}
%	\int_{0}^{\infty}dx\, \cos(kx) = \frac{1}{2} \int_{-\infty}^{\infty}dx\, \cos(kx) = \frac{1}{2} \text{Re} \int_{-\infty}^{\infty}dx\, e^{i k x} = \pi \delta(k)
%\end{equation}
Putting this all together we finally obtain
\begin{equation}\label{CC-D}
	\Psi_0\left[\Phi(x)\right] = \Psi_0\left[a(k)\right] = \mathcal{N} \exp\left[-\frac{1}{\pi}\int_{0}^{\infty} dk \frac{k\, a(k)^2}{\tanh(kL)}\right]. %\quad \text{where} \quad \Phi(x)= \f2\pi \int_0^\infty dk\, a(k) \cos(kx) .
\end{equation}
The normalization is proportional to the determinant of quadratic fluctuations around this classical solution. For the corresponding density matrix in the thermal state see \eqref{thermal-state}.

%%%%%%%%%%%%%%%%%%%%
\subsection{Real time evolution} \label{app:coupling}

At $t \ge 0$ we turn on interactions between the bath and the SYK model, which are in their respective specified initial states. Consider the Lorentzian evolution till time $T$ with the interactions turned on, 
\begin{equation}
	\ket{\Psi(T)} = e^{-i H T} \left(\ket{B_s(l)} \otimes  \ket{\Psi_0(L)}\right) %, \quad \text{with} \quad \ket{\Psi(T)} 
	= \int D\tilde\Phi \ d\phi_T\  \Psi_T[\tilde\Phi,\phi_T] \ket{\phi_T,\tilde \Phi} ,
\end{equation}
\begin{align}
	\Psi_T[\tilde\Phi,\phi_T] = \int_{\phi_0}^{\phi_{T}} D\phi \, e^{i S_{Sch}[\phi(t)]}\  \rchi[\tilde\Phi(x),\phi(t)] ,
\end{align}
with 
\begin{align}
	\rchi[\tilde\Phi(x),\phi(t)] = \int_{\Phi(x)}^{\tilde\Phi(x)} D\Phi \ \Psi_0\left[a(k)\right]\ \exp\left[i \int_0^T dt\int_0^\infty dx \mathcal{L}_{bath}[\Phi(t,x)] + i\int_{0}^T dt \mathcal{L}_{int} [\Phi(t,x=0),F(\phi(t))]\right] , %\\
	%=& \int_{B^{(l)}_s} D\phi \, e^{i S_{Sch}[\phi]} \Psi_T\left(\{c_k\},F(\phi(t))\right)
\end{align}
where the lower boundary condition is $\Phi(x) = 2\int dk \cos(kx)a(k)/\pi$,  $F(\phi(t))$ is defined in \eqref{F-final}, and $\phi_0 = 2 \log(\pi/\beta J)$ \cite{Kourkoulou:2017zaj}. Similar expression can be obtained for the bra part $\left(\bra{B_s(l)} \otimes \bra{\Psi_0(\Phi)}\right) e^{i H T}$.\\

\subsubsection*{Comment on the coupling}	
From the point of view of the bath fields, $F(\phi)$ simply acts as an external source. As in the Euclidean evolution we can use saddle point method here as well. We will now solve the equations of motion for the field in the presence of non-trivial sources. Here the choice of boundary condition for the bath field at $x=0$ becomes important. To appreciate this, consider a free scalar field on half-line ($x\ge 0$), with a source localized at the boundary at $x=0$
\begin{equation}
	S_{half}=\int dt \int_{x\ge 0} dx \frac{1}{2}\left(\partial \Phi\right)^2 + \int dt J(t) \Phi(t,x=0).
\end{equation}
Its variation leads to
\begin{align}
	\delta S_{half} &= \int_{x\ge 0} d^2x \left[-\partial^2 \Phi \delta \Phi + \partial_{\mu}\left(\partial^{\mu}\Phi \delta \Phi\right) \right] + \int dt J(t) \delta \Phi(x=0) \nonumber \\
	&= \int_{x\ge 0} d^2x \left(-\partial^2\Phi \right) \delta \Phi + \int dt \left[J(t) + \partial_{x} \Phi \right] \delta \Phi(x=0) .
\end{align}
To minimize the action, together with the EOM $\partial^2\Phi=0$, we impose Neumann boundary condition $J(t) + \partial_{x} \Phi(t,x=0) = 0$. Our model \eqref{main-model} is essentially the same with $J(t) = -g F(\phi(t))$, but in our case it is no longer a classical source, it is dynamical. The choice of Neumann boundary condition allows for interesting interacting dynamics between $F(\phi(t))$ and the bath field. If we impose Dirichlet boundary condition $\delta \Phi(t,x)|_{x=0}=0$, then the bath decouples from the Schwarzian theory and can be integrated out trivially. This is essentially driving the dynamics of the system mechanically through the boundary condition. The exchange of energy-momentum at $x=0$ boundary is quantified by the stress-tensor
\begin{equation}
	T_{tx}(t,0) = \partial_t \Phi \partial_x \Phi(t,0) = g \, \dot{\Phi}(t,0) F(t) \qquad \text{(Neumann BC)} .
\end{equation} 
The Neumann BC allows for exchange of energy-momentum which in general is non-zero. This exchange of energy is necessary for the black hole to evaporate. 

%%%%%%%%%%%%
\subsubsection*{Computation of $\rchi[\tilde\Phi(x),\phi(t)]$ via saddle point}
In this subsection, we will simply denote $gF(\phi(t))$ as $F(t)$ to reduce clutter. We will bring back the full notation in the next subsection \ref{app:effective_action}. To accommodate the modified Neumann boundary condition we have the following mode expansion
\begin{equation}
	\Phi(x,t) = \frac{2}{\pi} \int_{0}^{\infty} dk\, \alpha(k,t) \cos(kx) + \frac{2}{\pi} \int_{0}^{\infty} dk\, s(k) F(t) \sin(kx) ,
\end{equation} 
such that
\begin{equation}
	\partial_x \Phi(t,0) = F(t) \frac{2}{\pi} \int_{0}^{\infty} dk\, k\, s(k) = s'(0)\ F(t) = F(t).
\end{equation}
A function which satisfies this is
\begin{equation}
	s(x) = x \exp(-a x^2), \quad a>0.
\end{equation}
We have in mind $a\rightarrow \infty$, such that it quickly decays for $x>0$. For this choice of the shift function
\begin{equation}
	s(k) = \int_{0}^{\infty} dk \sin(kx) s(x) = \sqrt{\frac{\pi}{2}} \frac{k}{a^{3/2}} e^{-k^2 /2a}.
\end{equation}
Now we evaluate
\begin{equation}
	\rchi[\tilde\Phi(x),\phi(t)] = \int_{\Phi(x)}^{\tilde\Phi(x)} D\Phi \ \Psi_0\left[a(k)\right]\ \exp\left[i \int_0^T dt\int_0^\infty dx \mathcal{L}_{bath}[\Phi(t,x)] + i\int_{0}^T dt \mathcal{L}_{int} [\Phi(t,x=0),F(t)]\right] .
\end{equation}
As in the Euclidean part, we evaluate the  on-shell action. Lets deal with each term separately. First, just the bath action 
%These are calculated as follows.
%\begin{align}
%	\dot{\Phi}(x,t) =& \frac{2}{\pi} \int_{0}^{\infty} dk\, \dot{\alpha}_k(t) \cos(kx) + \frac{2}{\pi} \int_{0}^{\infty} dk\, s_k \dot{F}(\phi(t)) \sin(kx) \nonumber \\
%	\int_{0}^{\infty} dx\, \dot{\Phi}^2(x,t) =& \left(\frac{2}{\pi}\right)^2 \int_{0}^{\infty} dk\, dq\, \dot{\alpha}_k \dot{\alpha}_q \int_{0}^{\infty} dx \cos(kx) \cos(qx) \nonumber \\
%	+& \left(\frac{2}{\pi}\right)^2 \int_{0}^{\infty} dk\, dq\, s_k s_q \dot{F}^2 \int_{0}^{\infty} dx \sin(kx) \sin(qx) \nonumber \\
%	+& 2 \left(\frac{2}{\pi}\right)^2 \int_{0}^{\infty} dk\, dq\, \dot{\alpha}_k s_q \dot{F} \int_{0}^{\infty} dx \cos(kx) \sin(qx)
%\end{align}
\begin{equation}
	\int_{0}^{\infty} dx\, \dot{\Phi}^2(x,t) = \frac{2}{\pi}\int_{0}^{\infty} dk \left[\dot{\alpha}_k^2(t) + s_k^2 \dot{F}^2(t) + 2\dot{\alpha}_k(t) \dot{F}(t) \frac{1}{a} I(k/\sqrt{a})\right] ,
\end{equation}
where
\begin{equation}
	\frac{1}{a} I(k/\sqrt{a}) \equiv \frac{2}{\pi} \int_0^\infty dq \frac{q}{q^2-k^2}\,s_q = \frac{1}{a} (1-2 D_F(k/\sqrt{2a})) ; \quad D_F(x)= e^{-x^2} \int_{0}^{x} dy\, e^{y^2} .
\end{equation}
In obtaining this expression, we have used $\int_{0}^{\infty}dx \cos(kx)= \frac{1}{2} \text{Re}\int_{0}^{\infty}dx \exp(ikx)= \pi\delta(k)$ and $\int_{0}^{\infty}dx \sin(kx) =1/k$. There is a similar expression for $	\int_{0}^{\infty} dx\, \Phi'^2(x,t)$.
%\begin{align}
%	\Phi'(x,t) =& -\frac{2}{\pi} \int_{0}^{\infty} dk\, \alpha_k(t) k \sin(kx) + \frac{2}{\pi} \int_{0}^{\infty} dk\, s_k F(\phi(t)) k \cos(kx) \nonumber \\
%	\int_{0}^{\infty} dx\, \Phi'^2(x,t) =& \left(\frac{2}{\pi}\right)^2 \int_{0}^{\infty} dk\, dq\, k\,q\, \alpha_k \alpha_q \int_{0}^{\infty} dx \sin(kx) \sin(qx) \nonumber \\
%	+& \left(\frac{2}{\pi}\right)^2 \int_{0}^{\infty} dk\, dq\, k\,q\, s_k s_q F^2 \int_{0}^{\infty} dx \cos(kx) \cos(qx) \nonumber \\
%	-& 2\left(\frac{2}{\pi}\right)^2 \int_{0}^{\infty} dk\, dq\, k\,q\, \alpha_k s_q F \int_{0}^{\infty} dx \sin(kx) \cos(qx) \nonumber \\
%	=& \frac{2}{\pi}\int_{0}^{\infty} dk\, k^2 \left[\alpha_k^2(t) + s_k^2 F^2(t) + \alpha_k(t) F(t) \frac{1}{a} I(k/\sqrt{a}) \right]
%\end{align}
Combining, we get
\begin{align}
	S_{bath} =& \frac{1}{2} \int_{0}^{T} dt \int_{0}^{\infty} dx \left(\dot{\Phi}^2(t,x) - \Phi'^2(t,x) \right) \nonumber \\
	%=& \frac{1}{2} \int dt \frac{2}{\pi}\int_{0}^{\infty} dk \left[ \left(\dot{\alpha}_k^2(t) - k^2 \alpha_k^2 \right) + s_k^2 \left(\dot{F}^2(t) - k^2 F^2(t)\right) + \frac{2}{a} I(k/\sqrt{a}) \left(\dot{\alpha}_k(t) \dot{F}(t) - k^2 \alpha_k(t) F(t)\right) \right] \\
	=& \frac{1}{2} \int dt \frac{2}{\pi}\, \int_{0}^{\infty} dk \left[ \left(\dot{\gamma}_k^2(t) - k^2 \gamma_k^2 \right) + \left(\dot{F}^2(t) - k^2 F^2(t)\right) \left\{s_k^2 - \left(\frac{1}{a} I(k/\sqrt{a})\right)^2\right\} \right] ,
\end{align}
where we have completed squares in $\alpha_k$ and defined $\gamma_k= \alpha_k + F(t)\frac{1}{a}I(\frac{k}{\sqrt{a}})$. The interaction term is simple
\begin{align}
	S_{int} =& -\int dt\, \Phi(t,0) F(t) = -\int dt\, \frac{2}{\pi}\int dk\, \alpha_k(t) F(t) \nonumber \\
	=& -\int dt\, \frac{2}{\pi}\int dk \left[\gamma_k(t) F(t) - \frac{1}{a} I(k/\sqrt{a}) F^2(t) \right] = -\int dt\, \frac{2}{\pi}\int dk\, \gamma_k(t) F(t).
\end{align}
The last equality follows because $\int_{0}^{\infty} dk I(k/\sqrt{a}) =0$. Together, we have
\begin{equation}
	S_{bath}+S_{int} = \frac{1}{2} \int dt \frac{2}{\pi}\, \int_{0}^{\infty} dk \left[ \left(\dot{\gamma}_k^2(t) - k^2 \gamma_k^2 -2\gamma_k(t) F(t) \right) + \left(\dot{F}^2(t) - k^2 F^2(t)\right) \left\{s_k^2 - \left(\frac{1}{a} I(k/\sqrt{a})\right)^2\right\} \right] .
\end{equation}
The last term vanishes in the $a\rightarrow \infty$ limit. Numerical analysis shows that it vanishes even at finite $a$. Henceforth we will ignore this $\gamma_k$ independent term. The variation leads to the following equation for the modes $\gamma_k$
\begin{equation}
	\ddot{\gamma}_k(t) + k^2 \gamma_k + F(t) =0 ,
\end{equation}
with the solution
\begin{equation}
	\gamma_k(t) = \gamma_k(0) \cos(kt) + \dot{\gamma}_k(0)\frac{\sin(kt)}{k} - \int_{0}^{t} dt' \frac{\sin[k(t-t')]}{k} F(t') .
\end{equation}
This is better expressed in terms of the initial and final values of the modes $\gamma_k(0)=a_k$ and $\gamma_k(T)=c_k$
\begin{align}
	\gamma_k(t) =& a_k A(k,t) + B(k,t) + c_k C(k,t) , \\
	A(k,t) =& \frac{\sin[k(T-t)]}{\sin(kT)}, \quad C(k,t) = \frac{\sin(kt)}{\sin(kT)} , \nonumber \\
	B(k,t) =& C(k,t) \int_{0}^{T} dt' \frac{\sin[k(T-t')]}{k} F(t') -\int_{0}^{t} \frac{\sin[k(t-t')]}{k} F(t') . \nonumber
\end{align}
The on-shell value of the action evaluated on this solution is
\begin{align}
	\left(S_{bath}+S_{int}\right)\bigg\rvert_{on-shell} =& \frac{1}{\pi} \int_{0}^{\infty} dk \left\{\gamma_k(T)\dot{\gamma}_k(T) - \gamma_k(0)\dot{\gamma}_k(0) \right\} - \int dt\,F(t) \frac{1}{\pi} \int_{0}^{\infty} dk \gamma_k(t) \\
	=& \frac{1}{\pi} \int_{0}^{\infty} dk \Bigg[-a_k^2 \dot{A}(k,t) + a_k \left\{c_k \left(\dot{A}(k,t)-\dot{C}(k,t)\right) - \dot{B}(k,t) - \int_{0}^{\infty} dt A(k,t)F(t) \right\} \nonumber \\
	&+ c_k^2 \dot{C}(k,t) + c_k \left\{\dot{B}(k,T)-\int_{0}^{\infty}dt C(k,t)F(t) \right\} -\int_{0}^{\infty}dt B(k,t)F(t) \Bigg] .
\end{align}
Now integrating wrt to the initial bath wavefunctional $\Psi_0[a(k)]$ gives
\begin{align}
	\rchi[c_k,\phi(t)] &= \int \left(\prod_{k} da_k\right) \Psi_0\left[\{a_k\}\right] \int_{\{a_k\}} D\Phi \, \exp\left[i \left(S_{bath}+S_{int}\right)\bigg\rvert_{on-shell} \right] \\
	%&=\mathcal{N} \exp\left[\int_{0}^{\infty} \frac{dk}{\pi} \left\{\frac{j^2(k)}{4\eta(k)} + i c_k^2 \dot{C}(k,t) + i c_k \left(\dot{B}(k,T)-\int_{0}^{\infty}dt C(k,t)F(t) \right) -i \int_{0}^{\infty}dt B(k,t)F(t) \right\} \right] \nonumber \\
	&=\mathcal{N} \exp\left[-\int_{0}^{\infty} \frac{dk}{\pi} \kappa_1(k)\, c_k^2 + \int_0^T dt \int_{0}^{\infty} \frac{dk}{\pi} \kappa_2(k,t) F(t)\, c_k +\int_0^T dt \int_0^T dt' F(t) \kappa_3(t,t') F(t') \right] , \nonumber
\end{align}
where
\begin{align}\label{k123}
	\kappa_1(k) =& k \coth[k(L+ i T)] , \\
	\kappa_2(k,t) =& -2 i \csch[k(L+ i T)] \sinh[k(L+ i t)] , \nonumber \\
	\kappa_3(t,t') =& \frac{i}{4}-\frac{i}{\pi} \int_0^\infty \frac{dk}{k} \sinh\left[k (L+i t)\right] \csch\left[k (L+i T)\right] \sin\left[k (T-t')\right] . \nonumber
\end{align}
\subsection{The Feynman-Vernon influence functional} \label{app:effective_action}
Now that we have the state $\ket{\Psi(T)} = e^{-i H T} \left(\ket{B_s(l)} \otimes  \ket{\Psi_0(L)}\right)$, it is easy to construct the state 
\begin{equation}
	e^{-iHT}\hat{\rho}_{sch}\otimes \hat{\rho}_{bath}e^{iHT} = \underbrace{\ket{\Psi(T)}}_{+ contour}\underbrace{\bra{\Psi(T)}}_{- contour} \, .
\end{equation}
As indicated we differentiate the fields on the upper contour from those on the lower contour by adding a superscript $\pm$ appropriately.
\begin{align}
	\ket{\Psi(T)}\bra{\Psi(T)} = \int Dc_k^+\ Dc_k^-\ d\phi_T^+\  d\phi_T^-\ \left(\Psi_T[c^+_k,\phi^+_T] \Psi_T[c^-_k,\phi^-_T]^*\right)\ (\ket{c^+_k,\phi^+_T}\bra{c^-_k,\phi^-_T}) ,
\end{align}
with
\begin{align}
	\Psi_T[c^+_k,\phi^+_T] \Psi_T[c^-_k,\phi^-_T]^*= \int_{\phi_0}^{\phi^-_{T}} D\phi^-   \int_{\phi_0}^{\phi^+_{T}} D\phi^+ \,e^{i S_{Sch}[\phi^+(t)]}\ e^{-i S_{Sch}[\phi^-(t)]}\  \rchi[c^+_k,\phi^+(t)]\  \rchi[c^-_k,\phi^-(t)]^* .
\end{align}
Obtaining $RDM$ from here is straight forward
\begin{equation}
	\hat{\rho}(T) = \int Dc_k\ d\phi_T^+\  d\phi_T^-\ \left(\Psi_T[c_k,\phi^+_T] \Psi_T[c_k,\phi^-_T]^*\right)\ (\ket{\phi^+_T}\bra{\phi^-_T}).
\end{equation}
Finally we identify the bath field modes on the $+$ and $-$ contours, and integrate.
\begin{equation}
	\int Dc_k\ 
	\Psi_T[c_k,\phi^+_T] \Psi_T[c_k,\phi^-_T]^*= \int_{\phi_0}^{\phi^-_{T}} D\phi^-   \int_{\phi_0}^{\phi^+_{T}} D\phi^+ \,e^{i S_{Sch}[\phi^+(t)]}\ e^{-i S_{Sch}[\phi^-(t)]}\ \int Dc_k\  \rchi[c_k,\phi^+(t)]\  \rchi[c_k,\phi^-(t)]^*.
\end{equation}
Let us focus on
\begin{align}
	&\int Dc_k\  \rchi[c_k,\phi^+(t)]\  \rchi[c_k,\phi^-(t)]^* \nonumber\\
	=& |\mathcal{N}|^2 \int \left(\prod_{k} dc_k\right) \exp \Bigg[-\int_{0}^{\infty} c_k^2 \left(\kappa_1(k) + \kappa_1^*(k) \right) + \int_{0}^{\infty} c_k g \left(\int_{0}^{T} dt \kappa_2(k,t) F(\phi^+(t)) + \int_{0}^{T}dt \kappa_2^*(k,t) F(\phi^-t)\right)\nonumber\\
	&+ g^2\int_{0}^{T}dt\int_{0}^{T}dt' F(\phi^+(t))\kappa_3(t,t')F(\phi^+(t')) + g^2\int_{0}^{T}dt\int_{0}^{T} dt'  F(\phi^-(t)) \kappa_3(t,t') F(\phi^-(t'))\Bigg] .
\end{align}
Similar to the $a_k$ integral, this is also quadratic and is easily performed Let,
\begin{align}
	\int \left(\prod_{k} dc_k\right)& \int Dc_k\  \rchi[c_k,\phi^+(t)]\  \rchi[c_k,\phi^-(t)]^*=\mathcal{N}' \exp(W[F(\phi^+),F(\phi^-)]) \\
	W[F(\phi^+),F(\phi^-)] &= g^2\int_0 ^{\infty} \frac{dk}{4\pi} \frac{\left(\int_0^T dt\, \kappa_2(k,t) F(\phi^+(t)) + \int_0^T dt\, \kappa_2^*(k,t) F(\phi^-(t)) \right)^2}{\kappa_1(k)+\kappa_1^*(k)} \nonumber \\
	&+ g^2\int_0^T dt \int_0^T dt' F(\phi^+(t)) \kappa_3(t,t') F(\phi^+(t')) + g^2\int_0^T dt \int_0^T dt' F(\phi^-(t)) \kappa_3^*(t,t') F(\phi^-(t')),
\end{align}
with following definitions
\begin{align}\label{kernels-pure-1}
	\kappa_{++}(t,t') =& \kappa^{*}_{--}(t,t') = \kappa_3(t,t') + \int_{0}^{\infty} \frac{dk}{4\pi} \frac{\kappa_2(k,t) \kappa_2(k,t')}{\kappa_1(k) + \kappa^{*}_1(k)} , \\
	\kappa_{+-}(t,t') =& \kappa^{*}_{-+}(t,t') = \int_{0}^{\infty} \frac{dk}{4\pi} \frac{\kappa_2(k,t) \kappa_2^{*}(k,t')}{\kappa_1(k) + \kappa^{*}_1(k)} ,
	\label{kernels-pure-2}
%\kappa_1(k) =& k \coth[k(L+ i T)] \nonumber \\
%\kappa_2(k,t) =& -2 i \csch[k(L+ i T)] \sinh[k(L+ i t)] \nonumber \\
%\kappa_3(t,t') =& \frac{i}{4}-\frac{i}{\pi} \int_0^\infty \frac{dk}{k} \sinh\left[k (L+i t)\right] \csch\left[k (L+i T)\right] \sin\left[k (T-t')\right] \nonumber
\end{align}
where $\kappa_1,\kappa_1$ and $\kappa_3$ were already defined in \eqref{k123} above. 
We can rearrange above equation to obtain \eq{FVIF}
\begin{align}
	W[F(\phi^+),F(\phi^-)] = g^2 & \Bigg[\int_0^T dt\, dt' F(\phi^+(t)) \kappa_{++}(t,t') F(\phi^+(t')) + \int_0^T dt\, dt' F(\phi^-(t)) \kappa_{--}(t,t') F(\phi^-(t')) \nonumber \\
&+ 2\int_0^T dt\, dt' F(\phi^+(t)) \kappa_{+-}(t,t') F(\phi^-(t')) \Bigg].
\end{align}
For the corresponding expressions of kernels where the bath is in a thermal state, see Appendix \ref{app:thermal}.\\

Hence the final form of the $RDM$ at time $t=T$ is 
\begin{equation}
	\hat{\rho}(T) = \mathcal{N}' \int d\phi_T^+\  d\phi_T^-\ \left(\int_{\phi_0}^{\phi^-_{T}} D\phi^- \int_{\phi_0}^{\phi^+_{T}} D\phi^+ \,e^{i S_{Sch}[\phi^+(t)]}\, e^{-i S_{Sch}[\phi^-(t)]} \, e^{W[F(\phi^+(t)),F(\phi^-(t))]}\right) \ket{\phi^+_T}\bra{\phi^-_T} .
\end{equation}
			
%%%%%%%%%%%%%%%%%%%%
\subsection{Bath at a finite cutoff $\Lambda$}\label{UVkernel}

At a finite cutoff $\Lambda$ the kernel in last line of \eqref{kappa-relations} is UV finite and reads
\begin{align}
	K(t,t') &= -\frac{2}{\pi L} \int_{0}^{\Lambda} dk \frac{\sin[k t]\sin[k t']}{k^2} -\frac{2L}{\pi} \sum_{n=1}^{\infty} \int_{0}^{\Lambda} dk \left\{\frac{\cos[k(t-t')]}{n^2 \pi^2 + k^2 L^2} -\frac{\cos[k(t+ t')]}{n^2 \pi^2 + k^2 L^2} \right\} \nonumber \\
	& -\frac{2L}{\pi} \sum_{n=1}^{\infty} \int_{0}^{\Lambda} dk \left\{\frac{\cos[k(t-t')]}{\left[(2n-1\pi/2)\right]^2 + k^2 L^2} +\frac{\cos[k(t+ t')]}{\left[(2n-1\pi/2)\right]^2 + k^2 L^2} \right\} \nonumber\\
	& =\frac{2 \sin (\Lambda  t) \sin (\Lambda t')}{\pi \Lambda L} + \frac{1}{L\pi} \left[(t-t') \text{Si}((t-t') \Lambda )+(t+t') \text{Si}((t+t') \Lambda )\right] \nonumber \\
	& + \sum_{n=1}^{\infty} \left[ \kappa_t(t-t') +\kappa_t(t+t') + \kappa_c(t-t') - \kappa_c(t+t') \right] ,
\end{align}
where
\begin{align*}
	\kappa_c(y) =& -\frac{2L}{\pi} \int_{0}^{\Lambda} dk \frac{\cos(k y)}{n^2 \pi^2 + k^2 L^2} \\
	= &\frac{1}{\pi^2 n} \Bigg\{\cosh \left[\frac{\pi n y}{L}\right] \left[-i \text{Ci}\left(\left(\frac{i \pi  n}{L}+\Lambda \right) y \right)+i \text{Ci}\left(\left(\Lambda -\frac{i n\pi}{L}\right) y \right) -\pi \right] \\ & +\sinh \left[\frac{\pi n y}{L}\right] \left[\text{Si}\left(\left(\frac{i \pi n}{L}+\Lambda \right) y \right)+\text{Si}\left(\left(\Lambda -\frac{i n\pi }{L}\right) y\right)\right] \Bigg\} ,
\end{align*}
and
\begin{align*}
	\kappa_t(y) =& -\frac{2L}{\pi} \int_{0}^{\Lambda} dk \frac{\cos(k y)}{\left[(2n-1\pi/2)\right]^2 + k^2 L^2} \\
	=& \frac{1}{\pi^2(2 n-1)} \Bigg\{ 2 \sinh \left[\frac{\pi (2 n-1)  y}{2 L}\right] \Big[-i \text{Shi}\left(\frac{(2 n-1) \pi y}{2 L}\right)-i \text{Shi}\left(\frac{(1-2 n) \pi y}{2 L}\right) \\
	& +\text{Si}\left(\Lambda y +\frac{i (2 n-1) \pi y}{2 L}\right) +\text{Si}\left(\Lambda y +\frac{i (1-2 n) \pi y}{2 L}\right) \Big] \\
	& -2 i \cosh \left[\frac{\pi (2 n-1) y}{2 L}\right] \Big[ \text{Ci}\left(\Lambda y +\frac{i (2 n-1) \pi y}{2 L}\right) -\text{Ci}\left(\Lambda y +\frac{i (1-2 n) \pi y}{2 L}\right) \\
	&-\text{Ci}\left(\frac{i (2 n-1) \pi y}{2 L}\right)+\text{Ci}\left(\frac{i (1-2 n) \pi y}{2 L}\right) \Big] \Bigg\} .
\end{align*}
Here Si and Ci sine-integral and cosine-integral functions respectively. We have also used
\begin{align*}
	-\frac{2}{\pi L} \int_{0}^{\Lambda} dk \frac{\sin[k t]\sin[k t']}{k^2} = \frac{2 \sin (\Lambda  t) \sin (\Lambda t')}{\pi \Lambda L} + \frac{1}{L\pi} \left[(t-t') \text{Si}((t-t') \Lambda )+(t+t') \text{Si}((t+t') \Lambda )\right] .
\end{align*}			

%%%%%%%%%%%%%%%%%%%%
\subsection{Thermal bath}\label{app:thermal}
For the bath in a thermal state, the initial density matrix (at $t=0$) reads
\begin{equation}\label{thermal-state}
	\rho_0[a_k, \bar a_k] = \mathcal{N} \exp\left[-\frac{1}{\pi}\int_{0}^{\infty} dk\,k \frac{(a_k^2+ \bar a_k^2)\cosh(k\beta) -2a_k \bar a_k}{\sinh(kL)}\right] .
\end{equation}
This is the analogue of the pure state \eqref{CC-D}. Following the same procedure as in the pure state, one can integrate out the bath field $\Phi$ to get the effective action. The structure is exactly the same as \eqref{FVIF} in the main text except the kernels are different from \eqref{kernels-pure-1},\eqref{kernels-pure-2}
\begin{align}\label{kernels-thermal}
	\kappa_{++}(t,t') =& \kappa^*_{--}(t,t') =\kappa_{FF}(t,t') + \int_{0}^{\infty} \frac{dk}{4\pi} \frac{\kappa_F(k,t) \kappa_F(k,t')}{\kappa_\beta(k)} , \\
	\kappa_{+-}(t,t') =& \frac{1}{2}\kappa_{F \bar F}(t,t') + \int_{0}^{\infty} \frac{dk}{4\pi} \frac{\kappa_F(k,t) \kappa_{\bar F}(k,t')}{\kappa_\beta(k)} , \nonumber \\
	\kappa_{FF}(t,t') =& \kappa^*_{\bar F \bar F}(t,t') = \frac{i}{4} -\int_0^\infty \frac{dk}{\pi k} \csch(k\beta) \sin\left[k(T-t')\right] \sin\left[k(T-t-i\beta)\right] , \nonumber \\
	\kappa_{F \bar F}(t,t') =& 2\int_0^\infty \frac{dk}{\pi k} \csch(k\beta) \sin\left[k(T-t)\right] \sin\left[k(T-t')\right] , \nonumber \\
	\kappa_F(k,t) =& \kappa_{\bar F}^* =-2 i \cos[k(T-t+i\beta/2)] \sech[k\beta/2] , \nonumber \\
	\kappa_\beta(k) =& 2k \tanh[k\beta/2] . \nonumber
\end{align}
		
%%%%%%%%%%%%%%%%%%%%%%%%%%%%%%%%%%%%%%%%
\section{Initial conditions \eq{init-phi-cl}\label{app:init-cond}}

%\iffalse{
As explained in \cite{Kourkoulou:2017zaj}, for operators which are ``flip-symmetric'', large $N$ expectation values in a pure state $|{B_s(l)}\ran$ become thermal expectation values corresponding to an inverse temperature $\beta=2l$. The $Z_2$ symmetry of the original Euclidean configuration, $\tau \leftrightarrow - \tau$, is retained ($-l$ and $l$ being identified). The consequence of this is as follows. Consider the Euclidean equation of motion on the thermal circle (with the marked point $\tau=0$)
%\comment{
\begin{align}
	\ddot{\phi}(\tau) - V'(\phi(\tau)) =0 .
	\label{ph-cl-eqn-b}
\end{align}
%}\fi
The solution is
\[ \phi =  \log\left( \fr{e_1}{4 J^2} \sec^2(\fr{\sqrt{e_1}}2 ( \tau + e_2)) \right) . \]
Imposing
\[ \phi(\tau)= \phi(- \tau), \ \phi(\tau)= \phi(\tau+\b) , \]
%}\fi
gives the initial condition; we get $e_1= \fr{4\pi^2}{\b^2}$, $e_2=0$, leading to
%}\fi
\begin{align}
	\phi(\tau) = \log\left(\fr{\pi^2}{\beta^2 J^2}\sec^2(\fr{\pi}{\beta}\tau)\right) .
	\label{euclid}
\end{align}
This gives
\begin{align}
	\phi(0)= \log\left(\fr{\pi^2}{\beta^2 J^2}\right) = \phi_0 .
	\label{kour-mal-init}
\end{align}
Let us now include the Lorentzian parts of the Schwinger-Keldysh contour. We imagine $\tau=0$ to coincide with $t=0$. Then \eq{kour-mal-init} gives the first of the conditions in \eq{init-phi-cl}. %(see fig \ref{fig:Z2}) 
%\begin{figure}[H]
%\centering
%\includegraphics[width=.6\linewidth, height=.5\linewidth]{Z2.pdf}
%\caption{SK for KM state}
%\label{fig:Z2}
%\end{figure}
\cite{Kourkoulou:2017zaj} assumes that the initial time slice $t=0$ is a moment of time reversal symmetry, which implies
\[ \dot\phi(0)=0 . \]
In the full SK contour, this implies $\phi_\pm(0)=\phi_0$ and $\dot\phi_\pm(0)=0$. 

\section{Other bath couplings}\label{app:other-bath-couplings}

%%%%%%%%%%%%%%%%%%%%
\subsection{Interaction with just the marginal operator ($\Delta=1$)} \label{sec:marginal}
Here we turn on only the marginal interaction ($O_{_{m=2}}$ in the language of \eq{om}), i.e. in \eq{main-model} we specialize to
\begin{align}
	g F(\phi(t)) &= g' J \exp[\phi(t)], \;\; g'= -\f{\hat g_2}{4\pi} \cos^2\theta 
\end{align}

%%%%%%%%%%
\subsubsection{Numerical solution}
After setting $gF(\phi)=g' J e^{\phi}$, the 3rd order equation \eqref{third-order-general} now reads (after setting $J=1$, which can be reinserted simply by replacing $\beta \rightarrow \beta J, t\rightarrow t J$, etc.)
\begin{equation}\label{third-order-marginal}% with J=1
	\dddot{\phi} - \ddot\phi \dot\phi -\tilde g'^2 e^{2\phi} =0, \qquad \tilde{g}'^2 \equiv n_f g'^2/\alpha_s
\end{equation}
The solution $\phi$ and the (Schwarzian) energy are shown in figures \ref{fig:phi-marg} and \ref{fig:energy-marg}. Here the coupling $\tilde{g}'^2$ must be of the same order as $\beta J \gg 1$ for any significant evaporation to happen. This is evident from the plot of energy in Figure \ref{fig:energy-marg}.

Here also the solution shows a runaway behavior if the coupling is made very large. More precisely this happens at $\tilde g_{\rm crit}' \approx 0.86\,\sqrt{\beta J}\, \gg 1$. We will disregard these pathological solutions when the coupling $g'$ exceeds the critical value $\tilde g_{\rm crit}'$.
\begin{figure}[]
\centering
\begin{subfigure}{.45\textwidth}
	\centering
	\includegraphics[width=\linewidth]{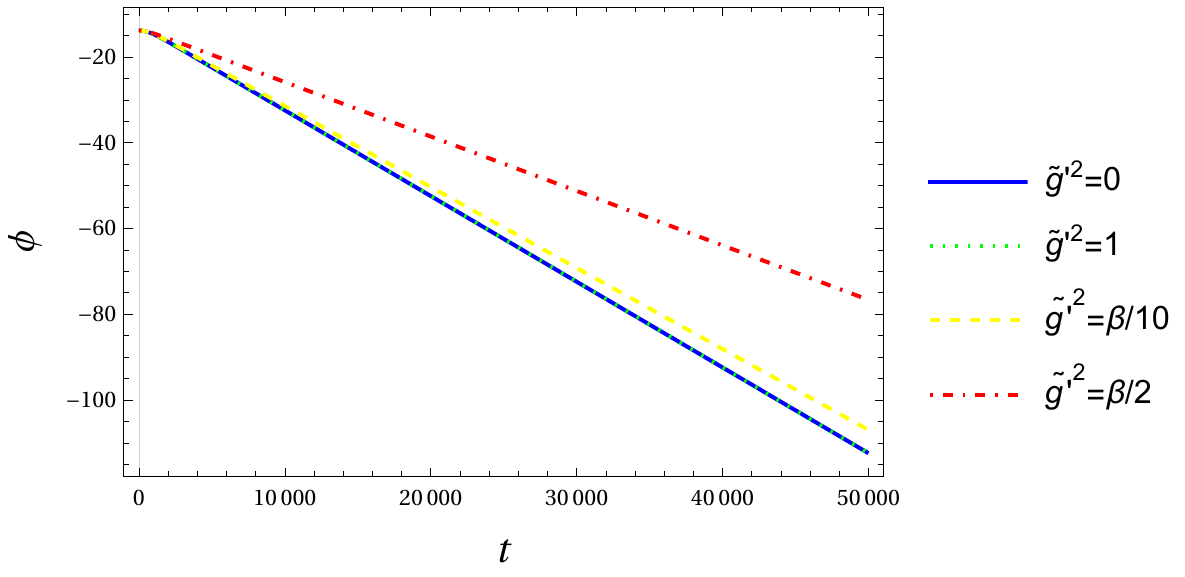}
	\caption{$\phi(t)$}
	\label{fig:phi-marg}
\end{subfigure}\hfill
\begin{subfigure}{.45\textwidth}
	\centering
	\includegraphics[width=\linewidth]{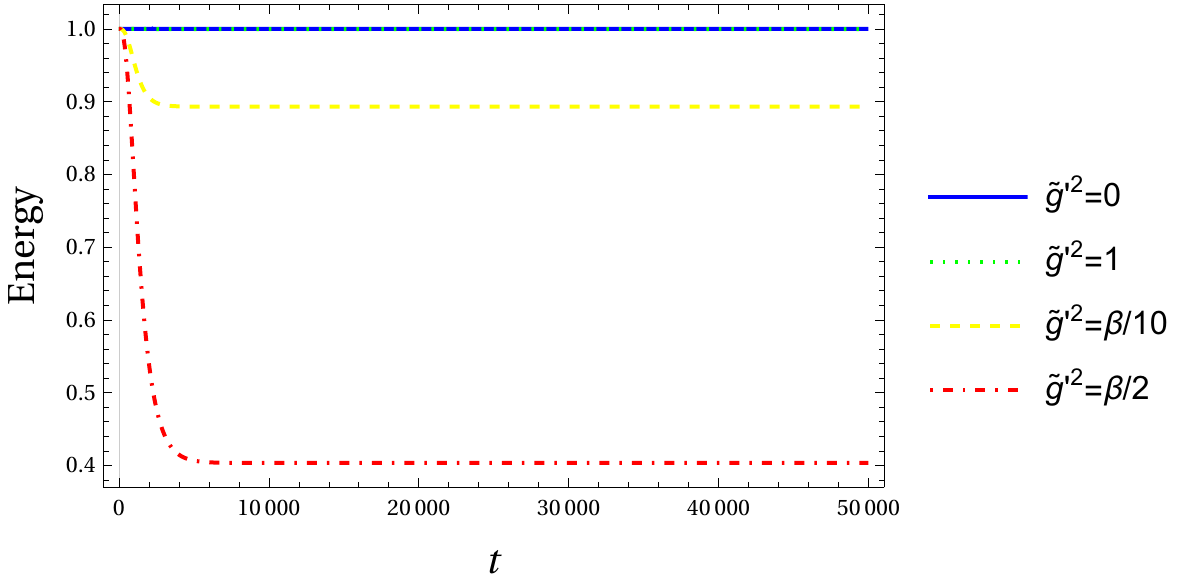}
	\caption{$E_{sch}(t)$}
	\label{fig:energy-marg}
\end{subfigure}
\caption{Plot of $\phi(t)$ and (Schwarzian) energy for various values of the marginal coupling $\tilde g'$. Note that here the coupling $\tilde g'^2$ has to be very large ($\sim \beta J$) for any significant evaporation. Numerics done for $\beta=1000\pi$ with $J=1$.}
\end{figure}

%%%%%%%%%%
\subsubsection{Analytic solution}
After setting $g F(\phi)=g' Je^{\phi}$, the equation of motion \eq{ph-cl-eqn-a} becomes (after setting $J=1$)
\begin{equation}
	\left(\ddot{\phi}(t) + 2 e^{\phi(t)}\right) -{\tilde g}'^2 e^{\phi(t)}\int_0^t e^{\phi(t')} dt' =0, \qquad \tilde g'^2= n_f g'^2/\alpha_s
	\label{ph-cl-eqn-a-one}
\end{equation}
We can solve this equation perturbatively in $\tilde g'^2$. We find
\begin{equation}
	\phi(t) = \log\left(\fr{\pi^2}{\beta^2} \sech^2 \fr{\pi t}{\beta}\right) - \frac{\pi \tilde g'^2}{3 \beta ^2} \left[\beta  \log 2-\beta  \log \left(e^{\frac{2 \pi  t}{\beta}}+1\right)+\pi  t\right] \tanh \left(\frac{\pi t}{\beta}\right) + O(\tilde g'^4)
\end{equation}
%\begin{align}
%  \phi(t)= 6\beta~  ln \left(\frac{2 \pi}\beta \right)-\pi  {\tilde g}^2 \ln~2\, +t \left(\frac{\pi^2 {\tilde g}^2}{\beta  }-6 \pi \right) + O(e^{-2\pi t/\beta})
%\end{align}
The long time solution is of the form
\begin{equation}
	\phi(t) = \left(\frac{\pi^2}{3\beta^2}\tilde{g}'^2 -\frac{2\pi}{\beta}\right)t -2\left[\log\fr{\beta}{2\pi} + \fr\pi{6\beta} \tilde g'^2 \log 2 \right] + O\left(\exp[-\pi t/\beta]\right) % \label{asym-phi}
\end{equation}
Clearly the dominant term is linear
\begin{equation}\label{asym-phi-2}
	\phi=-a t +\ldots, \qquad a= \left(\frac{2\pi}{\beta}-\frac{\pi^2}{3\beta^2} \tilde{g}'^2 \right).
\end{equation}
$a>0$ gives a sensible solution for the original variable $\dot f= e^{\phi}$; this happens in the range $\tilde g' < \tilde g'^A_{\rm crit} \equiv \sqrt{6\beta J/\pi}$. In this range $\dot f \to 0$, which signals an asymptotic solution which is a black hole, with asymptotic energy $E_{sch}=E_\Delta = N\frac{\alpha_{s}}{J} \frac{a^2}{2}$.

%At the critical value, the solution for $\phi$ becomes asymptotically constant, and the energy becomes zero, which signals complete evaporation (zero temperature). Since this critical value is very large in the IR regime $\b J \gg 1$, such an asymptotic solution is not accessible in perturbation theory.

This is what we also find in the numerical solutions except the critical value obtained numerically $\tilde g_{\rm crit}' \approx 0.86\,\sqrt{\beta J}$ is smaller than the analytical value $\tilde g'^A_{\rm crit} = \sqrt{6\beta J/\pi} \approx 1.38 \sqrt{\beta J}$.

%%%%%%%%%%%%%%%%%%%%
\subsection{Relevant Interaction + KM term}\label{sec:relevant+KM}

%As we saw in the previous subsection that depending on the coupling the final black hole mass could be made very small, but it was not possible to achieve complete evaporation. In this subsection and the next we overcome this challenge.

Here to our model \eqref{main-model}, we add the Kourkoulou-Maldacena term with a fixed (non-dynamical) coupling
\begin{equation}
	S_{KM} = \frac{N\alpha_s}{J} \int dt\, \hat{\epsilon} J^2 e^{\phi/2}; \qquad \hat{\epsilon} < \hat\epsilon_{cr}=\frac{2\pi}{\beta J}
\end{equation}
It was shown in [KM] that for $\hat{\epsilon} > \hat\epsilon_{cr}$, the new geometry has no horizon. Therefore the condition $\hat{\epsilon} < \hat\epsilon_{cr}$ is imposed such that there is a black hole in absence of the bath. The (modified) Schwarzian energy now includes the KM potential term
\begin{equation}\label{energy-KM}
	E= N\frac{\alpha_{s}}{J}\left[ \frac{1}{2}\dot{\phi}^2 + 2 J^2 e^{\phi} -\frac{\hat{\epsilon}}{2} J2 e^{\phi/2} \right]
\end{equation}
The equation of motion is modified to
\begin{equation}
	\ddot{\phi}(t) + 2 e^{\phi(t)} -\frac{\hat{\epsilon}}{2} e^{\phi(t)/2} -{\tilde g}^2 e^{\phi(t)/2}\int_0^t e^{\phi(t')/2} dt' =0, \quad {\tilde g}^2\equiv \fr{n_f g^2}{2\alpha_s}
	\label{including-KM-term}
\end{equation}
where we have put $J=1$. Now after coupling to the bath at $t=0$, it is possible for the (modified) Schwarzian energy, given by \eqref{energy-KM}, to become negative for range of the coupling $g$. This however happens for $g\ge g_*$,\footnote{The critical coupling here is sensitive to the KM potential term and differs from \eqref{g-crit-n}.} when the solution shows a runaway behaviour as discussed in Section \ref{sec:relevant}. There is however a window of time when the energy remains negative (see Figure \ref{fig:energy-window}), during which if we decouple the bath we get oscillating and bounded solution for $\phi$. This means the horizon has disappeared and we are in a non-black hole phase. In Figure \ref{fig:switchoff} the numerical solution\footnote{In presence of the KM potential term, the initial condition for $\ddot{\phi}$ is modified to $\ddot{\phi}(0)=-J^2 \left(2e^{\phi(0)} - \frac{\hat\epsilon}{2} e^{\phi(0)/2} \right)$.} is shown where we achieve complete black hole evaporation.
\begin{figure}[]
	\centering
	\begin{subfigure}{.4\textwidth}
		\centering
		\includegraphics[width=\linewidth]{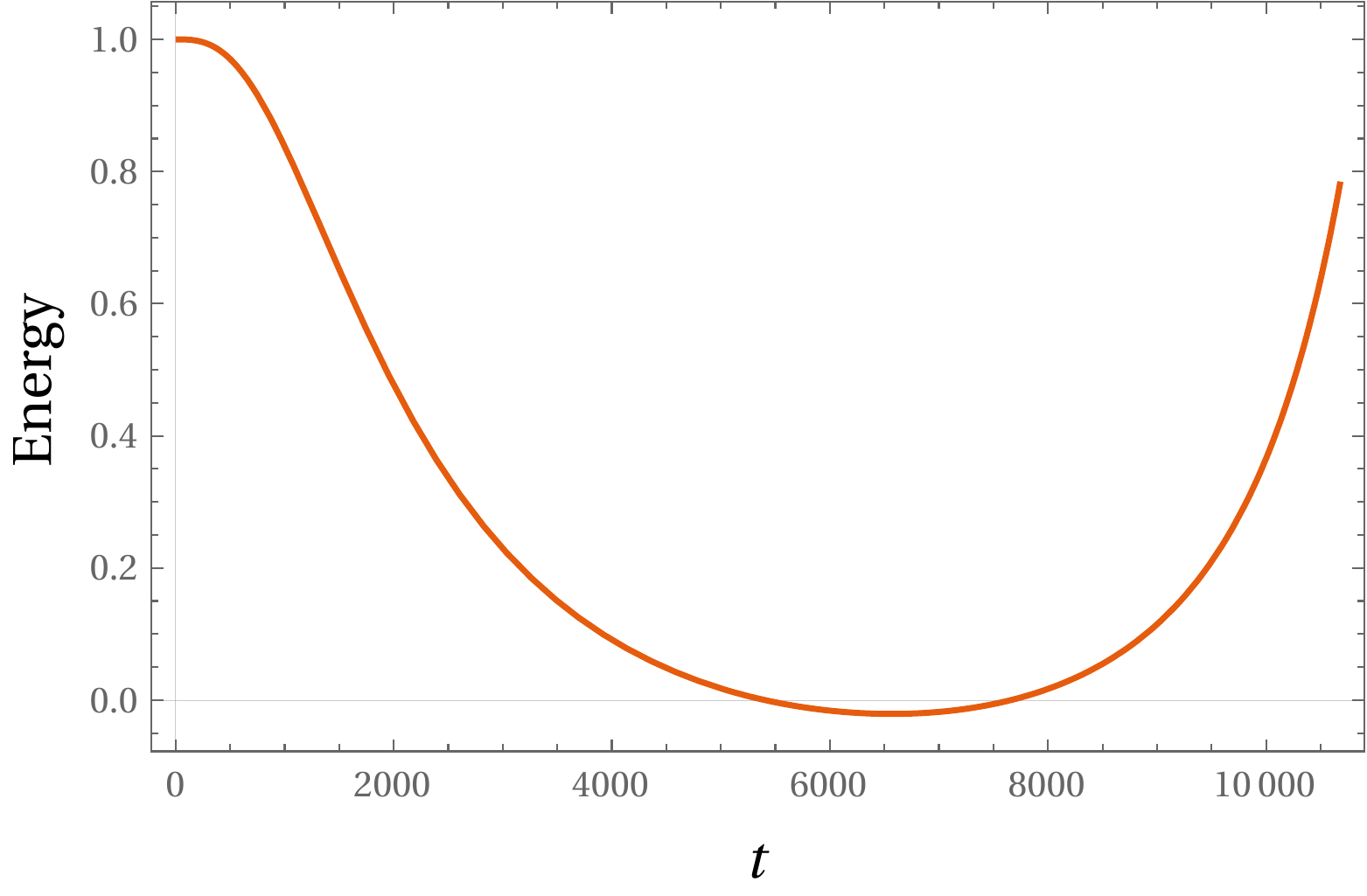}
		%\caption{$\phi(t)$}
	\end{subfigure}\hspace{10ex}
	\begin{subfigure}{.4\textwidth}
		\centering
		\includegraphics[width=1.04\linewidth]{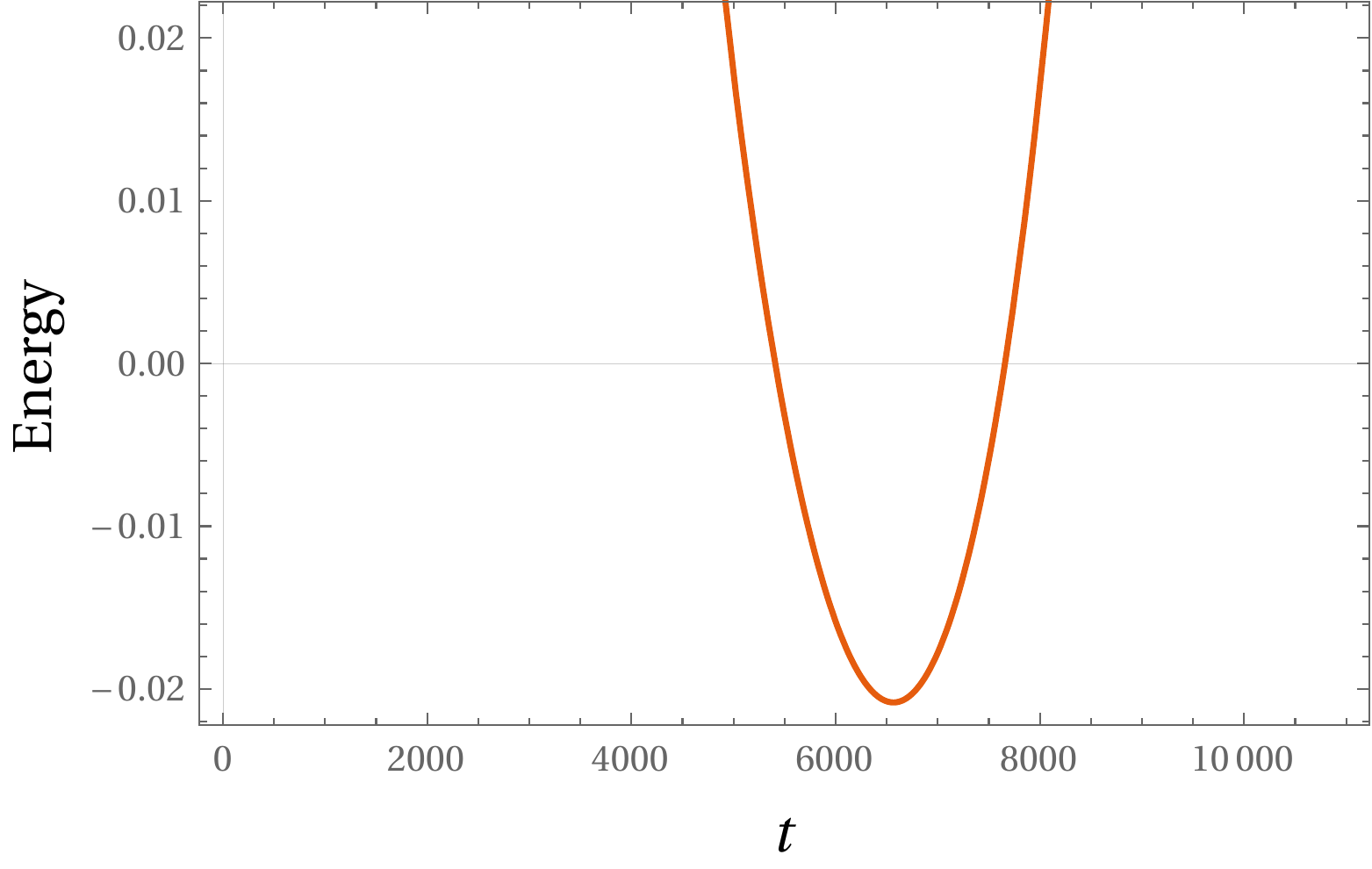}
		%\caption{$E(t)$}
	\end{subfigure}
	\caption{The (modified) Schwarzian energy first decreases and then starts increasing. For a finite time window it is negative allowing us to decouple the bath and go to non black hole geometry. On the right we see the same energy zoomed to make this point clear. Here $J=1, \beta=1000\pi, \tilde{g}^2=0.00075,\hat\epsilon=\frac{1}{4}\hat\epsilon_{cr}$.} \label{fig:energy-window}
\end{figure}
\begin{figure}[]
	\centering
	\begin{subfigure}{.4\textwidth}
		\centering
		\includegraphics[width=\linewidth]{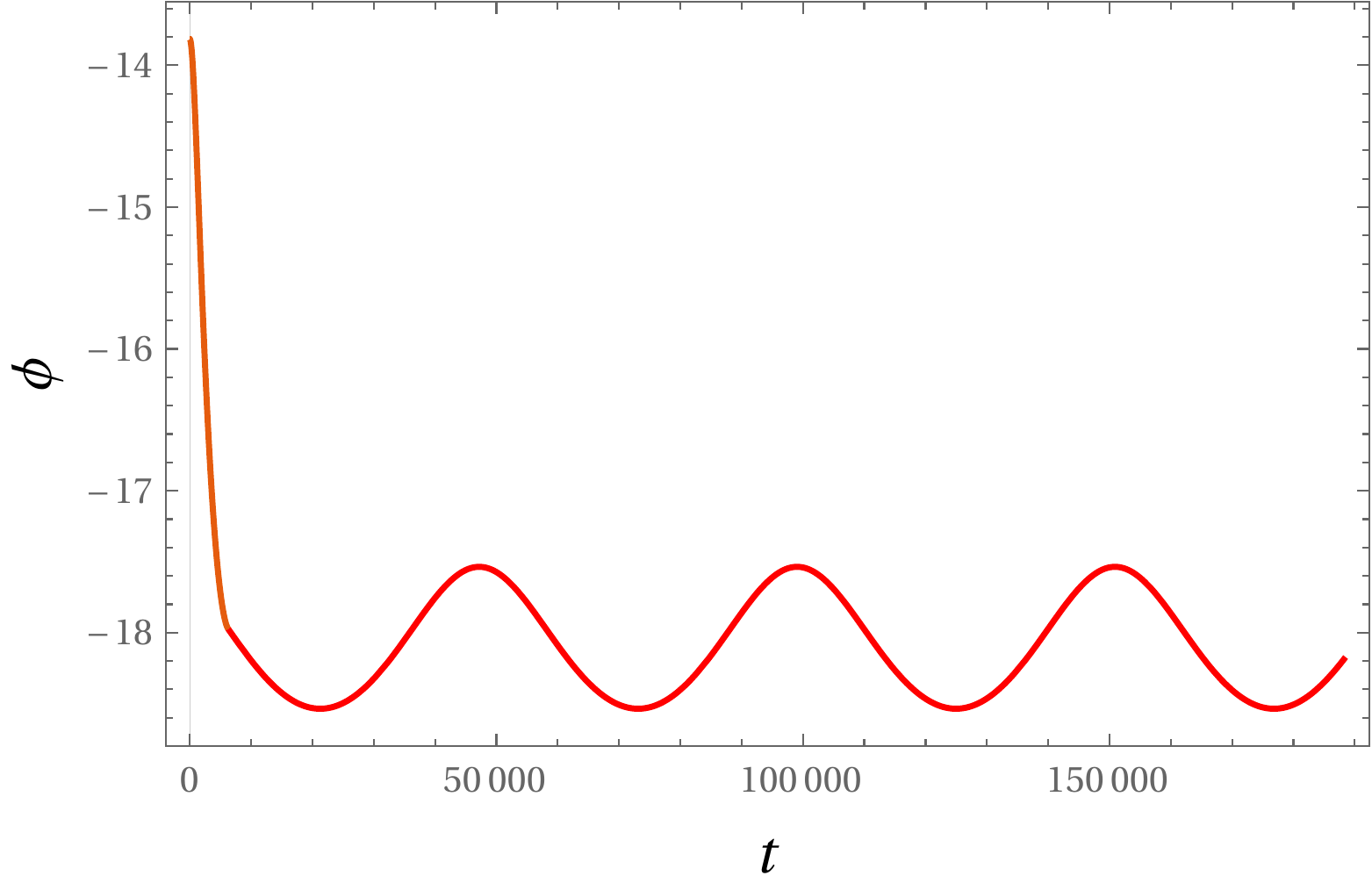}
		\caption{$\phi(t)$}
		%\label{phi-switchoff}
	\end{subfigure}\hspace{10ex}
	\begin{subfigure}{.4\textwidth}
		\centering
		\includegraphics[width=\linewidth]{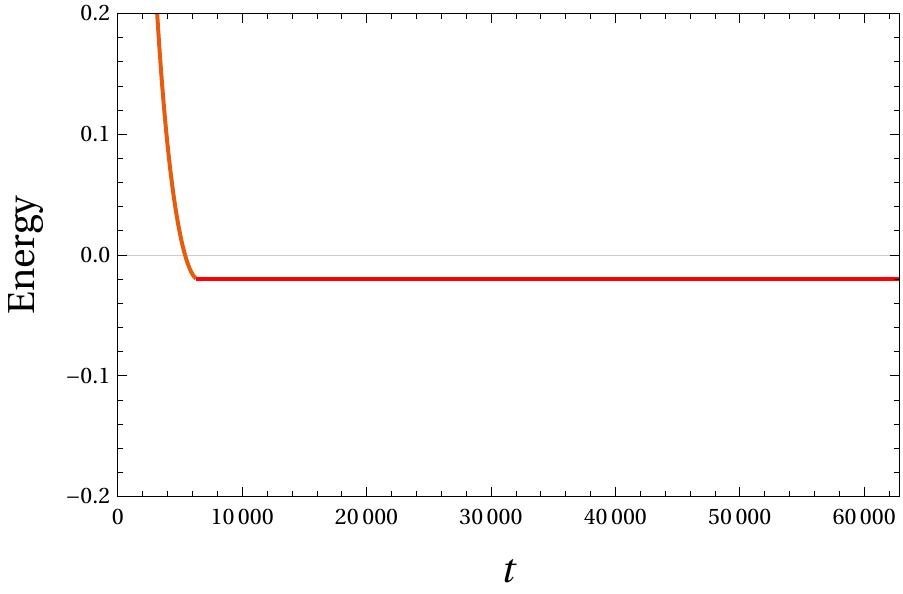}
		\caption{$E(t)$}
		%\label{energy-switchoff}
	\end{subfigure}
	\caption{The coupling to the bath is switched off at $t=2000\pi$. The solution after that (in red) is found by demanding continuity of $\phi$ and $\dot{\phi}$ and is shown on the left. The black hole loses energy initially but after it is decoupled, the energy is constant as seen on the right. Here $J=1, \beta=1000\pi, \tilde{g}^2=0.00075, \hat\epsilon=\frac{1}{4}\hat\epsilon_{cr}$.} \label{fig:switchoff}
\end{figure}
It was crucial for the coupling to be switched off at $O(\beta)$ time as one can see from the plot of the (modified) Schwarzian energy as a function of time.

Note that even though we have added the KM term, without also coupling to the bath it was not possible to go to the non-BH geometry (see Figure \ref{fig:KMnoKM}), since $\hat{\epsilon} < \frac{2\pi}{\beta J}$. It is due to the fact that the BH dumps energy into the bath thereby effectively going to the bound state of the potential.
\begin{figure}[]
	\centering
	\includegraphics[width=.4\linewidth]{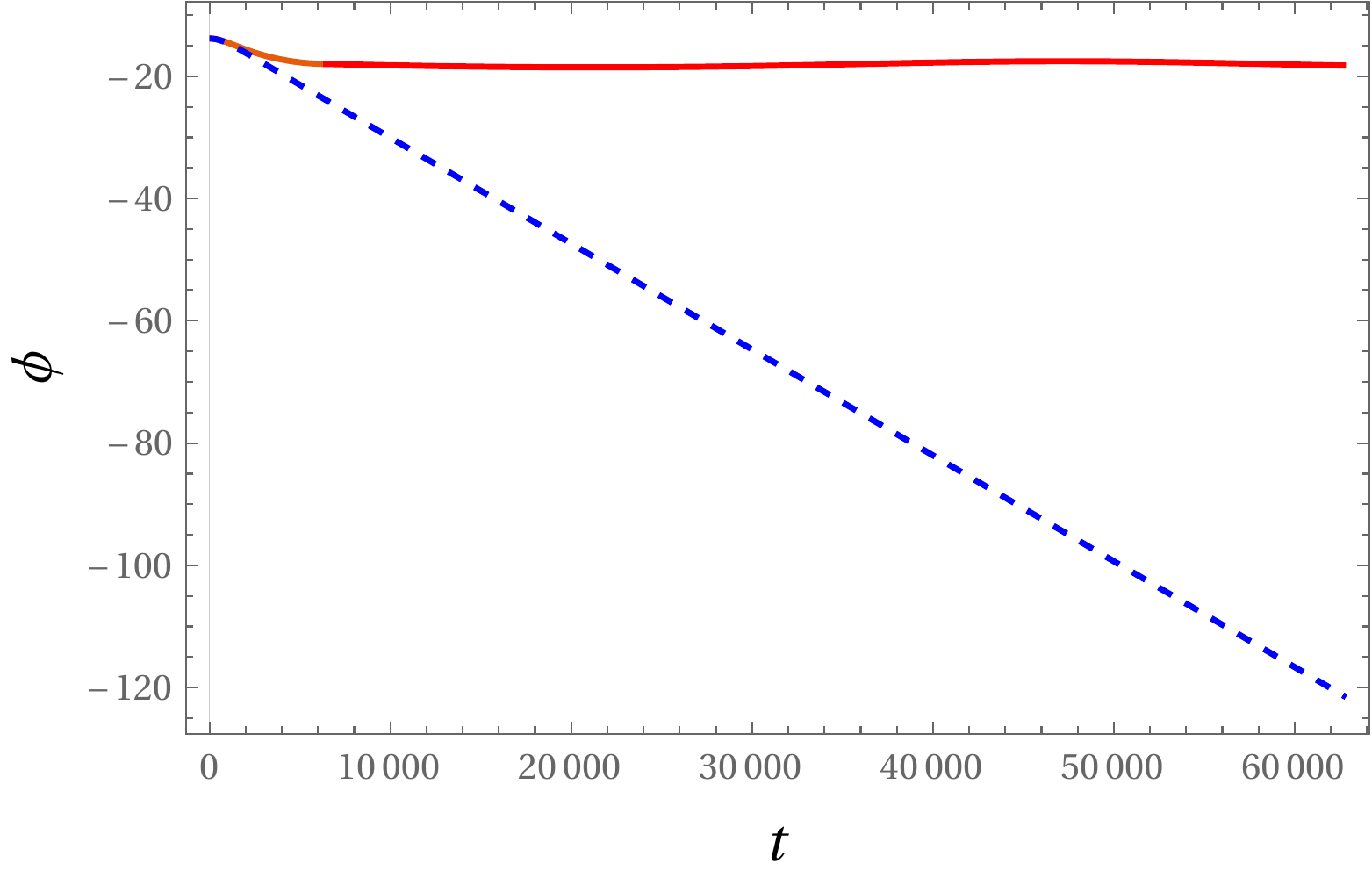}
	\caption{The blue dashed curve shows the effect of the KM term by itself without coupling to the bath. Clearly $\phi(t)\rightarrow -\infty$ as $t\rightarrow \infty$. For comparison the non-black hole solution obtained after coupling to bath is also shown in red. Here $J=1, \beta=1000\pi, \tilde{g}^2=0.00075, \hat\epsilon=\frac{1}{4}\hat\epsilon_{cr}$.} \label{fig:KMnoKM}
\end{figure}

%%%%%%%%%%%%%%%%%%%%%%%%%%%%%%%%%%%%%%%%			
\section{Finding spin of the initial state in Model (a) \label{app:Stirling}}

In this appendix we show that as claimed in Section \ref{sec:recovery-a}, $\binom{N/2}{2}$ spins are enough to determine $s$ up to a sign. Let us call the different trial spins as $s'= s^{(\alpha)}, \alpha=1,2,..., \binom{N/2}{2}$. They satisfy
\begin{align}
	(s^{(\alpha)}\cdot s)^2 = \left(\sum_{k=1}^{N/2} s^{(\alpha)}_k s_k \right)^2 = \left(\frac{N}{2} \cos \theta^{(\alpha)} \right)^2, \quad \alpha=1,2,..., \binom{N/2}{2} \label{model-a-eq}
\end{align}
where the value $(\cos\theta^{(\alpha)})^2$ for each choice of $s^{(\alpha)}$ is `experimentally determined'.

Since the above equations are difficult to work with, we proceed by by linearizing this problem. The above set of equations can be rewritten as
\begin{equation}\label{model-a-eq-linear}
	\sum_{1\le i< j\le N/2}^{N/2} s^{(\alpha)}_i s^{(\alpha)}_j s_i s_j = \frac{N}{4} \left[\frac{N}{2} \left(\cos \theta^{(\alpha)} \right)^2 -1 \right] ,
\end{equation}
where we have used 
\begin{equation}
	(s^{(\alpha)}\cdot s)^2 = \sum_{k=1}^{N/2} \left(s^{(\alpha)}_k\right)^2 s_k^2 + 2\sum_{1\le i< j\le N/2}^{N/2} s^{(\alpha)}_i s^{(\alpha)}_j s_i s_j = \frac{N}{2} + 2\sum_{1\le i< j\le N/2}^{N/2} s^{(\alpha)}_i s^{(\alpha)}_j s_i s_j ,
\end{equation}
and plugged this into the equation \eqref{model-a-eq}.

Now note that the equations \eqref{model-a-eq-linear} represents a system of linear equations in $s_i s_j$ as the variables which are $\binom{N/2}{2}$ in number since $1\le i< j\le N/2$. These can be solved exactly provided the vectors $\left( s^{(\alpha)}_1 s^{(\alpha)}_2, s^{(\alpha)}_1 s^{(\alpha)}_3,..., s^{(\alpha)}_{N/2-1} s^{(\alpha)}_{N/2} \right)$ for $\alpha=1,2,..., \binom{N/2}{2}$, are linearly independent (as $\binom{N/2}{2}$ dimensional vectors). This can always be ensured by choosing $s^{(\alpha)}$'s to be the spins vectors with $2$ minuses and $N/2 - 2$ pluses which are exactly $\binom{N/2}{2}$ in number. We have verified this explicitly in Mathematica up to $N=100$. The fact that a choice of $\binom{N/2}{2}$ spins always exists is not surprising since the total number of all spins grows exponentially large with $N$.

Thus we can always determine the variables $s_i s_j, 1\le i< j\le N/2$. Since this gives us $s_1 s_j$ for $j=1,2,..., N/2$, choosing $s_1=\pm 1$ uniquely determines the spin $s$ up to an overall sign.

%The mathematical problem now boils down to determining  the unknown variables $s_i$ ($N/2$ components of the spin vector $s$) from the $N/2$ non-linear equations \eq{model-a}. It is easy to show that such solutions for $s_i$ exist and are unique up to an overall sign. The overall sign cannot be determined from the analysis of Model (a) since the asymptotic expressions such as \eq{energy-final-a} involve the square of $g$.

\begin{comment}
	The null space of a given spin $s$ is the space of solutions of the equation
\begin{equation}
	s\cdot s' =0
\end{equation}
Without loss of generality, we can take $s=(+,+,...,+)$. Clearly any spin $s'$ with equal number of pluses and minuses (we take $N/2$ to be even) will have vanishing inner product with $s$. The total number of such spins is given by
\begin{equation}
	n_O = \binom{N/2}{N/4}.
\end{equation}
By Stirling's approximation
\begin{align}
	\log n_O &\approx \left[\frac{N}{2}\log \frac{N}{2} -\frac{N}{2} +\frac{1}{2}\log(\pi N) \right] -\frac{1}{2} \left[\frac{N}{4}\log \frac{N}{4} -\frac{N}{4} +\frac{1}{2}\log(\frac{\pi N}{2}) \right] \nonumber \\
	&=\frac{N}{2}\log 2 -\frac{1}{2} \log\frac{\pi N}{4}
\end{align}
therefore the number of such spins is
\begin{equation}
	n_O \approx \frac{2}{\sqrt{\pi N}} 2^{N/2}
\end{equation}
which is a small fraction of all possible spins $2^{N/2}$. The number of spins in the complement is given by
\begin{equation}
	n_C \approx 2^{N/2} \left(1 - \frac{2}{\sqrt{\pi N}} \right)
\end{equation}
\end{comment}

%%%%%%%%%%%%%%%%%%%%%%%%%%%%%%%%%%%%%%%%			
\section{Two-point function in Model (a) \label{app:2pt}}

The integral on the RHS of equation \eqref{2ptF} can be evaluated explicitly. For this we use the Mittag-Leffler pole expansions for $\coth$ and $\csch$ functions in the expression \eqref{2ptF}, perform the integral term by term and then sum them up to get
\begin{align}\label{eq:2pt-full}
	\langle \delta\hat{\phi}_c(t) \delta\hat{\phi}_c\left(t'\right)\rangle &= \frac{16 C}{a^4 L}|t^-| +\frac{8C}{a^4} e^{-\frac{1}{2} a |t^-|} \left [\left(a|t^-|+4\right) \cot(aL) + 2aL\csc^2(a L)\right] \\
	& +\frac{16 C}{\pi^2 a^4} e^{-\frac{\pi |t^-|}{2 L}} 
	\Bigg[a L \left\{\Phi_L \left(e^{-\frac{\pi |t^-|}{2 L}},2,\frac{a L}{\pi }+1\right)-\Phi_{HL} \left(e^{-\frac{\pi |t^-|}{2 L}},2,1-\frac{a L}{\pi }\right)\right\} \nonumber \\
	&+2 \pi \left\{\Phi_{HL} \left(e^{-\frac{\pi |t^-|}{2 L}},1,1-\frac{a L}{\pi }\right)+ \Phi_L \left(e^{-\frac{\pi |t^-|}{2 L}},1,\frac{a L}{\pi }+1\right) \right\} \Bigg] + \frac{64 C}{\pi a^4} \log \left(1-e^{\frac{\pi |t^-|}{2 L}}\right) \nonumber \\
	&-\frac{16 C}{a^4 L} t^+ -\frac{2^7 C}{a^4 L} + \frac{64 C}{\pi^4 a^4} \Bigg[a L e^{-\frac{\pi t^+}{2 L}} \Bigg\{a L \left[\pi \Phi_L \left(-e^{-\frac{\pi t^+}{2 L}},3,\frac{a L}{\pi}+1 \right) 
	+a L \Phi_L \left(-e^{-\frac{\pi t^+}{2 L}},4,\frac{a L}{\pi }+1\right)\right] \nonumber \\
	&+\pi^2 \Phi_L \left(-e^{-\frac{\pi t^+}{2 L}},2,\frac{a L}{\pi }+1\right) \Bigg\} -\pi^3 e^{\frac{1}{2} a (t^+ -2 i L)} B_{-e^{-\frac{\pi t^+}{2 L}}}\left(\frac{a L}{\pi }+1,0\right)-\pi ^3 \log \left(e^{-\frac{\pi t^+}{2 L}}+1\right) \Bigg] .  \nonumber
\end{align}
We have defined $\delta\hat{\phi}_c(t)= \delta\phi_c(t)/F'(\phi(t))$ and $t^\pm = t\pm t'$. The answer is expressed terms of the (Hurwitz)-Lerch transcendent $\Phi_{(H)L}$ and the incomplete beta function $B_z(a,b)$.

\begin{section}{A geometrical appendix}\label{app:geom}
Solutions of JT gravity are completely specified by the Schwarzian mode. The metric can be described in two separate but equivalent ways. 
	
In the first,  the Schwarzian mode shows up as a large diffeomorphism of the AdS$_2$ metric. The metric has  the following form \cite{Mandal:2017thl}\footnote{This form is the 2D counterpart of the Banados metric in 3D \cite{Banados:1998gg}.} in the Fefferman-Graham coordinates:
\begin{equation}
	ds^2 = \frac{dz^2}{z^2} -\frac{dt^2}{z^2} \left(1+\frac{z^2}{2}\left\{f(t), t\right\}\right)^2 .
	\label{AAdS}
\end{equation}
Since in the boundary theory, conformal symmetry is explicitly broken by choosing a large, but finite, value of the dimensionful parameter $J$, the bulk counterpart is to truncate the above geometry by a small, non-zero boundary $z=\delta$ or, more invariantly, by a choice of a boundary value of the dilaton $\Phi= \frac{a}{\delta}$ which is related to $J$.
	
In the second viewpoint \cite{Maldacena:2016upp}, the bulk metric is fixed to Poincar\'e $AdS_2$ 
\begin{equation}\label{PoincareAdS}
	ds^2 = \frac{-d\hat t^2+ d\hat z^2}{\hat z^2} .
\end{equation}
The Schwarzian mode now shows up as a wiggle of the boundary curve which now reads, $(\hat t,\hat z)= (f(t), \delta \dot f(t))$ (see \eq{roberts} for more detail) and the entire dynamics is then contained in the motion of the boundary curve. 
	
The above two viewpoints are related to each other by a coordinate transformation 
%\subsection{Roberts}
\begin{align}
	& \hat t=f(t) +  \frac{2 z^2 f'(t)^2 f''(t)}{\rm Denominator} , \nonumber\\
	& \hat z=  \frac{4 z f'(t)^3}{\rm Denominator} , \qquad {\rm Denominator= 4 f'(t)^2-z^2 f''(t)^2} .
	\label{roberts}
\end{align}
It is easy to check that the metric \eq{PoincareAdS} transforms into \eq{AAdS} under \eq{roberts}, and that the boundary curves also transform into each other.

%\pagebreak

\paragraph{A topological remark:} It is important to note that although $f(t)$ is a one-to-one diffeomorphism acting on the real line,\footnote{We do not consider here the situation in which the {\it domain} of the map is a proper subset of the real line, reflecting the physics of the SYK model.} it need not be "onto", i.e. its image can be a proper (open) subset $U$ of ${\bf R}$. In Figure \ref{fig:topology}, the case (A) represents $f:{\bf R} \to {\bf R}$, while in the other cases we have $f:{\bf R} \to U \subset {\bf R}$. In case (B), $U=(-\infty, \hat t_{\infty})$, in case (C) $U=(-\hat t_{-\infty}, \infty)$, whereas in case (D), $U=(-\hat t_{-\infty}, \hat t_\infty)$. It is important to note that in order for $f(t)$ to asymptote to a finite value as $t \to \infty$, e.g. in (B), we must have $f'(t) \to 0$, as $t \to \infty$, i.e. $f'(\infty) =0$. In (C), we will similarly have $f'(-\infty)=0$ whereas in (D) we will have $f'(\infty) =0, f'(-\infty) =0$. In fact, the converse is also trivially true; if $f'(t) \to 0$ as $t$ goes to $\infty$ and/or $-\infty$, the image of $f(t)$ will be an open subset.
%\begin{figure}[H]
%\centering\includegraphics[scale=.5]{horizon.pdf} 
%\caption{Various topological possibilties of $f(t)$. In panels (b,c,d) $f:{\bf R} -> U$ where $U$ is a proper open subset of $R$.}
%\label{fig:topology}
%\end{figure}
\begin{figure}[H]
	\begin{minipage}{0.24\textwidth}
		\centering
		\begin{tikzpicture}[scale=0.4, every node/.style={transform shape}]
			\begin{axis}[xmin=-5, xmax=10, ymin=-200, ymax=700, xlabel=\textbf{\huge $t$}, ylabel=\textbf{\huge $f(t)$}]
				\addplot[domain=-5:10, samples=500, smooth, thick] {(x-1)^3 + 2*tanh(x)};
			\end{axis}
			\node[] at (3.5,-2) {\huge (A)};
		\end{tikzpicture}
		%\subcaption{f1}
	\end{minipage}
	\begin{minipage}{0.24\textwidth}
		\centering
		\begin{tikzpicture}[scale=0.4, every node/.style={transform shape}]
			\begin{axis}[xmin=-1, xmax=5, ymin=-2, ymax=2, xlabel=\textbf{\huge $t$}, ylabel=\textbf{\huge $f(t)$}]
				\addplot[domain=-1:1, samples=200, smooth, thick] {(x-1)};
				\addplot[domain=1:5, samples=200, smooth, thick] {tanh(x-1)};
				\addplot[domain=-1:5, smooth, dashed, gray] {1};
			\end{axis}
			\node[below] at (1,5.1) {\LARGE $\hat t_{\infty}$};
			\node[] at (3.5,-2) {\huge (B)};
		\end{tikzpicture}
		%\subcaption{f2}
	\end{minipage}
	\begin{minipage}{0.24\textwidth}
		\centering
		\begin{tikzpicture}[scale=0.4, every node/.style={transform shape}]
			\begin{axis}[xmin=-3, xmax=3, ymin=-2, ymax=2, xlabel=\textbf{\huge $t$}, ylabel=\textbf{\huge $f(t)$}]
				\addplot[domain=1:3, samples=200, smooth, thick] {(x-1)};
				\addplot[domain=-3:1, samples=200, smooth, thick] {tanh(x-1)};
				\addplot[domain=-3:3, smooth, dashed, gray] {-1};
			\end{axis}
			\node[] at (6,1) {\LARGE $\hat t_{-\infty}$};
			\node[] at (3.5,-2) {\huge (C)};
		\end{tikzpicture}
		%\subcaption{f3}
	\end{minipage}
	\begin{minipage}{0.24\textwidth}
		\centering
		\begin{tikzpicture}[scale=0.4, every node/.style={transform shape}]
			%\draw[help lines, very thin, blue] (0,0) grid (7,6);
			\begin{axis}[xmin=-3, xmax=5, ymin=-2, ymax=2, xlabel=\textbf{\huge $t$}, ylabel=\textbf{\huge $f(t)$}]
				\addplot[domain=-3:5, smooth, dashed, gray] {-1};
				\addplot[domain=-3:5, samples=200, smooth, thick] {tanh(x-1)};
				\addplot[domain=-3:5, smooth, dashed, gray] {1};
			\end{axis}
			\node[] at (6,1) {\LARGE $\hat t_{-\infty}$};
			\node[below] at (1,5.1) {\LARGE $\hat t_{\infty}$};
			\node[] at (3.5,-2) {\huge (D)};
		\end{tikzpicture}
		%\subcaption{f4}
	\end{minipage}
	\caption{Various topological possibilities of $f(t)$. In panels (B,C,D) $f:{\bf R}\rightarrow U$ where $U$ is a proper open subset of $R$.}
	\label{fig:topology}
\end{figure}
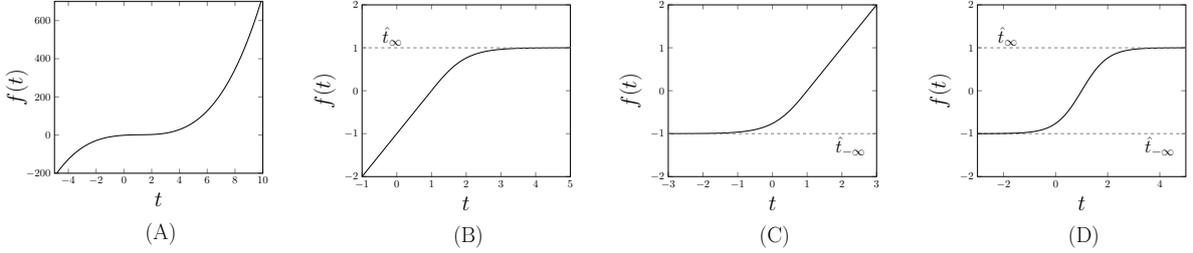
Examples of these classes:\\
(A) $f(t) = a(t-b)^{2n+1} + c\tanh(t)$, $n=0,1,2,...$ \\
(B) $f(t)= a(t-b),$ if $t\le b$, $=\tanh(a(t-b))$ if $t>b$ \\
(C) $f(t)= a(t-b),$ if $t>b$, $=\tanh(a(t-b))$ if $t\le b$ \\
(D) $f(t) = \tanh(a(t-b))$.\\
It is straightforward to argue, using these examples as well as generally,\footnote{The general proof involves two steps: (i) to locate the null line \eq{null} which ends up at $z=0$ at $t=\infty$. It is of the form $\hat t \pm \hat z= c$. If $c=\infty$, the null line coincides with the Poincare horizon. If $c<\infty$, it corresponds to the null line that last hits the boundary in terms of the $t$-clock. (ii) This null line is the boundary of the $(z,t)$ coordinates, since the null line has a coordinate singularity where the Denominator in \eq{roberts} vanishes. Note the similarity with the Schwarzschild geometry. \label{ftnt:general-pf}} that the image of the bulk diffeomorphism \eq{roberts} is the full space ${\bf R^2_+}= \{\hat z>0, \hat t \in {\bf R}\}$ in case (A), where in the other cases, the image of the map is a proper open subset $U_{\rm bulk}$ (see Figure \ref{fig:cutout}).\footnote{We should note that the {\it domain} of the map \eq{roberts} may not be the full range ${\bf R^2_+}= \{z>0, t \in {\bf R}\}$ in situations where the Denominator in \eq{roberts} vanishes. E.g the map may have singularities. E.g. for $f(t)= t^3$, the map is well-defined for $0 < z < t$, demarcating a {\it domain} that ends at the horizon. See footnote \ref{ftnt:general-pf}.} 

%\begin{figure}[H]
%	\centering\includegraphics[scale=.5]{cutout.pdf} 
%	\caption{The yellow regions are the open subsets $U_{\rm bulk}$, defined as the image of the map \eq{roberts}, in the four cases mentioned in Fig. \ref{fig:topology}. The boundary curve $z=\delta$ is shown in the $\hat t, \hat z$ coordinates in each case. In case (a), the boundary curve can have intermediate points where it touches $\hat z =0$ (e.g. for the choice $f(t)= a t^3$.) . The hallmark of appearance of a horizon is when the subset $U_{\rm bulk}$ is a {\bf proper} subset. The null boundary/boundaries of $U_{\rm bulk}$ turn out to be horizons. The labels ${\cal H}_+, {\cal H_-} $ represent future and past horizons respectively.}
%	\label{fig:cutout}
%\end{figure}
\begin{figure}[]
	\begin{minipage}{0.24\textwidth}
		\centering
		\begin{tikzpicture}[scale=0.8, every node/.style={transform shape}]
			%\draw[help lines, very thin, gray] (0,-2) grid (2,2);
			\fill[yellow] (0,0) -- (2,-2) -- (2,2) -- (0,0);
			\draw[thick] (0,0) -- (2,-2) -- (2,2) -- (0,0);
			\draw[red, thick] (2,-2) .. controls (1.5,-0.5) and (1.5,0.5) .. (2,2); 
			%\draw[red, thick] (2,-2) .. controls (1.7,-1) .. (2,0);
			%\draw[red, thick] (2,0) .. controls (1.7,1) .. (2,2);
			\node[right] at (1.5,-2.5) {(A)};
			\node[] at(2.5,1) {$\hat z=0$};
			\node[] at(0.2,1) {$\hat z=\infty$};
		\end{tikzpicture}
	\end{minipage}
	%\hspace{0.02\textwidth}
	\begin{minipage}{0.24\textwidth}
		\centering
		\begin{tikzpicture}[scale=0.8, every node/.style={transform shape}]
			%\draw[help lines, very thin, gray] (0,-2) grid (2,2);
			\fill[yellow] (0.5,-0.5) -- (2,-2) -- (2,1) -- (0.5,-0.5);
			\draw[thick] (0,0) -- (2,-2) -- (2,2) -- (0,0);
			\draw[thick] (2,1) -- (0.5,-0.5);
			\draw[red, thick] (2,-2) .. controls (1.6,-0.5) .. (2,1);
			\node[right] at (1.5,-2.5) {(B)};
			\node[] at (2.5,1) {$\hat z=0$};
			\node[] at (0.2,1) {$\hat z=\infty$};
			\node[above] at (1.5,0.5) {$\mathcal{H^+}$};
		\end{tikzpicture}
	\end{minipage}
	%\hspace{0.02\textwidth}
	\begin{minipage}{0.24\textwidth}
		\centering
		\begin{tikzpicture}[scale=0.8, every node/.style={transform shape}]
			%\draw[help lines, very thin, gray] (0,-2) grid (2,2);
			\fill[yellow] (0.5,0.5) -- (2,-1) -- (2,2) -- (0.5,0.5);
			\draw[thick] (0,0) -- (2,-2) -- (2,2) -- (0,0);
			\draw[thick] (2,-1) -- (0.5,0.5);
			\draw[red, thick] (2,-1) .. controls (1.6,0.5) .. (2,2);
			\node[right] at (1.5,-2.5) {(C)};
			\node[] at (2.5,1) {$\hat z=0$};
			\node[] at (0.2,1) {$\hat z=\infty$};
			\node[below] at (1.5,-0.5) {$\mathcal{H^-}$};
		\end{tikzpicture}
	\end{minipage}
	%\hspace{0.02\textwidth}
	\begin{minipage}{0.24\textwidth}
		\centering
		\begin{tikzpicture}[scale=0.8, every node/.style={transform shape}]
			%\draw[help lines, very thin, gray] (0,-2) grid (2,2);
			\fill[yellow] (1,0) -- (2,-1) -- (2,1) -- (1,0);
			\draw[thick] (0,0) -- (2,-2) -- (2,2) -- (0,0);
			\draw[thick] (2,1) -- (1,0) -- (2,-1);
			\draw[red, thick] (2,-1) .. controls (1.7,0) .. (2,1);
			\node[right] at (1.5,-2.5) {(D)};
			\node[] at(2.5,1) {$\hat z=0$};
			\node[] at(0.2,1) {$\hat z=\infty$};
			\node[above] at (1.5,0.5) {$\mathcal{H^+}$};
			\node[below] at (1.5,-0.5) {$\mathcal{H^-}$};
		\end{tikzpicture}
	\end{minipage}
	\caption{The yellow regions are the open subsets $U_{\rm bulk}$, defined as the image of the map \protect\eqref{roberts}, in the four cases mentioned in Figure \ref{fig:topology}. The boundary curve $z=\delta$ is shown in the $\hat t, \hat z$ coordinates in each case. The hallmark of appearance of a horizon is when the subset $U_{\rm bulk}$ is a {\bf proper} subset. The null boundary/boundaries of $U_{\rm bulk}$ turn out to be horizons. The labels $\mathcal{H}^+, \mathcal{H}^- $ represent future and past horizons respectively.} 
	\label{fig:cutout}
\end{figure}
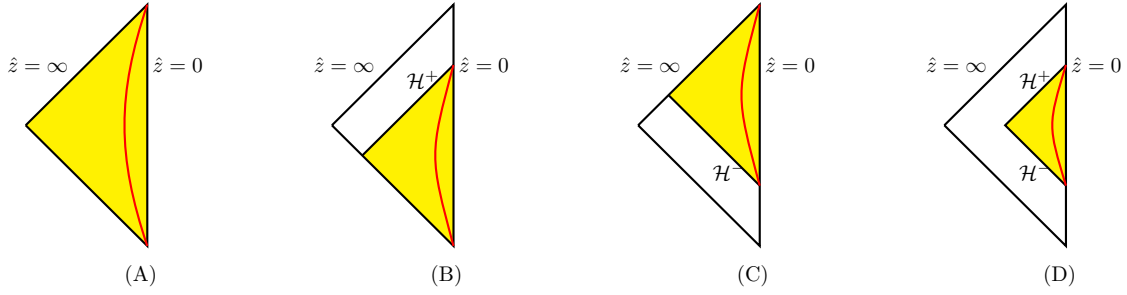
	
\subsection{Condition for existence of a horizon and how to locate it}\label{app:horizon}
	
The null lines in our geometry are obviously given by 
\begin{align}
	& \hat t \pm \hat z = c, \hbox{or equivalently} \nonumber\\
	& G_\pm = \frac{2 z^2 f'(t)^2 f''(t)}{4 f'(t)^2-z^2 f''(t)^2}+f(t) \pm  \frac{4 z f'(t)^3}{4 f'(t)^2-z^2 f''(t)^2} - c=0 .
\label{null}
\end{align}
The top line is the obvious representation; the second line represents the same surfaces in the coordinates $(t,z)$,  using \eq{roberts}. 
It can be explicitly verified that the null surface condition $(\del_t G)^2 g^{tt} +
2 \del_t G \del_z G g^{tz} + (\del_z G)^2 g^{zz} =0$ is satisfied for $G=G_\pm$, as it must.

It is easy to see that the boundaries of the subset $U_{\rm bulk}$ are null lines. In case (A), these are just the Poincar\'e horizons. In the other cases,  when the subset $U_{\rm bulk}$ is a proper open subset, the boundaries involve horizons. E.g. in case (B), the upper boundary is the null line $\hat t + \hat z = c$, with $c= \hat t_\infty \equiv f(t=\infty)$; the  value of $c$ can be determined from the fact that the this is the last line of $U_{\rm bullk}$ must meet the $z=0$ line at the boundary value of the $t$-coordinate, i.e. $t=\infty$.  It is obvious that this null line is the horizon ${\cal H}_+$. To elaborate this, we note that the yellow regions in Figure \ref{fig:cutout}, covered by the $(t,z)$ coordinates, are analogous to the region covered by the exterior Schwarzschild  coordinates $(t,r>2m)$ whose boundary $r=2m$ corresponds to $t=\infty$. The white region in panel (B) cannot send signals into the yellow region; hence the boundary of the yellow region is the horizon. The horizon in this case is a ``future" horizon (hence denoted as ${\cal H}_+$). By a similar reasoning, case (C) where $f(t)$ reaches as a finite value as $t\to -\infty$ (ensured by $f'(-\infty)=0$), has a horizon which is a past horizon (one cannot past evolve from the white region to the yellow region). In (D), we have both a future horizon and a past horizon. 

An alternative way of constructing the horizon is the following (see \cite{Kourkoulou:2017zaj} for a detailed discussion).  Consider the behaviour of the boundary curve (image of $z=\delta$) in the $(\hat t, \hat z)$ coordinates, namely
	\begin{equation}
		(\hat t(t),\hat z(t)) = (f(t), \delta\, f'(t))
		\label{boundary-hat}
	\end{equation}
which can be obtained by plugging in $z=\delta$ for in \eq{roberts} and taking $\delta$ small. The entire lifetime  of the boundary observer $-\infty < t < \infty$ is represented by the full extent of this boundary curve. In the case (B) of Figure \ref{fig:cutout}, the condition that $\hat t$ ends at a finite value $\hat t_\infty$ as $t\to \infty$ is the same as $\dot f(\infty)=0$ (see the analysis in Figure \ref{fig:topology}) and the related comments). This is equivalent to saying that the the boundary curve \eq{boundary-hat} must asymptotically reach $\hat z=0$. 

\paragraph{Condition for existence of a horizon} 
Thus, from both viewpoints the condition of existence of a future horizon is that in the limit $t\to\infty$, $f'(t) \to 0$, and $f(t)$ reaches a finite value. This is the condition we apply in Section \ref{sec:horizon}.

Note that the above condition is gauge invariant, i.e it remains unchanged if one makes a transformation $f(t) \to \tilde f(t)=\frac{a f(t) + b}{c f(t) + d}, \; ad-bc=1$. 

\paragraph{Equation of the horizon}
As explained above, the equation of the future horizon is $\hat t + \hat z = c$, with $c= \hat t_\infty \equiv f(t=\infty)$. Unlike the condition above, this equation depends on the SL(2,R) gauge, since the value of $f(t=\infty)$ changes under the SL(2,R) transformation mentioned above. In the language of the bulk, this SL(2,R) transformation reflects the isometry of the Poincar\'e metric \eq{PoincareAdS}.

\end{section}

\end{appendices}
		
%\enlargethispage{10pt}
%{\normalsize 
\bibliography{references.bib}

\providecommand{\href}[2]{#2}\begingroup\raggedright\begin{thebibliography}{10}

\bibitem{Sachdev:1992fk}
S.~Sachdev and J.-w. Ye, \emph{{Gapless spin fluid ground state in a random,
  quantum Heisenberg magnet}},
  \href{https://doi.org/10.1103/PhysRevLett.70.3339}{\emph{Phys. Rev. Lett.}
  {\bfseries 70} (1993) 3339},
  [\href{https://arxiv.org/abs/cond-mat/9212030}{{\ttfamily
  cond-mat/9212030}}].

\bibitem{Kitaev-talks:2015}
A.~Kitaev, \emph{{`A simple model of quantum holography', Talks at KITP, April
  7,and May 27, 2015}}, .

\bibitem{Maldacena:2016hyu}
J.~Maldacena and D.~Stanford, \emph{{Remarks on the Sachdev-Ye-Kitaev model}},
  \href{https://doi.org/10.1103/PhysRevD.94.106002}{\emph{Phys. Rev. D}
  {\bfseries 94} (2016) 106002},
  [\href{https://arxiv.org/abs/1604.07818}{{\ttfamily 1604.07818}}].

\bibitem{Sachdev:2015efa}
S.~Sachdev, \emph{{Bekenstein-Hawking Entropy and Strange Metals}},
  \href{https://doi.org/10.1103/PhysRevX.5.041025}{\emph{Phys. Rev. X}
  {\bfseries 5} (2015) 041025},
  [\href{https://arxiv.org/abs/1506.05111}{{\ttfamily 1506.05111}}].

\bibitem{Maldacena:2016upp}
J.~Maldacena, D.~Stanford and Z.~Yang, \emph{{Conformal symmetry and its
  breaking in two dimensional Nearly Anti-de-Sitter space}},
  \href{https://doi.org/10.1093/ptep/ptw124}{\emph{PTEP} {\bfseries 2016}
  (2016) 12C104}, [\href{https://arxiv.org/abs/1606.01857}{{\ttfamily
  1606.01857}}].

\bibitem{Sarosi:2017ykf}
G.~S\'arosi, \emph{{AdS$_{2}$ holography and the SYK model}},
  \href{https://doi.org/10.22323/1.323.0001}{\emph{PoS} {\bfseries Modave2017}
  (2018) 001}, [\href{https://arxiv.org/abs/1711.08482}{{\ttfamily
  1711.08482}}].

\bibitem{Trunin:2020vwy}
D.~A. Trunin, \emph{{Pedagogical introduction to the
  Sachdev\textendash{}Ye\textendash{}Kitaev model and two-dimensional dilaton
  gravity}}, \href{https://doi.org/10.3367/UFNe.2020.06.038805}{\emph{Usp. Fiz.
  Nauk} {\bfseries 191} (2021) 225--261},
  [\href{https://arxiv.org/abs/2002.12187}{{\ttfamily 2002.12187}}].

\bibitem{Stanford:2017thb}
D.~Stanford and E.~Witten, \emph{{Fermionic Localization of the Schwarzian
  Theory}}, \href{https://doi.org/10.1007/JHEP10(2017)008}{\emph{JHEP}
  {\bfseries 10} (2017) 008},
  [\href{https://arxiv.org/abs/1703.04612}{{\ttfamily 1703.04612}}].

\bibitem{Iliesiu:2020qvm}
L.~V. Iliesiu and G.~J. Turiaci, \emph{{The statistical mechanics of
  near-extremal black holes}},
  \href{https://doi.org/10.1007/JHEP05(2021)145}{\emph{JHEP} {\bfseries 05}
  (2021) 145}, [\href{https://arxiv.org/abs/2003.02860}{{\ttfamily
  2003.02860}}].

\bibitem{Bekenstein:1972tm}
J.~D. Bekenstein, \emph{{Black holes and the second law}},
  \href{https://doi.org/10.1007/BF02757029}{\emph{Lett. Nuovo Cim.} {\bfseries
  4} (1972) 737--740}.

\bibitem{Bekenstein:1973ur}
J.~D. Bekenstein, \emph{{Black holes and entropy}},
  \href{https://doi.org/10.1103/PhysRevD.7.2333}{\emph{Phys. Rev. D} {\bfseries
  7} (1973) 2333--2346}.

\bibitem{Hawking:1974rv}
S.~W. Hawking, \emph{{Black hole explosions}},
  \href{https://doi.org/10.1038/248030a0}{\emph{Nature} {\bfseries 248} (1974)
  30--31}.

\bibitem{Hawking:1975vcx}
S.~W. Hawking, \emph{{Particle Creation by Black Holes}},
  \href{https://doi.org/10.1007/BF02345020}{\emph{Commun. Math. Phys.}
  {\bfseries 43} (1975) 199--220}.

\bibitem{Strominger:1996sh}
A.~Strominger and C.~Vafa, \emph{{Microscopic origin of the Bekenstein-Hawking
  entropy}}, \href{https://doi.org/10.1016/0370-2693(96)00345-0}{\emph{Phys.
  Lett. B} {\bfseries 379} (1996) 99--104},
  [\href{https://arxiv.org/abs/hep-th/9601029}{{\ttfamily hep-th/9601029}}].

\bibitem{David:2002wn}
J.~R. David, G.~Mandal and S.~R. Wadia, \emph{{Microscopic formulation of black
  holes in string theory}},
  \href{https://doi.org/10.1016/S0370-1573(02)00271-5}{\emph{Phys. Rept.}
  {\bfseries 369} (2002) 549--686},
  [\href{https://arxiv.org/abs/hep-th/0203048}{{\ttfamily hep-th/0203048}}].

\bibitem{Maldacena:1997re}
J.~M. Maldacena, \emph{{The Large N limit of superconformal field theories and
  supergravity}}, \href{https://doi.org/10.1023/A:1026654312961}{\emph{Adv.
  Theor. Math. Phys.} {\bfseries 2} (1998) 231--252},
  [\href{https://arxiv.org/abs/hep-th/9711200}{{\ttfamily hep-th/9711200}}].

\bibitem{Gubser:1998bc}
S.~S. Gubser, I.~R. Klebanov and A.~M. Polyakov, \emph{{Gauge theory
  correlators from noncritical string theory}},
  \href{https://doi.org/10.1016/S0370-2693(98)00377-3}{\emph{Phys. Lett. B}
  {\bfseries 428} (1998) 105--114},
  [\href{https://arxiv.org/abs/hep-th/9802109}{{\ttfamily hep-th/9802109}}].

\bibitem{Witten:1998qj}
E.~Witten, \emph{{Anti-de Sitter space and holography}},
  \href{https://doi.org/10.4310/ATMP.1998.v2.n2.a2}{\emph{Adv. Theor. Math.
  Phys.} {\bfseries 2} (1998) 253--291},
  [\href{https://arxiv.org/abs/hep-th/9802150}{{\ttfamily hep-th/9802150}}].

\bibitem{Witten:1998zw}
E.~Witten, \emph{{Anti-de Sitter space, thermal phase transition, and
  confinement in gauge theories}},
  \href{https://doi.org/10.4310/ATMP.1998.v2.n3.a3}{\emph{Adv. Theor. Math.
  Phys.} {\bfseries 2} (1998) 505--532},
  [\href{https://arxiv.org/abs/hep-th/9803131}{{\ttfamily hep-th/9803131}}].

\bibitem{Penington:2019npb}
G.~Penington, \emph{{Entanglement Wedge Reconstruction and the Information
  Paradox}}, \href{https://doi.org/10.1007/JHEP09(2020)002}{\emph{JHEP}
  {\bfseries 09} (2020) 002},
  [\href{https://arxiv.org/abs/1905.08255}{{\ttfamily 1905.08255}}].

\bibitem{Almheiri:2019psf}
A.~Almheiri, N.~Engelhardt, D.~Marolf and H.~Maxfield, \emph{{The entropy of
  bulk quantum fields and the entanglement wedge of an evaporating black
  hole}}, \href{https://doi.org/10.1007/JHEP12(2019)063}{\emph{JHEP} {\bfseries
  12} (2019) 063}, [\href{https://arxiv.org/abs/1905.08762}{{\ttfamily
  1905.08762}}].

\bibitem{Almheiri:2019hni}
A.~Almheiri, R.~Mahajan, J.~Maldacena and Y.~Zhao, \emph{{The Page curve of
  Hawking radiation from semiclassical geometry}},
  \href{https://doi.org/10.1007/JHEP03(2020)149}{\emph{JHEP} {\bfseries 03}
  (2020) 149}, [\href{https://arxiv.org/abs/1908.10996}{{\ttfamily
  1908.10996}}].

\bibitem{Almheiri:2019yqk}
A.~Almheiri, R.~Mahajan and J.~Maldacena, \emph{{Islands outside the horizon}},
   \href{https://arxiv.org/abs/1910.11077}{{\ttfamily 1910.11077}}.

\bibitem{Rozali:2019day}
M.~Rozali, J.~Sully, M.~Van~Raamsdonk, C.~Waddell and D.~Wakeham,
  \emph{{Information radiation in BCFT models of black holes}},
  \href{https://doi.org/10.1007/JHEP05(2020)004}{\emph{JHEP} {\bfseries 05}
  (2020) 004}, [\href{https://arxiv.org/abs/1910.12836}{{\ttfamily
  1910.12836}}].

\bibitem{Penington:2019kki}
G.~Penington, S.~H. Shenker, D.~Stanford and Z.~Yang, \emph{{Replica wormholes
  and the black hole interior}},
  \href{https://arxiv.org/abs/1911.11977}{{\ttfamily 1911.11977}}.

\bibitem{Almheiri:2019qdq}
A.~Almheiri, T.~Hartman, J.~Maldacena, E.~Shaghoulian and A.~Tajdini,
  \emph{{Replica Wormholes and the Entropy of Hawking Radiation}},
  \href{https://doi.org/10.1007/JHEP05(2020)013}{\emph{JHEP} {\bfseries 05}
  (2020) 013}, [\href{https://arxiv.org/abs/1911.12333}{{\ttfamily
  1911.12333}}].

\bibitem{Chen:2020jvn}
H.~Z. Chen, Z.~Fisher, J.~Hernandez, R.~C. Myers and S.-M. Ruan,
  \emph{{Evaporating Black Holes Coupled to a Thermal Bath}},
  \href{https://doi.org/10.1007/JHEP01(2021)065}{\emph{JHEP} {\bfseries 01}
  (2021) 065}, [\href{https://arxiv.org/abs/2007.11658}{{\ttfamily
  2007.11658}}].

\bibitem{Almheiri:2020cfm}
A.~Almheiri, T.~Hartman, J.~Maldacena, E.~Shaghoulian and A.~Tajdini,
  \emph{{The entropy of Hawking radiation}},
  \href{https://doi.org/10.1103/RevModPhys.93.035002}{\emph{Rev. Mod. Phys.}
  {\bfseries 93} (2021) 035002},
  [\href{https://arxiv.org/abs/2006.06872}{{\ttfamily 2006.06872}}].

\bibitem{Geng:2020qvw}
H.~Geng and A.~Karch, \emph{{Massive islands}},
  \href{https://doi.org/10.1007/JHEP09(2020)121}{\emph{JHEP} {\bfseries 09}
  (2020) 121}, [\href{https://arxiv.org/abs/2006.02438}{{\ttfamily
  2006.02438}}].

\bibitem{Geng:2021hlu}
H.~Geng, A.~Karch, C.~Perez-Pardavila, S.~Raju, L.~Randall, M.~Riojas et~al.,
  \emph{{Inconsistency of islands in theories with long-range gravity}},
  \href{https://doi.org/10.1007/JHEP01(2022)182}{\emph{JHEP} {\bfseries 01}
  (2022) 182}, [\href{https://arxiv.org/abs/2107.03390}{{\ttfamily
  2107.03390}}].

\bibitem{Krishnan:2020fer}
C.~Krishnan, \emph{{Critical Islands}},
  \href{https://doi.org/10.1007/JHEP01(2021)179}{\emph{JHEP} {\bfseries 01}
  (2021) 179}, [\href{https://arxiv.org/abs/2007.06551}{{\ttfamily
  2007.06551}}].

\bibitem{Ghosh:2021axl}
K.~Ghosh and C.~Krishnan, \emph{{Dirichlet baths and the not-so-fine-grained
  Page curve}}, \href{https://doi.org/10.1007/JHEP08(2021)119}{\emph{JHEP}
  {\bfseries 08} (2021) 119},
  [\href{https://arxiv.org/abs/2103.17253}{{\ttfamily 2103.17253}}].

\bibitem{Krishnan:2020oun}
C.~Krishnan, V.~Patil and J.~Pereira, \emph{{Page Curve and the Information
  Paradox in Flat Space}},  \href{https://arxiv.org/abs/2005.02993}{{\ttfamily
  2005.02993}}.

\bibitem{Krishnan:2021ffb}
C.~Krishnan and V.~Mohan, \emph{{Interpreting the Bulk Page Curve: A Vestige of
  Locality on Holographic Screens}},
  \href{https://arxiv.org/abs/2112.13783}{{\ttfamily 2112.13783}}.

\bibitem{Raju:2021lwh}
S.~Raju, \emph{{Failure of the split property in gravity and the information
  paradox}}, \href{https://doi.org/10.1088/1361-6382/ac482b}{\emph{Class.
  Quant. Grav.} {\bfseries 39} (2022) 064002},
  [\href{https://arxiv.org/abs/2110.05470}{{\ttfamily 2110.05470}}].

\bibitem{DeVuyst:2022bua}
J.~De~Vuyst and T.~G. Mertens, \emph{{Operational islands and black hole
  dissipation in JT gravity}},
  \href{https://arxiv.org/abs/2207.03351}{{\ttfamily 2207.03351}}.

\bibitem{Bahiru:2022oas}
E.~Bahiru, A.~Belin, K.~Papadodimas, G.~Sarosi and N.~Vardian,
  \emph{{State-dressed local operators in AdS/CFT}},
  \href{https://arxiv.org/abs/2209.06845}{{\ttfamily 2209.06845}}.

\bibitem{Almheiri:2019jqq}
A.~Almheiri, A.~Milekhin and B.~Swingle, \emph{{Universal Constraints on Energy
  Flow and SYK Thermalization}},
  \href{https://arxiv.org/abs/1912.04912}{{\ttfamily 1912.04912}}.

\bibitem{Maldacena:2019ufo}
J.~Maldacena and A.~Milekhin, \emph{{SYK wormhole formation in real time}},
  \href{https://doi.org/10.1007/JHEP04(2021)258}{\emph{JHEP} {\bfseries 04}
  (2021) 258}, [\href{https://arxiv.org/abs/1912.03276}{{\ttfamily
  1912.03276}}].

\bibitem{Chen:2020wiq}
Y.~Chen, X.-L. Qi and P.~Zhang, \emph{{Replica wormhole and information
  retrieval in the SYK model coupled to Majorana chains}},
  \href{https://doi.org/10.1007/JHEP06(2020)121}{\emph{JHEP} {\bfseries 06}
  (2020) 121}, [\href{https://arxiv.org/abs/2003.13147}{{\ttfamily
  2003.13147}}].

\bibitem{Teitelboim:1983ux}
C.~Teitelboim, \emph{{Gravitation and Hamiltonian Structure in Two Space-Time
  Dimensions}}, \href{https://doi.org/10.1016/0370-2693(83)90012-6}{\emph{Phys.
  Lett.} {\bfseries 126B} (1983) 41--45}.

\bibitem{Jackiw:1984je}
R.~Jackiw, \emph{{Lower Dimensional Gravity}},
  \href{https://doi.org/10.1016/0550-3213(85)90448-1}{\emph{Nucl. Phys.}
  {\bfseries B252} (1985) 343--356}.

\bibitem{Kourkoulou:2017zaj}
I.~Kourkoulou and J.~Maldacena, \emph{{Pure states in the SYK model and
  nearly-$AdS_2$ gravity}},  \href{https://arxiv.org/abs/1707.02325}{{\ttfamily
  1707.02325}}.

\bibitem{Engelsoy:2016xyb}
J.~Engels\"oy, T.~G. Mertens and H.~Verlinde, \emph{{An investigation of
  AdS$_{2}$ backreaction and holography}},
  \href{https://doi.org/10.1007/JHEP07(2016)139}{\emph{JHEP} {\bfseries 07}
  (2016) 139}, [\href{https://arxiv.org/abs/1606.03438}{{\ttfamily
  1606.03438}}].

\bibitem{Calabrese:2007rg}
P.~Calabrese and J.~Cardy, \emph{{Quantum Quenches in Extended Systems}},
  \href{https://doi.org/10.1088/1742-5468/2007/06/P06008}{\emph{J. Stat. Mech.}
  {\bfseries 0706} (2007) P06008},
  [\href{https://arxiv.org/abs/0704.1880}{{\ttfamily 0704.1880}}].

\bibitem{Cardy:2014rqa}
J.~Cardy, \emph{{Thermalization and Revivals after a Quantum Quench in
  Conformal Field Theory}},
  \href{https://doi.org/10.1103/PhysRevLett.112.220401}{\emph{Phys. Rev. Lett.}
  {\bfseries 112} (2014) 220401},
  [\href{https://arxiv.org/abs/1403.3040}{{\ttfamily 1403.3040}}].

\bibitem{Dhar:2018pii}
A.~Dhar, A.~Gaikwad, L.~K. Joshi, G.~Mandal and S.~R. Wadia,
  \emph{{Gravitational collapse in SYK models and Choptuik-like phenomenon}},
  \href{https://doi.org/10.1007/JHEP11(2019)067}{\emph{JHEP} {\bfseries 11}
  (2019) 067}, [\href{https://arxiv.org/abs/1812.03979}{{\ttfamily
  1812.03979}}].

\bibitem{Papadodimas:2013jku}
K.~Papadodimas and S.~Raju, \emph{{State-Dependent Bulk-Boundary Maps and Black
  Hole Complementarity}},
  \href{https://doi.org/10.1103/PhysRevD.89.086010}{\emph{Phys. Rev. D}
  {\bfseries 89} (2014) 086010},
  [\href{https://arxiv.org/abs/1310.6335}{{\ttfamily 1310.6335}}].

\bibitem{Papadodimas:2015jra}
K.~Papadodimas and S.~Raju, \emph{{Remarks on the necessity and implications of
  state-dependence in the black hole interior}},
  \href{https://doi.org/10.1103/PhysRevD.93.084049}{\emph{Phys. Rev. D}
  {\bfseries 93} (2016) 084049},
  [\href{https://arxiv.org/abs/1503.08825}{{\ttfamily 1503.08825}}].

\bibitem{Brown:2019rox}
A.~R. Brown, H.~Gharibyan, G.~Penington and L.~Susskind, \emph{{The
  Python\textquoteright{}s Lunch: geometric obstructions to decoding Hawking
  radiation}}, \href{https://doi.org/10.1007/JHEP08(2020)121}{\emph{JHEP}
  {\bfseries 08} (2020) 121},
  [\href{https://arxiv.org/abs/1912.00228}{{\ttfamily 1912.00228}}].

\bibitem{Mandal:1995qb}
G.~Mandal and S.~R. Wadia, \emph{{Black hole geometry around an elementary BPS
  string state}},
  \href{https://doi.org/10.1016/0370-2693(96)00056-1}{\emph{Phys. Lett. B}
  {\bfseries 372} (1996) 34--44},
  [\href{https://arxiv.org/abs/hep-th/9511218}{{\ttfamily hep-th/9511218}}].

\bibitem{Dhar:1996vu}
A.~Dhar, G.~Mandal and S.~R. Wadia, \emph{{Absorption versus decay of black
  holes in string theory and T symmetry}},
  \href{https://doi.org/10.1016/0370-2693(96)01127-6}{\emph{Phys. Lett. B}
  {\bfseries 388} (1996) 51--59},
  [\href{https://arxiv.org/abs/hep-th/9605234}{{\ttfamily hep-th/9605234}}].

\bibitem{Calabrese:2005in}
P.~Calabrese and J.~L. Cardy, \emph{{Evolution of entanglement entropy in
  one-dimensional systems}},
  \href{https://doi.org/10.1088/1742-5468/2005/04/P04010}{\emph{J. Stat. Mech.}
  {\bfseries 0504} (2005) P04010},
  [\href{https://arxiv.org/abs/cond-mat/0503393}{{\ttfamily
  cond-mat/0503393}}].

\bibitem{Bagrets:2016cdf}
D.~Bagrets, A.~Altland and A.~Kamenev,
  \emph{{Sachdev\textendash{}Ye\textendash{}Kitaev model as Liouville quantum
  mechanics}},
  \href{https://doi.org/10.1016/j.nuclphysb.2016.08.002}{\emph{Nucl. Phys. B}
  {\bfseries 911} (2016) 191--205},
  [\href{https://arxiv.org/abs/1607.00694}{{\ttfamily 1607.00694}}].

\bibitem{Almheiri:2018ijj}
A.~Almheiri, A.~Mousatov and M.~Shyani, \emph{{Escaping the Interiors of Pure
  Boundary-State Black Holes}},
  \href{https://arxiv.org/abs/1803.04434}{{\ttfamily 1803.04434}}.

\bibitem{Mandal:2015jla}
G.~Mandal, R.~Sinha and N.~Sorokhaibam, \emph{{Thermalization with chemical
  potentials, and higher spin black holes}},
  \href{https://doi.org/10.1007/JHEP08(2015)013}{\emph{JHEP} {\bfseries 08}
  (2015) 013}, [\href{https://arxiv.org/abs/1501.04580}{{\ttfamily
  1501.04580}}].

\bibitem{Mandal:2015kxi}
G.~Mandal, S.~Paranjape and N.~Sorokhaibam, \emph{{Thermalization in 2D
  critical quench and UV/IR mixing}},
  \href{https://doi.org/10.1007/JHEP01(2018)027}{\emph{JHEP} {\bfseries 01}
  (2018) 027}, [\href{https://arxiv.org/abs/1512.02187}{{\ttfamily
  1512.02187}}].

\bibitem{Banerjee:2019ilw}
P.~Banerjee, A.~Gaikwad, A.~Kaushal and G.~Mandal, \emph{{Quantum quench and
  thermalization to GGE in arbitrary dimensions and the odd-even effect}},
  \href{https://doi.org/10.1007/JHEP09(2020)027}{\emph{JHEP} {\bfseries 09}
  (2020) 027}, [\href{https://arxiv.org/abs/1910.02404}{{\ttfamily
  1910.02404}}].

\bibitem{Rocha:2008fe}
J.~V. Rocha, \emph{{Evaporation of large black holes in AdS: Coupling to the
  evaporon}}, \href{https://doi.org/10.1088/1126-6708/2008/08/075}{\emph{JHEP}
  {\bfseries 08} (2008) 075},
  [\href{https://arxiv.org/abs/0804.0055}{{\ttfamily 0804.0055}}].

\bibitem{kamenev2011field}
A.~Kamenev, \emph{Field theory of non-equilibrium systems}.
\newblock Cambridge University Press, 2011.

\bibitem{Haehl:2016pec}
F.~M. Haehl, R.~Loganayagam and M.~Rangamani, \emph{{Schwinger-Keldysh
  formalism. Part I: BRST symmetries and superspace}},
  \href{https://doi.org/10.1007/JHEP06(2017)069}{\emph{JHEP} {\bfseries 06}
  (2017) 069}, [\href{https://arxiv.org/abs/1610.01940}{{\ttfamily
  1610.01940}}].

\bibitem{Pawula1967GeneralizationsAE}
R.~F. Pawula, \emph{Generalizations and extensions of the fokker-
  planck-kolmogorov equations}, {\emph{IEEE Trans. Inf. Theory} {\bfseries 13}
  (1967) 33--41}.

\bibitem{Risken1979OnTA}
H.~Risken and H.~D. Vollmer, \emph{On the application of truncated generalized
  fokker-planck equations}, {\emph{Zeitschrift f{\"u}r Physik B Condensed
  Matter} {\bfseries 35} (1979) 313--315}.

\bibitem{jackson2012classical}
J.~Jackson, \emph{Classical Electrodynamics}.
\newblock Wiley, 2012.

\bibitem{Caldeira:1982iu}
A.~O. Caldeira and A.~J. Leggett, \emph{{Path integral approach to quantum
  Brownian motion}},
  \href{https://doi.org/10.1016/0378-4371(83)90013-4}{\emph{Physica A}
  {\bfseries 121} (1983) 587--616}.

\bibitem{Gao:2016bin}
P.~Gao, D.~L. Jafferis and A.~C. Wall, \emph{{Traversable Wormholes via a
  Double Trace Deformation}},
  \href{https://doi.org/10.1007/JHEP12(2017)151}{\emph{JHEP} {\bfseries 12}
  (2017) 151}, [\href{https://arxiv.org/abs/1608.05687}{{\ttfamily
  1608.05687}}].

\bibitem{Maldacena:2017axo}
J.~Maldacena, D.~Stanford and Z.~Yang, \emph{{Diving into traversable
  wormholes}}, \href{https://doi.org/10.1002/prop.201700034}{\emph{Fortsch.
  Phys.} {\bfseries 65} (2017) 1700034},
  [\href{https://arxiv.org/abs/1704.05333}{{\ttfamily 1704.05333}}].

\bibitem{Maldacena:2018lmt}
J.~Maldacena and X.-L. Qi, \emph{{Eternal traversable wormhole}},
  \href{https://arxiv.org/abs/1804.00491}{{\ttfamily 1804.00491}}.

\bibitem{pathria2011statistical}
R.~Pathria and P.~Beale, \emph{Statistical Mechanics}.
\newblock Academic Press. Butterworth-Heinemann, 2011.

\bibitem{HOBM}
O.~Contreras-Vergara, N.~Lucero-Azuara, N.~Sánchez-Salas and
  J.~Jiménez-Aquino, \emph{Harmonic oscillator brownian motion: Langevin
  approach revisited},
  \href{https://doi.org/10.31349/RevMexFisE.18.97}{\emph{Revista Mexicana de
  Física E} {\bfseries 18} (01, 2021) 97}.

\bibitem{Mandal:2017thl}
G.~Mandal, P.~Nayak and S.~R. Wadia, \emph{{Coadjoint orbit action of Virasoro
  group and two-dimensional quantum gravity dual to SYK/tensor models}},
  \href{https://doi.org/10.1007/JHEP11(2017)046}{\emph{JHEP} {\bfseries 11}
  (2017) 046}, [\href{https://arxiv.org/abs/1702.04266}{{\ttfamily
  1702.04266}}].

\bibitem{Banados:1998gg}
M.~Banados, \emph{{Three-dimensional quantum geometry and black holes}},
  \href{https://doi.org/10.1063/1.59661}{\emph{AIP Conf. Proc.} {\bfseries 484}
  (1999) 147--169}, [\href{https://arxiv.org/abs/hep-th/9901148}{{\ttfamily
  hep-th/9901148}}].

\end{thebibliography}\endgroup
\bibliographystyle{JHEP}

\end{document}